\documentclass[12pt]{article}
\usepackage{graphicx}
\usepackage{authblk}
\usepackage{verbatim}
\ifdefined\CurrentStyleIsCMP
\newcommand{\CMPorArticle}[2]{#1}

\makeatletter
 \newfont{\Bbb}{msbm10 scaled 1\@ptsize00}
 \newfont{\Bbbb}{msbm7 scaled 1\@ptsize00}
\makeatother

\else

\newcommand{\CMPorArticle}[2]{#2}

\def\hybrid{\topmargin 0pt      \oddsidemargin 0pt
        \headheight 0pt \headsep 0pt
        \voffset=-0.5cm
        \textwidth 6.25in       %
        \textheight 9.5in       %
        \marginparwidth 0.0in
        \parskip 5pt plus 1pt   \jot = 1.5ex}
\catcode`\@=11
\def\marginnote#1{}

\newcount\hour
\newcount\minute
\newtoks\amorpm
\hour=\time\divide\hour by60
\minute=\time{\multiply\hour by60 \global\advance\minute by-\hour}
\edef\standardtime{{\ifnum\hour<12 \global\amorpm={am}%
        \else\global\amorpm={pm}\advance\hour by-12 \fi
        \ifnum\hour=0 \hour=12 \fi
        \number\hour:\ifnum\minute<10 0\fi\number\minute\the\amorpm}}
\edef\militarytime{\number\hour:\ifnum\minute<10 0\fi\number\minute}

\def\draftlabel#1{{\@bsphack\if@filesw {\let\thepage\relax
   \xdef\@gtempa{\write\@auxout{\string
      \newlabel{#1}{{\@currentlabel}{\thepage}}}}}\@gtempa
   \if@nobreak \ifvmode\nobreak\fi\fi\fi\@esphack}
        \gdef\@eqnlabel{#1}}
\def\@eqnlabel{}
\def\@vacuum{}
\def\draftmarginnote#1{\marginpar{\raggedright\scriptsize\tt#1}}
\def\draftlabel#1{{\@bsphack\if@filesw {\let\thepage\relax
   \xdef\@gtempa{\write\@auxout{\string
      \newlabel{#1}{{\@currentlabel}{\thepage}}}}}\@gtempa
   \if@nobreak \ifvmode\nobreak\fi\fi\fi\@esphack}
        \gdef\@eqnlabel{#1}}
\def\@eqnlabel{}
\def\@vacuum{}
\def\draftmarginnote#1{\marginpar{\raggedright\scriptsize\tt#1}}

\def\draft{\oddsidemargin -.5truein
        \def\@oddfoot{\sl preliminary draft \hfil
        \rm\thepage\hfil\sl\today\quad\militarytime}
        \let\@evenfoot\@oddfoot \overfullrule 3pt
        \let\label=\draftlabel
        \let\marginnote=\draftmarginnote
   \def\@eqnnum{(\theequation)\rlap{\kern\marginparsep\tt\@eqnlabel}%
\global\let\@eqnlabel\@vacuum}  }

\def\numberbysection{\@addtoreset{equation}{section}
        \def\theequation{\thesection.\arabic{equation}}}

\def\underline#1{\relax\ifmmode\@@underline#1\else
        $\@@underline{\hbox{#1}}$\relax\fi}

\def\titlepage{\@restonecolfalse\if@twocolumn\@restonecoltrue\onecolumn
     \else \newpage \fi \thispagestyle{empty}\c@page\z@
        \def\thefootnote{\fnsymbol{footnote}} }

\def\endtitlepage{\if@restonecol\twocolumn \else  \fi
        \def\thefootnote{\arabic{footnote}}
        \setcounter{footnote}{0}}  %
\relax

\numberbysection
\hybrid

\newfont{\Bbb}{msbm10 scaled 1\@ptsize00}
\newfont{\Bbbb}{msbm7 scaled 1\@ptsize00}

\fi

\newcommand{\I}{\mathbb{I}}
\newcommand{\CC}{\mathbb{C}}

\newcommand{\DDD}{\raise-1pt\hbox{$\mbox{\Bbbb D}$}}

\newcommand{\UUU}{\raise-1pt\hbox{$\mbox{\Bbbb U}$}}

\newcommand{\ZZ}{\mathbb{Z}}
\newcommand{\z}{\raise-1pt\hbox{$\mbox{\Bbbb Z}$}}
\def\clap#1{\hbox to 0pt{\hss#1\hss}}

\def\mathclap{\mathpalette\mathclapinternal}

\def\mathclapinternal#1#2{%
\clap{$\mathsurround=0pt#1{#2}$}}

\def\beq{\begin{equation}}
\def\eeq{\end{equation}}
\def\p{\partial}

\def\s{\sigma}

\def\<{\langle}
\def\>{\rangle}

\def\cP{{\cal P}}
\def\hD{\hat D}

\def\i2{\frac{i}{2}}
\def\tr{{\rm tr~}}
\def\hcD{\hat{\cal D}}

\def\psistar{\psi^{*}}

\def\normord{ {\scriptstyle {{\bullet}\atop{\bullet}}} }
\def\normordbare{ {\scriptstyle {{ \times}\atop{ \times}}} }

\def\lbr{\left <}
\def\rbr{\right >}

\def\lvac{\left <0\right |}
\def\rvac{\left |0\right >}

\def\lvacn{\left <n\right |}
\def\rvacn{\left |n\right >}

\usepackage{amssymb}
\usepackage{amsbsy}
\usepackage{amsmath}
\usepackage[utf8]{inputenc}

\usepackage{varioref}

\usepackage{hyperref}

\date{December 2011}

\begin{document}

\title{Classical tau-function for quantum spin chains}

\CMPorArticle{
\author{Alexander Alexandrov\inst{1}\fnmsep\inst{2}\fnmsep
  \thanks{E-mail:  {\tt alexandrovsash at gmail.com}},Vladimir Kazakov
  \inst{3}\fnmsep\inst{4}\fnmsep\thanks{member of Institut
    Universitaire de France; E-mail:  {\tt kazakov at lpt.ens.fr }},
  Sebastien Leurent \inst{4}\fnmsep\thanks{E-mail:  {\tt
      sebastien.leurent at normalesup.org}}, Zengo Tsuboi \inst{5}\fnmsep\thanks{an additional post member at Osaka City University Advanced Mathematical Institute (until 31 March 2012);
E-mail:  {\tt ztsuboi at yahoo.co.jp}
}, Anton Zabrodin\inst{2}\fnmsep\inst{6}\fnmsep\inst{7}\fnmsep \thanks{E-mail:  {\tt zabrodin at itep.ru}}}

\institute{Mathematisches Institut, Albert-Ludwigs-Universit\"{a}t,
Freiburg, Germany
\and ITEP, Bol. Cheremushkinskaya str. 25, 117259 Moscow, Russia
\and
Université Paris 6, Place Jussieu, 75005
  Paris, France
\and
\'Ecole Normale Supérieure, 45 rue d'Ulm,
  75005 Paris, France
\and
Institut f\"{u}r Mathematik und Institut f\"{u}r Physik,
Humboldt-Universit\"{a}t zu Berlin,
Johann von Neumann-Haus,
Rudower Chaussee 25, 12489 Berlin, Germany
\and 
Institute of Biochemical Physics,
Kosygina str. 4, 119991 Moscow, Russia
\and
NRU-HSE, Vavilova str. 7, 117312 Moscow, Russia}

\authorrunning{A. Alexandrov, V. Kazakov, S. Leurent, Z.Tsuboi \& A. Zabrodin}

}{

\begin{titlepage}

\author[1,2]{Alexander Alexandrov \thanks{E-mail:  {\tt alexandrovsash at gmail.com}}}
\author[3,4]{Vladimir Kazakov \thanks{member of Institut Universitaire de France; E-mail:  {\tt kazakov at lpt.ens.fr }}}
\author[4]{Sebastien Leurent \thanks{E-mail:  {\tt sebastien.leurent at normalesup.org}}}
\author[5]{Zengo Tsuboi \thanks{an additional post member at Osaka City University Advanced Mathematical Institute (until 31 March 2012);
E-mail:  {\tt ztsuboi at yahoo.co.jp}
}}
\author[2,6,7]{Anton Zabrodin \thanks{E-mail:  {\tt zabrodin at itep.ru}}}

\affil[1]{\small{Mathematisches Institut, Albert-Ludwigs-Universit\"{a}t,
Freiburg, Germany} }

\affil[2]{ITEP, Bol. Cheremushkinskaya str. 25, 117259 Moscow, Russia}
\affil[3]{Université Paris 6, Place Jussieu, 75005
  Paris, France}
\affil[4]{\'Ecole Normale Supérieure, 45 rue d'Ulm,
  75005 Paris, France}
\affil[5]{Institut f\"{u}r Mathematik und Institut f\"{u}r Physik,
Humboldt-Universit\"{a}t zu Berlin,
Johann von Neumann-Haus,
Rudower Chaussee 25, 12489 Berlin, Germany}
\affil[6]{Institute of Biochemical Physics,
Kosygina str. 4, 119991 Moscow, Russia}
\affil[7]{NRU-HSE, Vavilova str. 7, 117312 Moscow, Russia}

}

\maketitle

\begin{abstract}

For an arbitrary generalized quantum integrable spin chain we introduce
a ``master \(T\)-operator'' which represents a generating
function for commuting quantum transfer matrices
constructed by means of the fusion procedure in the
auxiliary space.
We show that the functional relations for
the transfer matrices are equivalent to an infinite set of
model-independent bilinear equations of the Hirota form
for the master \(T\)-operator, which allows one
to identify it with \(\tau\)-function of an integrable
hierarchy of classical soliton equations.
In this paper we consider
spin chains with rational \(GL(N)\)-invariant \(R\)-matrices but
the result is independent of a particular functional form of
the transfer matrices and
directly applies to quantum integrable models
with more general (trigonometric
and elliptic) \(R\)-matrices and to supersymmetric spin chains.

\end{abstract}

Report number: HU-Mathematik-2011-25, HU-EP-11/59,
ITEP-TH-49/11, LPT-ENS-12/17

\CMPorArticle{}{

\vfill

\end{titlepage}
}

\newpage

\tableofcontents

\section{Introduction}

The aim of this paper is to make precise the long anticipated
connection between transfer matrices of quantum integrable models
and \(\tau\)-functions of classical integrable hierarchies of
non-linear partial differential equations. We show that
properly defined generating function for commuting Hamiltonians
of quantum models is a \(\tau\)-function in the sense that it
obeys exactly the same bilinear
identities of the Hirota form \cite{Hirota81,Miwa82}
as the classical \(\tau\)-function does.
Since the operator-valued generating functions commute for all
values of the auxiliary parameters (one of which
being the familiar spectral
parameter), there is no problem with their ordering in non-linear
equations.

A similarity between quantum transfer matrices and classical
\(\tau\)-functions was first pointed out in \cite{KLWZ97}
(see also \cite{Z97}),
where a discrete integrable dynamics in the space of
integrals of motion of a quantum integrable model was introduced.
This classical dynamics was identified with
the discrete 3-term Hirota equation with special
boundary conditions.
It is equivalent to the so-called \(T\)-system
which is a system of fusion relations among
mutually commuting transfer matrices for bosonic \cite{Pearce:1991ty}
or supersymmetric \cite{Tsuboi:1997iq} quantum integrable spin chains.
Solutions to the Hirota equation carry
full information about spectral properties
of the quantum model. The diagonalization of transfer matrices
by means of the nested Bethe
ansatz technique was shown to be
equivalent to an ``undressing'' chain of B\"acklund
transformations for the discrete Hirota equation.
Later this approach was extended to
supersymmetric integrable models \cite{KSZ08}.

An important further step was recently made in \cite{KLT10},
where an operator realization of the B\"acklund flow describing
the ``undressing'' process was suggested and a commuting family of
transfer matrices (\(T\)-operators on different levels of the
nested Bethe ansatz)
depending on a number of discrete variables
(labels of representation in the auxiliary space) and on the
spectral parameter was constructed. These \(T\)-operators
obey the discrete Hirota equation
not only for usual but also for
supersymmetric integrable models, as it was directly demonstrated in
\cite{Kazakov:2007na} for \(GL(M|N)\)-invariant
spin chains.

The question which remained open after these works was how to
embed these \(T\)-operators depending on discrete variables
into an infinite classical integrable
hierarchy with continuous time flows
compatible with
the discrete ones.
In the present paper we suggest an answer which is
rather simple and natural.
Below in the introduction we describe it in a concise form.

\begin{figure}
\centering
\CMPorArticle{
\includegraphics[scale=1.0]{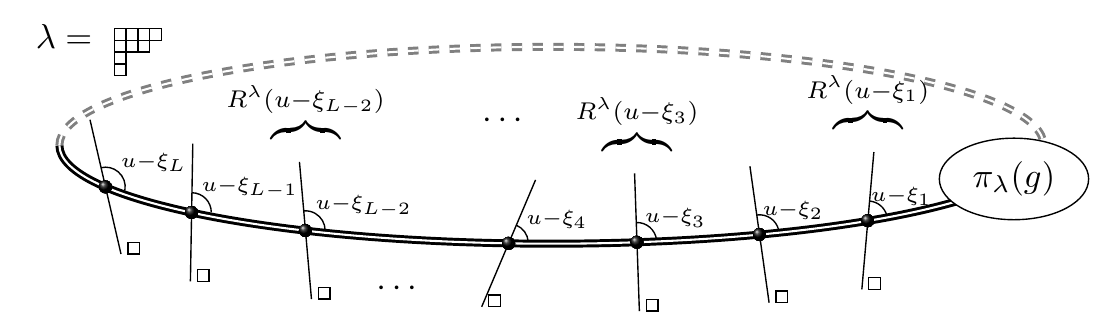}
}{
\includegraphics[scale=1.3]{figTop}}
\caption{T-operator for an integrable, inhomogeneous spin chain with
  twisted boundary condition.}
 \label{fig:Top}
\end{figure}

In this paper we
consider generalized quantum integrable spin chains with
rational \(GL(N)\)-invariant \(R\)-matrix \cite{KRS81}
\begin{equation}\label{QT101}
R^{\lambda}(u)=u \, \I \otimes \I
+\sum_{i,j=1}^{N}e_{ji}\otimes \pi_{\lambda}(e_{ij})
\end{equation}
which acts in the tensor product of the
\(N\)-dimensional vector representation \( \pi_{\Box}\)
and an arbitrary finite dimensional irreducible representation
\(\pi_{\lambda}\) of the group \(GL(N)\)
labeled by a Young diagram \(\lambda\).
Here \(e_{ij}\) are standard generators
of the algebra \(gl(N)\), \(\I \) is 
the unit matrix and \(u\in \CC\) is the spectral parameter.
A family of commuting operators
(quantum transfer matrices or simply \(T\)-operators), 
shown in figure \ref{fig:Top}, can be constructed as 
\begin{equation}\label{QT201}
T^{\lambda }(u)=\mbox{tr}_{\pi_\lambda}\left (
R^{\lambda}(u-\xi_L)\otimes \ldots \otimes R^{\lambda}(u-\xi_1)
(\I^{\otimes L} \otimes \pi_{\lambda}(g))\right ),
\end{equation}
where \(\xi_i\) are arbitrary parameters (inhomogeneities at the lattice sites)
and \(g\in GL(N)\) is called the twist matrix. 
The tensor product is taken over the quantum space (in the first 
space of \eqref{QT101}). 
The trace is taken in the
auxiliary space where the representation \(\pi_{\lambda}\)
is realized.
The \(T\)-operators act in the physical Hilbert space of the model
\({\cal H}=(\CC ^N)^{\otimes L}\) which is the tensor product of
\(L\) local spaces \(\CC ^N\) of the vector representations.
Formally, our setting includes also models with higher representations
at the sites because they can be obtained by fusing several vector
representations at the sites with properly chosen parameters
\(\xi_i\).
The Yang-Baxter equation for the
\(R\)-matrix implies that the \(T\)-operators with the same \(g\)
commute for all \(u\) and \(\lambda\) and can be diagonalized
simultaneously. They 
obey some functional relations which are given by
the Cherednik-Bazhanov-Reshetikhin (CBR) determinant formulas
\cite{Chered,BR90}:
\begin{equation}\label{tau4b-intro}
T^{\lambda}(u)
=\displaystyle{
\Bigl ( \prod_{k=1}^{\lambda_{1}'-1}
T^{\emptyset}(u\! -\! k)\Bigr )^{-1}\!\!
\det_{i,j =1,\ldots , \lambda_1^{\prime}}
T^{(\lambda_i  - i + j)}(u\! -\! j\! +\! 1)},
\end{equation}
where \(\lambda'_{1}\) is the height of the first column
of the Young diagram \(\lambda\) and \( \emptyset \) is the empty
diagram.
These relations mean that
the transfer matrices for arbitrary diagrams functionally depend
on the transfer matrices corresponding to the 1-row diagrams.

We call such models ``spin chains'' in a rather broad sense,
not implying existence of any local Hamiltonian of the Heisenberg type
(integrable local interactions in general do
not exist for inhomogeneous spin chains).
However, even in the general case of arbitrary parameters \(\xi _i\)
the model still makes sense as a generalized spin chain with
non-local interactions. The ``spin variables'' are
vectors from the spaces \(\CC ^N\) at each site.
Instead,
one may prefer to keep in mind integrable lattice models of statistical
mechanics rather than spin chains as such.
In either case the final goal of the theory is the diagonalization of the
transfer matrices of the type
(\ref{QT201}) which is usually achieved by the nested
Bethe ansatz method in one or another form
(see, e.g., \cite{Betheansatz}).

The key point of our approach is dealing with all the \(T\)-operators
simultaneously by
introducing their generating function of a special form
which we will call {\it the master \(T\)-operator}.
This generating function depends on an infinite number of
auxiliary parameters \({\bf t}=\{t_1, t_2, \ldots \}\)
which we call times\footnote{In a different form such an
operator was introduced in \cite{KLT10}}:
\begin{equation}\label{master-intro}
T(u,{\bf t})=\sum_{\lambda}s_{\lambda}({\bf t})T^{\lambda }(u).
\end{equation}
Here \(s_{\lambda}({\bf t})\) are the standard Schur functions \eqref{schur2}
and the sum goes over all Young diagrams \(\lambda\) including the
empty diagram.
By construction, the master
\(T\)-operators commute for different values of the
times:
 \begin{equation}\label{commT}
 [T(u,{\bf t}),\, T(u',{\bf t'})]=0.
\end{equation}

Our claim is that
{\it the master \(T\)-operator is the
\(\tau\)-function of the KP hierarchy}
with the times \(t_1, t_2, \ldots \, \).
In other words,
the transfer matrices \(T^{\lambda }(u)\)
are Pl\"ucker coordinates for the
element of the infinite dimensional
Grassmann manifold corresponding to the \(\tau\)-function \(T(u,{\bf t})\).
This means that
\(T(u,{\bf t})\) obeys the Hirota bilinear equation
\begin{equation}\label{bi2a01}
\begin{array}{l}
(z_2-z_3)T\left (u, {\bf t}+[z_{1}^{-1}]\right )T\left (u,
{\bf t} +[z_{2}^{-1}]+[z_{3}^{-1}]\right )
\\ \\
\quad +\, (z_3-z_1)T\left (u, {\bf t}+[z_{2}^{-1}]\right )T\left (u,
{\bf t} +[z_{1}^{-1}]+[z_{3}^{-1}]\right )
\\ \\
\quad \quad
+\, (z_1-z_2)T\left (u, {\bf t}+[z_{3}^{-1}]\right )T\left (u,
{\bf t} +[z_{1}^{-1}]+[z_{2}^{-1}]\right ) =0
\end{array}
\end{equation}
with respect to the \({\bf t}\)-variables with fixed \(u\).
Here the standard notation
\({\bf t}+[z^{-1}]=\{ t_1 +z^{-1},
t_2 +\frac{1}{2}\, z^{-2}, t_3 +\frac{1}{3}\, z^{-3}, \ldots \, \}\)
is used. This functional relation encodes infinitely many
partial differential equations for \(T\) obtained after expanding
the left hand side in inverse powers of \(z_i\).

Moreover,
when one incorporates the \(u\)-dependence
of the master \(T\)-operator, {\it the KP \(\tau\)-function
extends to the
\(\tau\)-function of the
modified KP (MKP) hierarchy}
with the times \(t_0=u, t_1, t_2, \ldots \, \).
This means that
\(T(u,{\bf t})\) obeys another Hirota bilinear equation
\begin{equation}\label{bi3a01}
\begin{array}{c}
z_2 T\left (u+1, {\bf t}+[z_{1}^{-1}]\right )
T\left (u, {\bf t}+[z_{2}^{-1}]\right )
-z_1 T\left (u+1, {\bf t}+[z_{2}^{-1}]\right )
T\left (u, {\bf t}+[z_{1}^{-1}]\right )
\\ \\
+\,\, (z_1 -z_2)T\left (u+1, {\bf t}+[z_{1}^{-1}] +[z_{2}^{-1}]\right )
T\Bigl (u,{\bf t} \Bigr )=0
\end{array}
\end{equation}
which encodes infinitely many differential-difference equations
of the MKP hierarchy.
In fact this statement is very general and is
independent of a specific functional form of
\(T^{\lambda }(u)\). It also does not depend on
particular features of
the quantum models in question: they can be
lattice or continuous, bosonic or supersymmetric, with rational,
trigonometric or elliptic \(R\)-matrices, etc.

Most part of the paper is devoted to the proof of this statement.
Our technical tools are the realization of the KP hierarchy
as a dynamical system on an infinite dimensional Grassmann manifold
\cite{Sato81} and the free fermionic construction of
the \(\tau\)-function \cite{DJKM83,JM83}
developed by the Kyoto school. The main idea of the proof
is to identify the generating series (\ref{master-intro}) for the
master \(T\)-operator with the Schur function expansion of the
\(\tau \)-function \(\tau _n ({\bf t})\) of the MKP hierarchy:
\begin{equation}\label{tau1-intro}
\tau_n ({\bf t})=\sum_{\lambda} s_{\lambda}({\bf t}) c_{\lambda}(n).
\end{equation}
The key fact proven in section 3 with the use of
the free fermionic formalism is that the coefficients
\(c_{\lambda}(n)\) (Pl\"ucker coordinates) obey the determinant
relations of the form
\begin{equation}\label{tau4a-intro}
c_{\lambda}(n)
=\displaystyle{
\Bigl ( \prod_{k=1}^{\lambda_{1}'-1}
c_{\emptyset}(n\! -\! k)\Bigr )^{-1}\!\!
\det_{i,j =1,\ldots , \lambda_1^{\prime}}
c_{(\lambda_i \! -\! i\! +\! j)}(n\! -\! j\! +\! 1)}.
\end{equation}
They mean that
the Pl\"ucker coordinates for arbitrary diagrams are expressed
through the basic ones, corresponding to the 1-row diagrams.
Identifying \(n\) with the spectral parameter \(u\)
and \(c_{\lambda}(n)\) with the \(T^{\lambda}(u)\),
one can see that (\ref{tau4b-intro}) coincides with
(\ref{tau4a-intro}). So, the statement that the master
\(T\)-operator is the \(\tau \)-function
appears to be equivalent to the
Cherednik-Bazhanov-Reshetikhin (CBR) determinant relations
for \(T^{\lambda }(u)\).

For spin chains with
rational \(R\)-matrices, one can say more.
In this case, the master \(T\)-operator has an
explicit realization in terms of the co-derivative
introduced in \cite{Kazakov:2007na}. It was also remarked
in \cite{Kazakov:2007na} that the
basic bilinear identity
for the \(gl(M|N) \)-characters\footnote{known as a particular
solution to the
discrete version of the KdV equation}
and their co-derivatives, which was the key part of the direct proof
of the CBR formulas, might be related to the
Hirota equation for a classical integrable hierarchy.
Indeed, now we can see that the ``master identity'' from \cite{KLT10}
is a particular case of
the bilinear equation (\ref{bi3a01}) for the master \(T\)-operator.
Therefore, the master \(T\)-operator contains all Baxter 
\(Q\)-operators and \(T\)-operators on all levels of the
nested Bethe ansatz. Presumably, this is also true for
models with other types of \(R\)-matrices.
The most important characteristic property of the master \(T\)-operators
for finite spin chains with rational \(R\)-matrices is
that they are polynomials in \(u\):
\begin{equation}\label{dyn101}
T(u,{\bf t})\propto \prod_{j=1}^{L}(u-u_j({\bf t})).
\end{equation}
The class of \(\tau\)-functions
with this property is well studied in the theory of soliton
equations and
can be characterized \cite{Krichever78,Krichever83}
in terms of algebraic geometry of curves with cusp
singularities.
Furthermore, the dynamics of
the roots \(u_j\) as functions of times \(t_i\), is
known \cite{KZ95} to be given by
the rational Ruijsenaars-Schneider model.

Sections 2-4 of the paper contain the necessary
material on the classical KP and MKP hierarchies
and their \(\tau\)-functions.
Although the main result can be proved without using the
formalism of free fermions, we found it instructive to
present all arguments in the fermionic operator language of
the Kyoto school.
In section 2 the free fermion operators and the vacuum states
are introduced and the vacuum expectation values are defined.
In section 3 the \(\tau\)-function is introduced as the vacuum
expectation value of a group-like element and bilinear
identities for it are derived.
In most cases we use the standard notation and conventions
from \cite{DJKM83,JM83}.
Section 4 is devoted to a detailed review of rational solutions
to the MKP hierarchy. It is precisely the class of solutions
relevant to quantum integrable models with rational \(R\)-matrices.
The core of the paper is section 5, where the master \(T\)-operator
is introduced and shown to satisfy the classical
bilinear identities and Hirota equations.

\section{Free fermions}

In this section we recall the formalism of
free fermions \cite{DJKM83,JM83}
used for construction of \(\tau\)-functions of integrable hierarchies.

\subsection{Fermionic operators}

Let \(\psi_n , \psistar_{n}\), \(n\in \ZZ\), be free fermionic
operators with usual anticommutation relations
\([\psi_n , \psi_m ]_+ = [\psistar_n, \psistar_m]_+=0\),
\([\psi_n , \psistar_m]_+=\delta_{mn}\). They generate
an infinite dimensional Clifford algebra. We also use their
generating series
\begin{equation}\label{ferm0}
\psi (z)=\sum_{k\in \z}\psi_k z^k, \quad \quad
\psistar (z)=\sum_{k\in \z}\psistar_k z^{-k}
\end{equation}
which can be regarded as free fermionic fields in the
complex plane of the variable \(z\).

Of particular
importance is the operator
\begin{equation}\label{ferm2}
J_+ = \sum_{k\geq 1}t_k J_k, \quad \quad
J_k =\sum_{j\in \z}\psi_j \psistar_{j+k},
\end{equation}
where \(t_k\) are arbitrary parameters (called times).
It is convenient to denote the set of times
by \({\bf t} = \{t_1, t_2 , \ldots \}\)
and to introduce their generating function
\[
\xi ({\bf t}, z)=\sum_{k\geq 1}t_{k}z^k\,.
\]

It is easy to check that the fields \(\psi (z)\), \(\psistar (z)\)
transform diagonally under the adjoint action
of the operator
\(e^{J_+}\):
\begin{equation}\label{ferm3}
e^{J_{+}}\psi (z)e^{-J_{+}}=e^{\xi ({\bf t} , z)}\psi (z)\,,
\quad \quad
e^{J_+}\psistar (z)e^{-J_{+}}=e^{-\xi ({\bf t},
z)}\psistar (z).
\end{equation}
In terms of the symmetric  Schur polynomials \(h_k({\bf t})\) defined by
\begin{equation}\label{schur1}
e^{\xi ({\bf t} ,\, z)}=\sum_{k\geq 0}h_k ({\bf t})z^k
\end{equation}
the corresponding formulas for \(\psi_n , \psistar_{n}\)
can be written as
\begin{equation}\label{ferm4}
e^{J_{+}}\psi _n e^{-J_{+}}=
\sum_{k\geq 0}\psi_{n- k}h_k({\bf t})\,, \quad \quad
e^{J_{+}}\psistar _n e^{-J_{+}}=
\sum_{k\geq 0}\psistar_{n+ k}h_k(-{\bf t}).
\end{equation}

\subsection{Dirac vacua and excited states}

Next, we introduce a vacuum state \(\left |0\rbr\) which is
a ``Dirac sea'' where all negative mode states are empty
and all positive ones are occupied:
\[
\psi_n \rvac =0, \quad n< 0; \quad \quad \quad
\psistar_n \rvac =0, \quad n\geq 0.
\]
(For brevity, we call the modes with indices \(n\geq 0\) as  {\it positive}).
With respect to this vacuum, the operators \(\psi_n\) with
\(n<0\) and \(\psistar_n\) with \(n\geq 0\) are annihilation operators
while the operators \(\psistar_n\) with \(n<0\) and
\(\psi_n\) with \(n\geq 0\) are creation operators.
The dual vacuum state has the properties
\[
\lvac \psistar_n  =0, \quad n< 0; \quad \quad \quad
\lvac \psi_n  =0, \quad n\geq 0.
\]
We also need the ``shifted'' Dirac vacua \(\rvacn\) and the
dual vacua \(\lvacn\)
defined as
\[
\rvacn = \left \{
\begin{array}{l}
\psi_{n-1}\ldots \psi_1 \psi_0 \rvac , \,\,\,\,\, n> 0
\\ \\
\psistar_n \ldots \psistar_{-2}\psistar_{-1}\rvac , \,\,\,\,\, n<0
\end{array} \right.
\]
\[
\lvacn = \left \{
\begin{array}{l}
\lvac \psistar_{0}\psistar_{1}\ldots \psistar_{n-1} , \,\,\,\,\, n> 0
\\ \\
\lvac \psi_{-1}\psi_{-2}\ldots \psi_{n} , \,\,\,\,\, n<0
\end{array} \right.
\]
Similarly to the properties of the Dirac vacuum,
for the shifted Dirac vacuum we have:
\begin{align}
 \psi_m \rvacn &=0, \quad m < n;
\qquad
\psistar_m \rvacn =0, \quad m \ge n,
\label{actfock1}
\\
\lvacn  \psi_{m}&=0 , \quad m \ge n;
\qquad
\lvacn  \psistar_{m}=0 , \quad m < n.
\end{align}
We will also use
\begin{align}
\psi_n \rvacn &= \left|n+1 \rbr,
\qquad  \psistar_n \left|n+1 \rbr = \left|n \rbr,
\\
 \lbr n+1 \right|\psi_n &= \lbr n \right|
\qquad
\lbr n \right|\psistar_n = \lbr n+1 \right| .
\label{actfock2}
\end{align}

Excited states are obtained by filling some empty states
and creating some holes. Let us introduce a convenient basis
of states in the fermionic Fock space \({\cal F}\).
The basis states \(\left |\lambda , n\rbr\) are
parameterized by Young diagrams \(\lambda\)
and numbers \(n\) of the Dirac vacua in the following way.
Given a Young diagram \(\lambda =
(\lambda_1 , \ldots , \lambda_{\ell})\) with \(\ell =\ell (\lambda )\)
nonzero rows, let
\((\vec \alpha |\vec \beta )=(\alpha_1, \ldots , \alpha_{d(\lambda )}|
\beta_1 , \ldots , \beta_{d(\lambda )})\) be the Frobenius notation
for the diagram \(\lambda\) \cite{Macdonald}.
Here \(d(\lambda )\) is the number of
boxes in the main diagonal and \(\alpha_i =\lambda_i -i\),
\(\beta_i =\lambda'_i -i\), where \(\lambda'\) is the transposed
(reflected about the main diagonal) diagram \(\lambda\). In other words,
\(\alpha_i\) is the length of the part of the
\(i\)-th row to the right from the
main diagonal and \(\beta_i\) is the length of the part of the \(i\)-th column
under the main diagonal (not counting the diagonal box). Then
\begin{equation}\label{lambda1}
\begin{array}{l}
\left |\lambda , n\rbr :=
\psistar_{n-\beta_1 -1}\ldots \psistar_{n-\beta_{d(\lambda )}\! -1}\,
\psi_{n+\alpha_{d(\lambda )}}\ldots \psi_{n+\alpha_1}\rvacn
\\ \\
\lbr \lambda , n \right |:=
\lvacn \psistar_{n+\alpha_1}\ldots \psistar_{n+\alpha_{d(\lambda )}}\,
\psi_{n-\beta_{d(\lambda )}\! -1}\ldots \psi_{n-\beta_1 -1}.
\end{array}
\end{equation}

The definition of the vacua
implies that \(e^{J_+}\rvacn =\rvacn\) while the dual coherent
states \(\lvacn e^{J_+}\) are expanded
as linear combinations of the basis vectors as follows:
\begin{equation}\label{lambda2}
\lvacn e^{J_+({\bf t})}=\sum_{\lambda} (-1)^{b(\lambda )}
s_{\lambda}({\bf t})\lbr \lambda , n \right |,
\end{equation}
where the sum runs over all Young diagrams \(\lambda\) including the
empty diagram,
\beq\label{boflambda}
b(\lambda )=\sum_{i=1}^{d(\lambda )}(\beta_i +1)
\eeq
and \(s_{\lambda}({\bf t})\) is the Schur polynomial
corresponding to the diagram \(\lambda \). It can be expressed
through the functions \(h_k({\bf t})\) defined in (\ref{schur1}) (which are
Schur polynomials for one-row diagrams) with the help of the
Jacobi-Trudi formula \cite{Macdonald}:
\begin{equation}\label{schur2}
s_{\lambda}({\bf t})=\det_{i,j=1, \ldots , \ell (\lambda )}
h_{\lambda_i -i +j}({\bf t}).
\end{equation}

\subsection{The expectation values and normal ordering}

The vacuum expectation value \(\lvacn \ldots \rvacn\) is a
Hermitian linear form on the Clifford algebra defined
on bilinear combinations of fermions
by the properties \(\left. \lvacn \! n\right > =1\),
\(\lvacn \psi_i \psi_j\rvacn = \lvacn \psistar_i \psistar_j \rvacn =0\)
for all \(i,j\) and
\[
\lvacn \psi_i \psistar_j\rvacn =\delta_{ij}\quad
\mbox{for \(j<n\)}, \quad \quad
\lvacn \psi_i \psistar_j\rvacn =0 \quad \mbox{for \(j\geq n\)}.
\]
One can see that the basis vectors (\ref{lambda1})
are orthogonal with respect to the
scalar product induced by the expectation value:
\[
\lbr \lambda , n\right | \left. \mu , m \rbr =
\delta_{mn}\delta_{\lambda \mu}.
\]

Expectation values of general products of
fermionic operators are given by the Wick's theorem.
Let \(v_i =\sum_j v_{ij}\psi_j\) be linear combinations
of \(\psi_j\)'s only and
\(w_i^*=\sum_j w^{*}_{ij}\psistar_j\)
be linear combinations of \(\psistar_j\)'s only,
then the standard Wick's theorem states that
\[
\lvacn v_1 \ldots v_M w_M^* \ldots w_1^* \rvacn =
\det_{i,j =1,\ldots , M}\lvacn v_i w_j^* \rvacn .
\]

One may define the normal ordering \(\normord (\ldots )\normord \)
with respect
to the Dirac vacuum \(\rvac\): all annihilation operators
are moved to the right and all creation operators are moved to
the left taking into account their anti-commutativity under mutual
permutations. For example,
\(\normord  \psistar_{1}\psi_{1}\normord =
-\psi_{1} \psistar_{1}=\psistar_{1}\psi_{1} -1\),
and
\(\normord \psistar_m \psi_n\normord =
\psistar_m \psi_n -\lvac \psistar_m \psi_n \rvac\).
More generally, for any linear combinations \(f_0, f_1,
\ldots , f_m\) of the fermion operators \(\psi_i, \psistar_j\)
we have the recursive formula
\beq\label{normord}
f_0 \normord  f_1 f_2 \ldots
f_m  \normord =\normord
f_0 f_1 f_2 \ldots  f_m \normord
+\sum_{j=1}^{m}(-1)^{j-1}
\lvac f_0 f_j \rvac \normord f_1 f_2 \ldots
\not{\!\!f}_{\!\!j} \ldots f_m\normord
\eeq
where \(\not{\!\!f}_{\!\!j}\) means that this factor should be omitted.

In fact a normal ordering can be defined with respect to
any vacuum state.
Another useful normal ordering is the one defined with respect to the
bare vacuum \( \left |\infty \right >\), which is the
absolutely empty state. We denote this normal ordering by
\(\normordbare (\ldots )\normordbare \). It means
that all \(\psistar \)'s are moved to the left
and all \(\psi \)'s to the right, with taking into account
the sign factors appeared each time one operator is permuted
with another. For example,
\(\normordbare  \psistar_{m}\psi_{n}\normordbare =
\psistar_{m}\psi_{n}\), \(\normordbare \psi_{n}\psistar_{m}\normordbare =
-\psistar_{m}\psi_{n}\) which is equivalent to
\(\normordbare  \psi_n \psistar_m\normordbare =
 \psi_n \psistar_m - 
\left < \infty \right |  \psi_n \psistar_m
\left | \infty \right >\). 
In general, for this normal ordering
equation (\ref{normord}) holds for the vacuum expectation taken
with respect to the bare vacuum.

\subsection{Group-like elements}

Neutral bilinear combinations
\(\sum_{mn} b_{mn}\psistar_m \psi_n\)
of the fermions, with certain conditions
on the matrix \( b = (b_{mn}) \), generate an
infinite-dimensional Lie algebra \cite{JM83}.
Exponentiating these expressions, one obtains
an infinite dimensional group (a version
of \(GL(\infty )\)).
Elements of this group can be represented
in the form
\begin{equation}\label{gl}
G=\exp \Bigl (\sum_{i, k \in {\mathbb Z}}b_{ik}\psistar_i \psi_k\Bigr )
\end{equation}
The inverse element can be written in the same way with the matrix
 \( (-b_{mn} ) \).
An important example
is provided by the operator \(e^{J_{+}}\) (\ref{ferm2}).

It is straightforward to see that the group elements of the form
(\ref{gl}) obey a rather special property that the adjoint
action of such elements preserves the linear space spanned
by the fermion operators \(\psi _n\) and the dual space
spanned by \(\psistar _n\). More precisely, we have:
\beq\label{rotation}
\psi_n G =  \sum_l R_{nl} G\psi_l\,, \quad \quad
G\psistar_n =\sum_{l} R_{ln} \psistar_l  G\,,
\eeq
where the matrix
\(R = (R_{nl}) \) of the induced linear transformation
is given by \(R=e^b\).

Let us note,
following the works
of the Kyoto school \cite{JM83,SMJ}, that the group elements
can be equivalently represented
as normal ordered exponents of bilinear forms.
For example, \(G\) given in (\ref{gl}) can be written as
\begin{equation}\label{ferm1}
G=
\normordbare \exp \Bigl (\sum_{i,k \in {\mathbb Z}}
B_{ik}\psistar_i \psi_k\Bigr )
\normordbare
=\det (I\! +\! P_+ B)\,
\normord \exp \Bigl (\sum_{i,k \in {\mathbb Z}}
A_{ik}\psistar_i \psi_k\Bigr )
\normord
\end{equation}
with the matrices \(A\), \(B\) determined by
the matrix \(b\) in (\ref{gl}) according to the formulas \cite{SMJ}
\beq\label{matrices}
B=e^b -I\,, \quad  \quad B-A =AP_+ B\,.
\eeq
Here \(I\) is the unity
matrix and \(P_+\) is the projector on the positive mode space
(\((P_+)_{ik}=\delta_{ik}\) for \(i,k\geq 0\) and 0 otherwise).
The product of two group-like elements is also a group-like
element:
\[
\normordbare \exp \Bigl (\sum_{i,k \in {\mathbb Z}}
A_{ik}\psistar_i \psi_k\Bigr )
\normordbare
\normordbare \exp \Bigl (\sum_{i,k \in {\mathbb Z}}
B_{ik}\psistar_i \psi_k\Bigr )
\normordbare =
\normordbare \exp \Bigl (\sum_{i,k \in {\mathbb Z}}
C_{ik}\psistar_i \psi_k\Bigr )
\normordbare ,
\]
where \(C=A+B+AB\).
As a slight generalization of (\ref{ferm1}),
one can obtain the formulas
\begin{equation}\label{group02a}
\lvacn
\normordbare e^{\sum_{ik} B_{ik}\psistar_i \psi_k}
\normordbare \rvacn =
\det \left (I+ B^{(n)} \right ),
\end{equation}
where \( B^{(n)}=(B_{ik})_{i, k \geq n } \) is the
half-infinite matrix obtained from \(B\)
by truncation up to  the
\(n\)-th row and \(n\)-th column and
\begin{equation}\label{group02}
\lvacn \psistar_s
\normordbare e^{\sum_{ik} B_{ik}\psistar_i \psi_k}
\normordbare \psi_r \rvacn
=\left \{ \begin{array}{l}
\det \left (I+ B^{(n)} \right )
\left [ \left (1+ B^{(n)} \right )^{-1}\right ]_{r, s}
\,\,\, \mbox{if}\,\, r,s\geq n
\\ \\
0\quad \mbox{otherwise}.
\end{array} \right.
\end{equation}
Note that the r.h.s of the last formula is
the \((rs)\)-minor of the matrix \(I+B^{(n)}\) (up to a sign).

The general theory of normal ordering and the
group elements in the
the Clifford algebra was given in \cite{SMJ}.
What is important here, is that the normal ordering allows one
to represent in the form (\ref{ferm1}) not only the group elements but
also some non-invertible elements of the Clifford algebra that satisfy
the same commutation relations (\ref{rotation}) with some matrix \(R\).

We call the elements \(G\) of the Clifford algebra, such that
the commutation relations (\ref{rotation}) hold
with some (not necessarily invertible) matrix \(R\), the 
{\it group-like elements}. If the matrix
\(R\) fails to be invertible, so does \(G\), as an element
of the Clifford algebra. In this case it can not be represented
in the exponential form (\ref{gl}). However, in general
it still admits a representation as a {\it normal ordered} exponent of
 bilinear forms in the fermionic operators.

Let us give an example. Let \(\Psi \), \(\Phi^*\) be arbitrary
linear combinations of the fermion operators \(\psi_n\),
\(\psistar_n\), respectively, and consider the element
\[
G=e^{\beta \Phi^* \Psi}=
\normordbare e^{\alpha \Phi^* \Psi}\normordbare
= 1+\alpha \Phi^* \Psi = 1+\alpha \gamma -\alpha \Psi \Phi^* ,
\]
where
\(\gamma := \lbr \infty \right | \Psi \Phi^* \left |\infty \rbr \)
and \(\alpha \), \(\beta \) are related as
\(e^{\gamma \beta} =1+\gamma \alpha \).
For general values of \(\alpha \) (and for
\(\gamma \neq 0\)) this element is invertible and the
two representations are equivalent. However, at \( \alpha =-1/\gamma \)
the invertibility breaks down and the element \(G\) becomes
\[
G= \frac{\Psi \, \Phi^*}{\lbr \infty
\right | \! \Psi \Phi^* \! \left |\infty \rbr}.
\]
which can not be written in the exponential form (\ref{gl}) but
can be represented as the normally ordered exponent.

From (\ref{rotation}) it easily follows that
any group-like element obeys
the commutation relation
\begin{align}
\sum_{k \in {\mathbb Z}} \psi_{k} G \otimes  \psi_{k}^{*} G =
\sum_{k \in {\mathbb Z}}G\psi_{k} \otimes G \psi_{k}^{*}
\label{commute}
\end{align}
which is equivalent to
the bilinear identity
\begin{align}
\sum_{k \in {\mathbb Z}} \lbr U \right|  \psi_{k} G \left|V \rbr
 \lbr U^{\prime} \right|  \psistar_{k} G \left|V^{\prime} \rbr =
\sum_{k \in {\mathbb Z}} \lbr U \right| G \psi_{k} \left|V \rbr
 \lbr U^{\prime} \right| G \psistar_{k} \left|V^{\prime} \rbr .
 \label{bilinear-fermi}
\end{align}
valid for any states \(\left|V \rbr , \left|V^{\prime} \rbr\),
\(\lbr U \right|,  \lbr U^{\prime} \right|\)
from the space
\({\cal F}\) and its dual. This is the basic identity for deriving
various bilinear equations for the \(\tau \)-function (see below).

\subsection{The generalized Wick's theorem}

Let \(v_i =\sum_j v_{ij}\psi_j\) be
linear combinations of \(\psi_j\)'s only and
\(w_i^*=\sum_j w^{*}_{ij}\psistar_j\)
be linear combinations of \(\psistar_j\)'s only,
as before.
A more general version of the Wick's theorem given in
\cite{DJKM83} is
\begin{equation}\label{Wick1}
\frac{\lvacn v_1 \ldots v_m w^{*}_m \ldots w^{*}_1 G\rvacn }{\lvacn
G\rvacn }=\det_{i,j =1,\ldots , m}
\frac{\lvacn v_j w^{*}_iG\rvacn }{\lvacn
G\rvacn },
\end{equation}
where \(G\) is any group-like element.
Here we imply that \(\lvacn G\rvacn \neq 0\) for all \(n\)
but in fact this restriction can be removed by multiplying
the both sides by an appropriate power of \(\lvacn G\rvacn \).
The similar formula
with the exchange \(v_j \leftrightarrow w_j\) also holds.
This statement can be proved by induction using the bilinear identity
(\ref{bilinear-fermi}) (see Appendix A).
The following particular case
is often useful in applications. Writing
\(\lbr n-m \right | = \lvacn \psi_{n-1}\ldots \psi_{n-m}\)
with \(m>0\),
we get from (\ref{Wick1}):
\begin{equation}\label{Wick2}
\frac{\lbr n-m \right | w^{*}_m \ldots w^{*}_1 G\rvacn }{\lvacn
G\rvacn }=\det_{i,j =1,\ldots , m}
\frac{\lvacn \psi_{n-j} w^{*}_i G\rvacn }{\lvacn
G\rvacn }.
\end{equation}

For our purpose, it is important to
note that there exists an alternative determinant
representation of the expectation value in the l.h.s. of
(\ref{Wick2}), which is another form of the generalized
Wick's theorem:
\begin{equation}\label{Wick3}
\frac{\lbr n-m \right | w^{*}_m \ldots w^{*}_1 G\rvacn }{\lvacn
G\rvacn }=\det_{i,j =1,\ldots , m}
\frac{\lbr n\! -\! j\right | w^{*}_i G\left | n\! -\! j\! +\! 1
\rbr }{\lbr n\! -\! j\! +\! 1\right |
G \left | n\! -\! j\! +\! 1\rbr }.
\end{equation}
Indeed, the first columns of the two matrices in the r.h.s.
of (\ref{Wick2}) and (\ref{Wick3}) coincide and one can show that
the \(j\)-th column of the matrix in (\ref{Wick2}) is equal to
the \(j\)-th column of
the matrix in (\ref{Wick3}) plus a linear combination of the
first \(j-1\) columns (see Appendix A). We also note an 
equivalent form of (\ref{Wick3}):
\begin{equation}\label{Wick3aa}
\frac{\lbr n \right | Gv_1 \ldots v_m 
\left |n-m\right > }{\lvacn G\rvacn }=\det_{i,j =1,\ldots , m}
\frac{\lbr n\! -\! j\! +\! 1\right | Gv_i \left | n\! -\! j
\rbr }{\lbr n\! -\! j\! +\! 1\right |
G \left | n\! -\! j\! +\! 1\rbr }.
\end{equation}
which is obtained from it by conjugation of the states and 
operators.

\section{The \texorpdfstring{$\tau$}{tau}-function and the Baker-Akhiezer functions}

As it has been established in the works of the Kyoto school,
the expectation values of group-like elements are tau-functions
of integrable hierarchies of nonlinear differential equations.
In particular,
\begin{equation}\label{ferm6}
\tau ({\bf t})=\lvac e^{J_+ ({\bf t})}G\rvac
\end{equation}
is the tau-function of the KP hierarchy depending on
the times \({\bf t}=\{t_1, t_2, \ldots \}\), which means that it obeys an
infinite set of Hirota bilinear equations \cite{Hirota81,Miwa82}.
More generally, introducing a ``discrete time'' \(n\) via
\begin{equation}\label{ferm6a}
\tau_n ({\bf t})=\lvacn e^{J_+ ({\bf t})}G\rvacn
\end{equation}
one obtains a \(\tau\)-function which solves the modified KP (MKP)
hierarchy
\cite{Kupershmidt,Tak01,TakTeo06}\footnote{In \cite{TakTeo06}
a slightly more general version, called there
a coupled modified KP (cMKP) hierarchy, was considered.}.
Equations of the MKP hierarchy are differential-difference equations
which include shifts of the variable \(n\).
At each fixed \(n\), the \(\tau\)-function (\ref{ferm6a})
is a \(\tau\)-function of the KP hierarchy, so the \(n\)-evolution
represents an infinite chain of B\"acklund transformations.
The meaning of the discrete variable \(n\) depends on the class
of solutions. For some classes of solutions \(n\) can be regarded as a continuous, or even complex, variable.
The last point will be of a prime
interest for us since in  section~\ref{sec:quantum}
\(n\) will be identified with the spectral parameter \(u\) of the quantum transfer matrix of a spin chain.

\subsection{Bilinear equations for \texorpdfstring{$\tau$}{tau}-function}

\subsubsection{The bilinear identity}

The \(\tau\)-function obeys a bilinear identity which is
a direct consequence of (\ref{bilinear-fermi}).
Setting \(\left|V \rbr =\rvacn\),
\(\left|V^{\prime} \rbr =\left |n'\right >\) with \(n\geq n'\),
where \(\rvacn\) and \(\left |n'\right >\) are two shifted
Dirac vacua, we have
\begin{equation}\label{B1}
\sum_{k \in {\mathbb Z}} \lbr U \right|  \psi_{k} G \left|n \rbr
\lbr U^{\prime} \right|  \psistar_{k} G \left|n' \rbr =0
\end{equation}
because either \(\psi_k\) or \(\psistar_k\) kills the state
in each term in the r.h.s. of (\ref{bilinear-fermi}).
Now, setting \(\lbr U \right| = \lbr n+1 \right | e^{J_{+}({\bf t})}\),
\(\lbr U' \right| = \lbr n'-1 \right | e^{J_{+}({\bf t'})}\),
we can write
\[
\begin{array}{lll}
0&=&\displaystyle{\sum_k
\lbr n+1 \right | e^{J_{+}({\bf t})} \psi_k G \rvacn
\lbr n'-1 \right | e^{J_{+}({\bf t'})}\psistar_k G\left |n'\right >}
\\ &&\\
&=& \displaystyle{
\mbox{res}_{z=\infty}\left [
z^{-1} \lbr n+1 \right | e^{J_{+}({\bf t})} \psi (z) G \rvacn
\lbr n'-1 \right | e^{J_{+}({\bf t'})}\psistar (z)
G\left |n'\right >\right ]}
\\ &&\\
&=& \displaystyle{
\mbox{res}_{z=\infty}\left [
e^{\xi ({\bf t}-{\bf t'},z)}z^{n-n'}
\lbr n \right | e^{J_+({\bf t}-[z^{-1}])}G\rvacn
\lbr n' \right | e^{J_+({\bf t'}+[z^{-1}])}G\left |n' \rbr \right ],}
\end{array}
\]
where we have used the commutation relations of the fermion
operators with \(e^{J_+({\bf t})}\) and the
bosonization rules
\begin{equation}\label{bosonization}
\begin{array}{l}
\lvacn \psi (z) e^{J_+({\bf t})}=z^{n-1}\left < n-1\right |
e^{J_{+}({\bf t}-[z^{-1}])}
\\ \\
\lvacn \psistar (z) e^{J_+({\bf t})}=z^{-n}\left < n+1\right |
e^{J_{+}({\bf t}+[z^{-1}])}
\end{array}
\end{equation}
established by the Kyoto school.
Here and below we use the standard notation
\[
{\bf t} \pm [z^{-1}]=\{t_1 \pm z^{-1}, \, t_2 \pm
\frac{1}{2}z^{-2}, t_3 \pm
\frac{1}{3}z^{-3}, \, \ldots \}.
\]
In this way we arrive at the
bilinear identity
\begin{equation}\label{bi1}
\oint_{{\mathcal C}} e^{\xi ({\bf t}-{\bf t'},z)}z^{n-n'}
\tau_n ({\bf t}-[z^{-1}])\tau_{n'}({\bf t'}+[z^{-1}])dz =0
\end{equation}
for the \(\tau\)-function valid for all \({\bf t}, {\bf t'}\)
and \(n\geq n'\). It encodes all the PDE's of the MKP hierarchy.

The choice of the integration contour \({\mathcal C}\)
depends on the type of the time evolution. Formally,
the evolution factor \(e^{\xi ({\bf t}-{\bf t'},z)}z^{n-n'}\)
has an essential singularity at \(\infty\), and the contour
should then encircle the whole complex plane 
(only the singularity at \(\infty\)
stays outside the contour).  This is the usual prescription and if
only a finite number of times \(t_k\) are nonzero, then it is
correct. However, in general, when the values of the times are
such that the factor \(e^{\xi ({\bf t}-{\bf t'},z)}z^{n-n'}\) has
singularities for finite values of \(z\), then the prescription is as
follows:
the contour \({\mathcal C}\)
must encircle all singularities of the
function \(\tau_n ({\bf t}-[z^{-1}])\tau_{n'}({\bf t'}+[z^{-1}])\),
leaving all the singularities of 
\(e^{\xi ({\bf t}-{\bf t'},z)}z^{n-n'}\) 
outside the contour.

\noindent
{\bf Remark.} There is a freedom to
multiply the \(\tau\)-function of the MKP hierarchy
by an exponent of any linear form of times
with constant coefficients and by an arbitrary function
of \(n\):
\begin{equation}\label{linear}
\tau_n ({\bf t})\rightarrow C(n)\exp \Bigl (
\sum_{k\geq 1}C_k t_k\Bigr )\tau_n ({\bf t}).
\end{equation}
Clearly, this transformation preserves the
form of the bilinear identity.
By noting that
\[
\normordbare
\exp \left (\sum_{j\in \z} d_j \psistar_j \psi_j \right )
\normordbare \rvacn =C(n)  \rvacn,
\quad \quad d_n =\frac{C(n)}{C(n+1)} -1,
\]
we see that the transformation 
\( \tau_n ({\bf t})\rightarrow C(n)\tau_n ({\bf t})\) means
\( G \rightarrow G \normordbare e^{
\sum_{j} d_j \psistar_j \psi_j}
\normordbare
\) for
the group-like element.
In the MKP hierarchy,
there are no restrictions on the form of \(C(n)\).
Such restrictions appear in the more general 2D Toda lattice
hierarchy, where this function is constrained to be of the form
\(C(n)=Ce^{C_0 n}\) with constant \(C, C_0\).

\subsubsection{3-term bilinear equations}

Setting \(n'=n\) and
\(t_k'=t_k-\frac{1}{k}(z_{1}^{-k}+
z_{2}^{-k}+z_{3}^{-k})\) in (\ref{bi1}), we see that the essential
singularity at \(\infty\) splits into 3 simple poles
at \(z_1, z_2, z_3\):
\[
e^{\xi ({\bf t}-{\bf t'},z)}=
\frac{1}{
(1-\frac{z}{z_1})(1-\frac{z}{z_2})(1-\frac{z}{z_3})}.
\]
Taking the residues,
we arrive at the 3-term relation
\begin{equation}\label{bi2}
(z_2-z_3)\tau _n\left ({\bf t}-[z_{1}^{-1}]\right )\tau _n
\left ({\bf t}-[z_{2}^{-1}]-[z_{3}^{-1}]\right )
+(231)+(312)=0,
\end{equation}
where the last two terms are obtained from the first one by
the cyclic permutations of the indices.
Another choice is \(n'=n\) and
\(t_k'=t_k+\frac{1}{k} z_{0}^{-k} -\frac{1}{k}(z_{1}^{-k}+
z_{2}^{-k}+z_{3}^{-k})\), then
\[
e^{\xi ({\bf t}-{\bf t'},z)}=
\frac{1-\frac{z}{z_0}}{
(1-\frac{z}{z_1})(1-\frac{z}{z_2})(1-\frac{z}{z_3})}.
\]
This leads to a slightly more general equation
\begin{equation}\label{bi201}
(z_0 -z_1)(z_2-z_3)\tau _n\left ({\bf t}-
[z_{0}^{-1}]-[z_{1}^{-1}]\right )\tau _n
\left ({\bf t}-[z_{2}^{-1}]-[z_{3}^{-1}]\right )
+(231)+(312)=0.
\end{equation}

Setting \(n'=n-1\),
\(t_k'=t_k-\frac{1}{k}(z_{1}^{-k}+
z_{2}^{-k})\), we see that the essential
singularity at \(\infty\) splits into simple poles
at \(z_1, z_2\):
\[
ze^{\xi ({\bf t}-{\bf t'},z)}=
\frac{z}{
(1-\frac{z}{z_1})(1-\frac{z}{z_2})}.
\]
Besides the residues at these points, there is also
a contribution from the residue at \(\infty\), so
we obtain the 3-term relation
\begin{equation}\label{bi3}
\begin {array}{c}
z_2\tau_{n+1}\left ({\bf t}-[z_{2}^{-1}]\right )
\tau_{n}\left ({\bf t}-[z_{1}^{-1}]\right )-
z_1\tau_{n+1}\left ({\bf t}-[z_{1}^{-1}]\right )
\tau_{n}\left ({\bf t}-[z_{2}^{-1}]\right )
\\ \\
+\, (z_1-z_2)\tau_{n+1}({\bf t})\tau_{n}
\left ({\bf t}-[z_{1}^{-1}]-[z_{2}^{-1}]\right )\, =0.
\end{array}
\end{equation}
It can be formally regarded as a particular case of
(\ref{bi201}) in the limit \(z_3\to \infty\), \(z_0\to 0\).
The limit \(z_3\to \infty\) is smooth while the other one
requires to assign a meaning to
\(\lim\limits_{z_0 \to 0} \tau_n ({\bf t}- [z_{0}^{-1}])\).
The form of the evolution multiplier, \( z^n e^{\xi ({\bf t}, z)}\),
suggests to set formally
\begin{equation}\label{adddef}
\lim_{z_0 \to 0} \tau_n ({\bf t}\pm [z_{0}^{-1}])=
\tau_{n\mp 1}({\bf t})
\end{equation}
which actually holds if the function
\(\tau_n ({\bf t}-[z^{-1}])\) for arbitrary \(n\) can be analytically
continued from a neighborhood of \(\infty \) to a
neighborhood of the point \(z=0\) and is regular there.
If it is singular, as in the case of solutions relevant to
quantum spin chains, (\ref{adddef}) should be substituted by a more general
prescription (for an example see (\ref{adddef1}) below) which, however, also allows
one to regard (\ref{bi3}) as a particular case of (\ref{bi201}).

\subsection{Examples of \texorpdfstring{$\tau$}{tau}-functions}

Here we consider some particular cases of the general
fermionic construction of \(\tau\)-functions which will be
important for us in section 5.

Let us start from a very simple example.
Set \( \Psi =\psi (p)+\alpha \psi (q)\) and consider
the operator \( \normord e^{\beta \Psi \psistar (r)}\normord\)
(with \(|r| < \mbox{min}\, (|p|, |q|)\)). 
It is a group-like
element for any \(\beta \) but
at \(\beta =\beta_0 =(\lvac \Psi \psistar (r)\rvac )^{-1}\) it is not
invertible and equals \(\beta_0 \Psi \psistar (r)\). Therefore,
\[
\tau_n ({\bf t}) = \lvacn e^{J_+({\bf t})}\Psi \psistar (r)\rvacn
\]
is a \(\tau \)-function of the MKP hierarchy.
Then we would like to consider the limit \(r\to 0\).
The limit is singular and requires some care.
From
\[
r^{n-1}
\psistar (r)\! \rvacn = \sum_l r^{n-l-1}\psistar_{l}\! \rvacn
=\sum_{l\leq n-1} r^{n-l-1}\psistar_{l}\! \rvacn
=\left | n-1\right > +O(r)
\]
we see that in order to get a well-defined limit,
one should multiply the \(\tau\)-function by \(r^{n-1}\)
before tending \(r\to 0\) (this is just a transformation
of the form (\ref{linear})). Then we come to the conclusion that
\CMPorArticle{
\begin{align*}
  \tau_n ({\bf t})&=\lvacn e^{J_+({\bf t})}\Psi \left |n-1\right > \\
 &= \lvacn e^{J_+({\bf t})} \Bigl (\psi (p)\! +\! \alpha \psi (q)\Bigr )
  \left |n\! -\! 1\right >= p^{n-1}e^{\xi ({\bf t}, p)}+\alpha
  q^{n-1}e^{\xi ({\bf t}, q)}
\end{align*}
}{
\[
\tau_n ({\bf t})=\lvacn e^{J_+({\bf t})}\Psi \left |n-1\right >
= \lvacn e^{J_+({\bf t})}
\Bigl (\psi (p)\! +\! \alpha \psi (q)\Bigr ) \left |n\! -\! 1\right >=
p^{n-1}e^{\xi ({\bf t}, p)}+\alpha q^{n-1}e^{\xi ({\bf t}, q)}
\]
}is a \(\tau\)-function.
Indeed, it is the \(\tau\)-function for the 1-soliton solution
of the MKP hierarchy.

This example can be generalized.
Fix a finite set of distinct points \(p_i \in \CC\),
\(i\in I\), and construct a (finite) linear combination of the operators
\(\psi (z)\) and their \(z\)-derivatives at the points \(p_i\):
\begin{equation}\label{I1}
\Psi_{I}:=\sum_{i\in I}
\sum _{m\geq 0} a_{im}\, \p_z^m \psi (z)\Bigr |_{z=p_i}
\end{equation}
Examples of the operators \(\Psi_{I}\) are
\(
\psi (p), \, \psi(p)+\alpha \psi ' (p), \,
\psi (p_1)+\alpha \psi (p_2), \,
\psi (p_1)+\alpha  \psi ' (p_2)+\beta \psi '' (p_2)
\,\,\, \mbox{etc}
\),
where \(\psi '(z)\equiv\p_z \psi (z)\).
Clearly,
\(
\lvacn e^{J_+({\bf t})}\normord e^{\beta \Psi_I \psistar (r)}\normord
\rvacn
\)
is a \(\tau\)-function for any fixed \(\beta\).
Let us take \(\beta =(\lvac \Psi_I \psistar (r)\rvac )^{-1}\), then
the group-like element is non-invertible but
\(
\lvacn e^{J_+({\bf t})}\Psi_I \psistar (r)
\rvacn
\)
is still a \(\tau\)-function. The limit  \(r\to 0\) can be performed
as in the previous example and we conclude that
\[
\tau_n ({\bf t})=\lvacn e^{J_+({\bf t})}\Psi_I \left |n-1\right >
\]
is a \(\tau\)-function.

This example can be further generalized
by considering products of operators of the form \(\Psi_I\).
Take \(N\)
operators of the form \(\Psi_I\) for different sets
\(I_1, \ldots , I_N\): \(\Psi_{I_1}, \ldots \, ,\Psi_{I_N}\). 
As before, we can construct the group-like
elements \(\normord e^{\beta_j \Psi_{I_j}\psistar (r_j)}\normord\)  and 
consider a \(\tau\)-function 
\(
\lvacn e^{J_+({\bf t})}
\normord e^{\beta_N \Psi_{I_N} \psistar (r_N)} \normord 
 \cdots
\normord e^{\beta_1 \Psi_{I_1} \psistar (r_1)} \normord 
 \rvacn
\).
Note that if the auxiliary points \(r_j\) are all distinct from the
points \(p_k\) labeled by the sets \(I_i\),
\( r_j \neq p_k\) for all \( j, k\in I_1 \cup \ldots \cup I_N\),
then these operators commute.
Again, we choose
\(\beta_j =(\lvac \Psi_{I_{j}} \psistar (r_j)\rvac )^{-1}\),
then each operator becomes the product
\(\beta_j \Psi_{I_j}\psistar (r_j)\)
(non-invertible). 
Thus the above \(\tau\)-function reduces to 
  a \(\tau\)-function of the form 
\(
(\beta_{1} \cdots \beta_{N})
 \lvacn e^{J_+({\bf t})}
\Psi_{I_1} \cdots \Psi_{I_N}
\psistar (r_N) \cdots \psistar (r_1)
 \rvacn 
\).
In order to implement the limit
\(r_j \to 0\), we redefine \(r_j \to \varepsilon r_j\) with
\(\varepsilon \to 0\), then it is not difficult to verify that
\begin{equation}\label{I2}
\begin{array}{l}
\varepsilon^{(n-1)N-\frac{1}{2}\, N(N-1)}
\psistar (\varepsilon r_N)\, \ldots \, \psistar (\varepsilon r_1)
\rvacn = (r_1 \ldots r_N)^{-n+1}
\Delta_{N}(r_i) \left |n-N\right > \, \CMPorArticle{ \\ ~\hspace{\stretch{1}}}{} + O(\varepsilon )
\\ \\
\varepsilon^{-nN-\frac{1}{2}\, N(N-1)}
\psi (\varepsilon r_N)\, \ldots \, \psi (\varepsilon r_1)
\rvacn \, = \, (r_1 \ldots r_N)^{n}
\Delta_{N}(r_i) \left |n+N\right > \, + \, O(\varepsilon ),
\end{array}
\end{equation}
where
\[
\Delta_N (r_i)=\det\limits_{i,j=1, \ldots , N}
r_{i}^{j-1}=\prod_{i>j}(r_i -r_j)
\]
is the Vandermonde determinant. The first equation
says that
in order to get a well-defined limit,
one should multiply the \(\tau\)-function by
\(\varepsilon^{(n-1)N-\frac{1}{2}\, N(N-1)}\)
before tending \(\varepsilon \to 0\).
The \(r_i \)-dependent factors \((r_1 \ldots r_N)^{-n+1}\Delta_N (r_i)\)
in the r.h.s. of (\ref{I2}) are irrelevant because they can be eliminated
by a transformation of the form (\ref{linear}).
We thus conclude that
\begin{equation}\label{I3}
\tau_n ({\bf t})=\lvacn e^{J_+({\bf t})}\Psi_{I_1}\ldots
\Psi_{I_N} \left |n-N\right >
\end{equation}
is a \(\tau\)-function.

In this case the prescription (\ref{adddef}) is not directly
applicable.
The behavior of the function \(\tau ({\bf t}\pm [z_{0}^{-1}])\)
at \(z_0 \to 0\) can be found
by using the bosonization formulae (\ref{bosonization})
in the opposite direction and moving the fermion operators
\(\psi (z_0)\) or \(\psistar (z_0)\) to the very right position.
In this way we obtain a more general
version of the prescription (\ref{adddef}):
\begin{equation}\label{adddef1}
\lim_{z_0\to 0}\left [
(-z_0)^{\pm N} \tau_n ({\bf t}\mp [z_{0}^{-1}])\right ]
=\tau_{n \pm 1} ({\bf t})
\end{equation}
which is equally enough to deduce the bilinear equation
(\ref{bi3}) from (\ref{bi201}) as a limiting case.

\subsection{Pl\"ucker coordinates}

One may expand the
\(\tau\)-function in the Schur polynomials:
\begin{equation}\label{tau1}
\tau_n ({\bf t})=\sum_{\lambda}c_{\lambda}(n) s_{\lambda}({\bf t}).
\end{equation}
The coefficients \(c_{\lambda}(n)\)
(``Pl\"ucker coordinates'') can be determined from the
formula (\ref{ferm6a}) by inserting the complete set of states
in between the operators \(e^{J_+}\) and \(G\) and using (\ref{lambda2}):
\[
\tau_n ({\bf t})=\sum_{\lambda}\lvacn e^{J_{+}({\bf t})}\left |
\lambda , n\rbr
\lbr \lambda , n \right | G\rvacn =
\sum_{\lambda}(-1)^{b(\lambda )} s_{\lambda}({\bf t})
\lbr \lambda , n \right | G\rvacn ,
\]
so
\begin{equation}\label{tau2}
\begin{array}{lll}
c_{\lambda}(n)&=&(-1)^{b(\lambda )}
\lbr \lambda , n \right | G\rvacn
\\ && \\
&=&(-1)^{b(\lambda )}\lvacn \psistar_{n+\alpha_1} \ldots
\psistar_{n+\alpha_{d(\lambda )}}
\psi_{n-\beta_{d(\lambda )}-1}  \ldots \psi_{n-\beta_1 -1}G\rvacn ,
\end{array}
\end{equation}
where \(b(\lambda )\) is defined in (\ref{boflambda}).
The Schur function expansions of \(\tau\)-function are also
discussed in \cite{Schexp,EH10}.

The coefficients \(c_{\lambda}(n)\) generalize the characters
of representations of the linear groups.
Remarkably, they admit determinant
representations of the Giambelli type
as well as of the Jacobi-Trudi type which express
\(c_{\lambda}(n)\) for arbitrary \(\lambda\)
through \(c_{\lambda}(n)\) for the hook and the one-row diagrams, 
respectively.

\subsubsection{Formulas of the Giambelli type}

Applying to (\ref{tau2})
the Wick's theorem in the form (\ref{Wick1}), we obtain
the Giambelli-like formula for \(c_{\lambda}(n)\):
\begin{equation}\label{tau3}
\begin{array}{lll}
c_{\lambda}(n)&=&\displaystyle{
(-1)^{b(\lambda )}
(\lvacn G \rvacn )^{-d(\lambda )+1}\,
\det_{i,j =1, \ldots , d(\lambda )}
\lvacn \psistar_{n+\alpha_i}
\psi_{n-\beta_j -1}G\rvacn }
\\ && \\
&=&\displaystyle{(c_{\emptyset}(n))^{-d(\lambda )+1}\,
\det_{i,j =1, \ldots , d(\lambda )}
c_{(\alpha _i |\beta _j)}(n)}.
\end{array}
\end{equation}
Here \(c_{(\alpha _i |\beta _j)}(n)\) are the
Pl\"ucker coordinates corresponding to
the hook diagrams \((\alpha _i |\beta _j)=(\alpha_i+1, 1^{\beta_j})\).
In the last expression, we have taken into account that
\[
\lvacn G \rvacn =\tau_n (0)= c_{\emptyset}(n).
\]
Formulas of the Giambelli type for the
expansion coefficients, or Pl\"ucker coordinates of the \(\tau\)-function
were given in \cite{EH10}. Using the Jacobi identity for
determinants, it is easy to see that they are equivalent to
the 3-term bilinear relation for the coefficients \(c_{\lambda}\)
given in \cite{JM83}.

\subsubsection{Formulas of the Jacobi-Trudi type}
\label{sec:Jacobi-Trudi}

Alternatively, we may apply to (\ref{tau2})
the Wick's theorem in the form (\ref{Wick3}). The details
are given in Appendix B. The result is
\begin{equation}\label{tau4a}
c_{\lambda}(n)
=\displaystyle{
\left ( \prod_{k=1}^{\lambda_{1}'-1}c_{\emptyset}(n-k)\right )^{-1}
\det_{i,j =1,\ldots , \lambda_1^{\prime}}
c_{\lambda_i -i+j}(n-j+1)},
\end{equation}
where
\(c_s (n):= c_{(s-1|0)}(n)=\left < n\! -\! 1\right |
\psistar_{n+s-1}G\rvacn \) are
the expansion coefficients for one-row diagrams and
\(\lambda'_{1}\) is the height of the first column
of the diagram \(\lambda \).

Therefore, the
Pl\"ucker coordinates of any MKP \(\tau\)-function satisfy the
determinant relations (\ref{tau4a}) of the Jacobi-Trudi type
sometimes called quantum Jacobi-Trudi formulas.
Note that the transformation \( c_{\lambda}(n) \rightarrow
C(n) c_{\lambda}(n)\) with arbitrary \(C(n)\) preserves
the determinant formula (\ref{tau4a}). Clearly, this freedom
corresponds to the possibility of multiplying the \(\tau \)-function
by any \(C(n)\), see (\ref{linear}).

In fact the inverse statement is also true: given any set of
quantities \(c_{\lambda}(n)\) that satisfy the determinant
relations (\ref{tau4a}), the function \(\sum_{\lambda} c_{\lambda}(n)
s_{\lambda}({\bf t})\) is a \(\tau\)-function of the MKP hierarchy.
In order to prove this, it is enough to show that for any given set
of quantities \(c_0(n)=c_{\emptyset}(n)\),
\(c_s(n)=c_{(s-1|0)}(n)\), \(s\geq 1\), one can find a
(\(n\)-independent) group-like element
\(G\) such that
\begin{equation}\label{group03}
c_s(n+1)=\lvacn \psistar_{n+s}G\left |n+1\right >=
\lvacn \psistar_{n+s}G \psi_n \rvacn \,,
\quad n\in \ZZ , \,\,\, s=0,1,2,\ldots
\end{equation}
Let us give a sketch of proof.
First of all we note that the element \(G\), if it exists,
is not unique because \(\normordbare e^{\sum_{i>k}A_{ik}\psistar_i\psi_k}
\normordbare \rvacn =\rvacn \) for any strictly lower triangular
matrix \(A\). This means that we can consider only upper triangular
matrices.
For simplicity we assume that
\(c_{\emptyset}(n)\neq 0\) for all \(n\).
Then we can (and from now on will)
put \(c_0(n)=1\) without any loss of generality
because this is the matter of normalization due to the
freedom to multiply the MKP \(\tau\)-function by an arbitrary
function of \(n\) as in (\ref{linear}). It is
convenient to formally put \(c_k(n)\equiv 0\)
for \(k<0\). Consider the infinite matrix
\(C_{nm}=c_{m-n}(n+1)\) (\(m,n \in \ZZ\)). It is an upper
triangular matrix with \(1\)'s on the main diagonal. One can see
that for any matrix \(C\)
of this form there exists a strictly upper triangular
matrix \(B\) such that
\[
C_{nm}=\lvacn \psistar_{m}
\normordbare \exp \Bigl (\sum_{i< k}B_{ik}\psistar_i \psi_k
\Bigr )\normordbare
\left |n+1\right >.
\]
Indeed, according to (\ref{group02}),
\( C_{nm}=\left [ (I+B^{(n)})^{-1}\right ]_{nm}\).
Set \(M=(I+B)^{-1}\), then for strictly
upper triangular \(B\) it holds \(M^{(n)}=(I+B^{(n)})^{-1}\), and so
the matrix \(B\)
is uniquely determined as \(B_{ik}=(C^{-1})_{ik}-\delta_{ik}\).

\subsubsection{Example: characters}

As an example, let us consider the \(\tau \)-function
corresponding to the coherent states
\[
G\rvacn =\exp \left ( \sum_{k\geq 1} x_{k}J_{-k}\right )\rvacn ,
\]
where \(J_{-k}\) are given by equation (\ref{ferm2}),
\(x_{k}=\frac{1}{k}\, \mbox{tr}\, g^k\), \(g\in GL(N)\).
The \(\tau\)-function (\ref{ferm6a}) is explicitly given by
\begin{equation}
\tau_n({\bf t}) =\exp \left ( \sum_{k\geq 1}kt_k x_{k}\right )
=\exp \left (\sum _{k\geq 1}\sum_{i=1}^{N}t_k p_{i}^{k}\right )\label{tausimpl}
\end{equation}
where \(p_i \) are eigenvalues of the matrix \(g\). 
This \( \tau \)-function
does not depend on \(n\). Therefore, the coefficients \(c_{\lambda}(n)\)
are \(n\)-independent and coincide with \(GL(N)\)-characters evaluated
for the element \(g\in GL(N)\):
\beq\label{characters}
c_{\lambda}(n)=\chi_{\lambda}(g)=s_{\lambda}({\bf x})=
\frac{\det_{1\leq i,j\leq N}
\left (p_{j}^{\lambda_i -i+N}\right )}{\prod_{1\leq l<k\leq N}
(p_l -p_k)}.
\eeq
The formulas (\ref{tau3}) and (\ref{tau4a}) become the well-known
determinant formulas for characters \cite{Macdonald}
(Giambelli and Jacobi-Trudi
respectively).

\noindent
{\bf Remark.} This simple example may be somewhat misleading because
in this case the \(\tau\)-function (\ref{tausimpl}) is
simultaneously the \(\tau\)-function with respect to the
``negative'' times \(t_{-k}=x_k\).
For more complicated solutions constructed below from the group element
\(g\in GL(N)\) with \(x_k=\frac{1}{k}\tr g^k\)
this is not true in general.

\subsection{Baker-Akhiezer functions and wave operators}

The Baker-Akhiezer (BA)
function and its adjoint are given by the Sato formulas
\begin{equation}\label{BA1}
\psi_n ({\bf t},z)=z^n e^{\xi ({\bf t},z)}
\frac{\tau_n ({\bf t}-[z^{-1}])}{\tau_n (t)}
=\frac{\lbr n+1\right | e^{J_{+}({\bf t})}
\psi (z) G\rvacn}{\lvacn e^{J_{+}({\bf t})}G\rvacn}
\end{equation}
\begin{equation}\label{BA1a}
\psi^*_n ({\bf t},z)=z^{-n} e^{-\xi ({\bf t},z)}
\frac{\tau_n ({\bf t}+[z^{-1}])}{\tau_n (t)}
=\frac{\lbr n-1\right | e^{J_{+}({\bf t})}
\psistar (z) G\rvacn}{z\lvacn e^{J_{+}({\bf t})}G\rvacn}.
\end{equation}
In terms of the BA function and its adjoint, the
bilinear identity (\ref{bi1}) acquires the form
\begin{equation}\label{bi1b}
\oint _{{\cal C}} \psi_{n}({\bf t},z)\psi^{*}_{n'}({\bf t'},z)\, dz =0.
\end{equation}

An important role in the theory \cite{UT84}
is played by the {\it wave operator} (or {\it dressing operator})
\(W(n,{\bf t})\) directly related to the BA functions. It is a pseudo-difference operator of the form
\begin{equation}\label{BA3}
W(n,{\bf t})=\sum_{k\geq 0} w_k(n,{\bf t})e^{-k\p_n},
\quad w_0(n,{\bf t})=1,
\end{equation}
and its formal inverse
\begin{equation}\label{BA3a}
W^{-1}(n,{\bf t})=\sum_{k\geq 0} e^{-k\p_n} w^*_k(n+1,{\bf t}),
\quad w_0^*(n,{\bf t})=1.
\end{equation}
The Baker-Akhiezer functions are obtained by applying the wave
operators to the bare exponent \(z^n e^{\xi ({\bf t},z)}\):
{\newcommand{\cnto}{\label{BA4} \psi_n ({\bf t},z)=W(n,{\bf t})z^n
    e^{\xi ({\bf t},z)}= 
\left ( 1+\frac{w_1(n,{\bf t})}{z}+
\frac{w_2(n,{\bf t})}{z^2}+
\ldots \right )z^n e^{\xi ({\bf t},z)}}\newcommand{\cntt}{\label{BA4a}
\psi^*_n ({\bf t},z)\CMPorArticle{&}{}=(W^{-1}(n\! -\! 1,{\bf t}))^{\dagger}
z^{-n} e^{-\xi ({\bf t},z)}\CMPorArticle{\\&}{}=\left ( 1+\frac{w^*_1(n,{\bf t})}{z}+
\frac{w^*_2(n,{\bf t})}{z^2}+
\ldots \right )z^{-n} e^{-\xi ({\bf
  t},z)}}\CMPorArticle{\begin{gather}
\cnto \\\begin{aligned} \cntt \end{aligned}
\end{gather}}{\begin{equation}
\cnto
\end{equation}
\begin{equation}
\cntt
\end{equation}}}
(by definition, the adjoint operator for \(A=f(n)e^{k\p_n}\)
is \(A^{\dagger}=e^{-k\p_n}f(n)\)).
Using formulas (\ref{BA1}), (\ref{BA1a}) we obtain
\begin{equation}\label{BA5}
w_k(n,{\bf t})=\frac{h_k(-\tilde \p )\tau_n({\bf t})}{\tau_n({\bf t})},
\quad \quad
w^*_k(n,{\bf t})=\frac{h_k(\tilde \p )\tau_n({\bf t})}{\tau_n({\bf t})},
\end{equation}
where the standard notation
\(
\tilde \p = \{\p_{t_1}, \, \frac{1}{2}\p_{t_2}, \,
\frac{1}{3}\p_{t_3}, \ldots \}
\)
is used. These formulas allow us to
express the wave operator and its inverse
in terms of the \(\tau\)-function as follows (cf. \cite{T90}):
\begin{equation}\label{BA6}
W(n,{\bf t})=\sum_{k\geq 0} \frac{h_k(-\tilde \p )
\tau_n({\bf t})}{\tau_n({\bf t})} \,
e^{-k\p_n}
\end{equation}
\begin{equation}\label{BA6a}
W^{-1}(n,{\bf t})=\sum_{k\geq 0} e^{-k\p_n} \,
\frac{h_k(\tilde \p )\tau_{n+1}({\bf t})}{\tau_{n+1}({\bf t})}
\end{equation}

The (pseudo-difference) Lax operator of the MKP hierarchy is
obtained by dressing the shift operator \(e^{\p_n}\) with the
wave operators:
\begin{equation}\label{Lax}
L=W(n,{\bf t})e^{\p_n}W^{-1}(n,{\bf t})=e^{\p _n}+a_0(n,{\bf t})+
a_1(n,{\bf t})e^{-\p_n}+\ldots
\end{equation}
The Lax operator obeys the Lax equations
\beq\label{Laxeq}
\p_{t_k}L=[(L^k)_{\geq 0},\, L],
\eeq
where \((L^k)_{\geq 0}\) means the (finite) part of the series
containing the shift operators \(e^{j \p_n} \) with \(j\geq 0\).

\section{Rational solutions of the KP and MKP hierarchy}
\label{sec:rati-solut-kp}

In this section we study rational solutions of the KP and
MKP hierarchies. For these solutions,
the \(\tau\)-function is a polynomial function of the times
\(n\) and \(t_i\) (possibly multiplied by an exponential function
of a linear combination of the times).
Stressing the fact that \(n\) for solutions
of this class is not necessarily integer but can be any
complex number \(u\in \CC\), we change the notation \(n\to u\).

\subsection{The construction of rational solutions}

The general construction of rational solutions
to the KP hierarchy was given by
Krichever in \cite{Krichever78}, see also
\cite{Krichever83}.
This construction can be directly extended to the
MKP hierarchy.
In this section we show how it works 
starting from the
bilinear identity (\ref{bi1}).
The solutions are
obtained in the Casorati determinant form
which is a discrete analog of the Wronskian determinant
(see, e.g. \cite{GS91}).

\subsubsection{Baker-Akhiezer functions and bilinear identity}

First of all, let us specify the bilinear identity for the case
at hand. Since in the solutions we are going to consider \(u\) is not
necessarily integer,
\(0\) and \(\infty\) are in general branch points
of the BA function.
In this case the appropriate form of the bilinear identity is
\begin{equation}\label{bi1c}
\oint _{C_{[0, \infty]}} \psi_{u}({\bf t},z)
\psi^{*}_{u'}({\bf t'},z)\, dz =0,
\end{equation}
where \([0, \infty]\) is an arbitrary cut between \(0\) and
\(\infty\) and the contour \(C_{[0, \infty]}\) is such that it
encircles the cut 
but does not
enclose any other singularity of the BA functions.

According to the Krichever's theory of rational and more
general algebro-geometric solutions, they can be characterized
and explicitly constructed by fixing certain analytic properties of
the BA function on a Riemann surface. For rational solutions,
the Riemann surface is the Riemann sphere (compactified complex plane),
which represents a genus zero algebraic curve with singularities.
Correspondingly, the BA function is, in this case, a rational function on the
complex plane multiplied by power-like and exponential factors
which give the required
asymptotics (\ref{BA4}) (the essential singularity at infinity).
Let us assume the following ansatz for the
BA function:
\begin{equation}\label{rat1}
\psi_u ({\bf t},z)=z^u e^{\xi ({\bf t},z)}\left (1+\frac{w_1(u, {\bf t})}{z}
+\ldots + \frac{w_N(u, {\bf t})}{z^N}\right ),
\end{equation}
where the coefficients \(w_i\) depend on the times \(u\), \(t_j\).
In fact this is just the truncated series (\ref{BA4}).
The multiplicity \(N\) of the pole at  \(z=0\) of the function
\( z^{-u}e^{-\xi ({\bf t},z)}\psi_u ({\bf t},z)\)
is a discrete parameter characterizing the class
of solutions to be constructed.

Given the BA function of the form (\ref{rat1}),
an important dynamical information
is contained in the adjoint BA function.
In general, it may have poles of arbitrary order at some
points \(p_i\in \CC\) (we assume that \(p_i\neq 0\)).
As we shall see soon, the bilinear identity is consistent when
the number of these points does not exceed \(N\), and, in
case of general position their number is just equal to \(N\).
Let us adopt this and
fix \(N\) distinct points \(p_i\in \CC\),
\(p_i\neq 0\), and also assume that the adjoint BA function
has poles of orders \(M_i +1\) at \(p_i\). 
Therefore, its general form is
\begin{equation}\label{rat101}
\psi_u ^{*}({\bf t},z)=z^{-u} e^{-\xi ({\bf t},z)}
\left (1+\sum_{i=1}^{N}
\sum_{m=0}^{M_i }\frac{m! \, \tilde a_{im}(u,{\bf t})}{(z-p_i)^{m+1}}
\right ),
\end{equation}
where \(\tilde 
a_{im}(u,{\bf t}) \) is set to be \(0\) for \(m \ge M_{i}+1 \).
As we shall see in section 5,
the \(\tau\)-functions
of our principal interest (master \(T\)-operators for spin
chains with rational \(R\)-matrices)
are, by construction, polynomials in \(u\).
This means that
the coefficients \(\tilde 
a_{im}(u,{\bf t})\) are rational functions of \(u\),
so we assume this in the present section.

\subsubsection{Krichever's conditions from the
bilinear identity}

Since the poles at \(p_i\) are the only
singularities of the product
\(\psi_{u}({\bf t},z)\psi^{*}_{u'}({\bf t'},z)\) outside the contour
\(C_{[0, \infty]}\),
the bilinear identity (\ref{bi1b}) then implies that
\[
\sum_{i=1}^{N} \mbox{res}_{z=p_i}\left (
\psi_{u} ({\bf t},z)\psi^{*}_{u'}({\bf t'}, z)\right )=0
\]
for all \(u,t_j, u', t_j'\). Let us rewrite this identically
in the form
\begin{equation}\label{rat102}
\sum_{i=1}^{N}p_{i}^{-u'}
\, R_{i}(u', {\bf t'}; u,{\bf t})=0,
\end{equation}
where the functions \(R_i\) are
\[
\left. R_{i}(u', {\bf t'}; u,{\bf t})=\sum_{m=0}^{M_i}
\tilde a_{im}(u', {\bf t'})
p_{i}^{u'}\p_z^m \left (
z^{-u'}e^{-\xi ({\bf t'}, z)}\psi_u ({\bf t},z)\right )\right  |_{z=p_i}.
\]
From the polynomiality of the \(\tau \)-function
it is clear that they are rational functions of \(u'\).
If so, the only way to satisfy (\ref{rat102})
for all \(u'\) is to put each term equal to zero separately
(because it is a sum of different
exponential functions multiplied by rational
functions which can be identically zero if and only if all
rational functions vanish identically).
Therefore, the bilinear identity is equivalent to \(N\)
conditions
\begin{equation}\label{rat103}
\mbox{res}_{z=p_i}\left (\psi_u({\bf t},z)
\psi^{*}_{u'}({\bf t'},z)\right )=0\,,
\quad \quad i=1,  \ldots , N,
\end{equation}
or, more explicitly,
\begin{equation}\label{rat2}
\left. \phantom{\int_a}
\sum_{m=0}^{M_i}\tilde a_{im}(u',{\bf t'})\,
\p_{z}^{m}\left (z^{-u'}
e^{-\xi ({\bf t'}, z)}\psi_u ({\bf t},z)\right )\right |_{z=p_i}\! =0\,,
\quad i=1, \ldots , N,
\end{equation}
which hold for any values of \(u, u'\) and \(t_i , t'_i\).
In particular, at \(u'=t'_i=0\) we have
\(\displaystyle{
\sum_{m=0}^{M_i}\tilde a_{im}(0,0)\,
\p_{z}^{m} \psi_u ({\bf t},z)\Bigr |_{z=p_i}\!\!\! =0}\) or
\begin{equation}\label{rat2a}
\sum_{m=0}^{M_i}a_{im}\,
\p_{z}^{m} \psi_u ({\bf t},z)\Bigr |_{z=p_i}\!\! =0\,,
\end{equation}
where \(a_{im} \equiv k_i \tilde a_{im}(0,0)\) with arbitrary
non-zero \(k_i\)
are parameters of the solution
in the Krichever's approach.
We call equations (\ref{rat2a}) the Krichever's conditions
\cite{Krichever78}.

\noindent
{\bf Remark.} In the Krichever's approach, the parameters
\(a_{im}\) can be taken arbitrary while in our case
\(\tilde a_{im}(0,0)\) appear to be constrained by
\(N\) conditions of the form (\ref{1001}), as we shall see below.
That is why we prefer to work with the set of independent
parameters \(a_{im}\). The initial values of the 
adjoint BA function coefficients 
\(\tilde a_{im}(0,0)\) are not independent and differ from
the \(a_{im}\) 
by the properly chosen \(N\) 
factors \(k_i\).

The set of conditions (\ref{rat2a})
yields a system of \(N\) linear equations for \(N\) coefficients
\(w_k\) which demonstrates the consistency of the bilinear identity
for BA functions of the assumed form and
allows one to fix them.
Then the \(\tau\)-function associated with
the \(\psi\)-function according to (\ref{BA1})
solves the MKP hierarchy.
The points \(p_i\) and the entries of the matrix \(a_{im}\) are free parameters
of the solution.
The coefficients \(w_k\) appear to be rational functions
of their arguments \(u, t_i\) while the \(\tau\)-function is a polynomial
(multiplied by the exponential function of a linear combination
of times).
From the algebro-geometric point of view this solution is associated
with a highly singular algebraic curve which is the Riemann sphere
with cusp singularities at the points \(p_i\).

\subsubsection{Solving the linear system}

It is easy to see that conditions (\ref{rat2a}) are equivalent to
the system of linear equations
\begin{equation}\label{rat3}
A_{i}(u,{\bf t})+\sum_{k=1}^{N}A_{i}(u-k,{\bf t})w_k=0,
\end{equation}
where
\begin{equation}\label{A}
\left. A_{i}(u,{\bf t})=\sum_{m=0}^{M_i}a_{im}
\p_{z}^{m}
\left (z^{u}e^{\xi ({\bf t},z)}\right ) \right |_{z=p_i}.
\end{equation}
These functions are polynomials in \(u, t_i\) multiplied by
the exponential  factor \(p_{i}^{u}e^{\xi ({\bf t}, p_i)}\).
In the case when all \(t_i\) are equal to \(0\), they admit 
a more explicit representation:
\beq\label{ratQ}
A_i(u,0)=p_{i}^{u}\Bigl ( a_{i0}+a_{i1}p_{i}^{-1}u+
a_{i2}p_{i}^{-2}u(u\! -\! 1)+ a_{i3}p_{i}^{-3}u(u\! -\! 1)(u\! -\! 2)
+\ldots \Bigr ).
\eeq
Note that \(A_i(0,0)=a_{i0}\).

The system (\ref{rat3}) can be solved by applying the
Cramer's rule. This results in the following determinant
representation for the BA function:
\begin{equation}\label{rat5}
\psi_u ({\bf t},z)=z^u e^{\xi ({\bf t},z )}\,
\frac{\left |\begin{array}{cccc}
1&z^{-1}&\ldots & z^{-N}\\
A_{1}(u,{\bf t})& A_{1}(u\! -\! 1,{\bf t})& \ldots &
A_{1}(u\! -\! N,{\bf t})\\
\vdots & \vdots &\ddots & \vdots \\
A_{N}(u,{\bf t})&A_{N}(u\! -\! 1,{\bf t})& \ldots
&A_{N}(u\! -\! N,{\bf t})
\end{array}\right |}{\left |\begin{array}{ccc}
A_{1}(u-1,{\bf t})&\ldots & A_{1}(u-N,{\bf t})\\
\vdots &\ddots & \vdots \\
A_{N}(u-1,{\bf t})& \ldots & A_{N}(u-N,{\bf t})
\end{array}\right |}.
\end{equation}
From the definition (\ref{A}) we have
\begin{equation}\label{1002}
\CMPorArticle{ \begin{aligned}
    A_{k}(u,{\bf t}-[z^{-1}])=&\left. \sum_{m=0}^{M_k}a_{km}
      \p_{\zeta}^{m} \left (\zeta^{u}e^{\xi ({\bf t},\zeta )}\Bigl
        (1-\frac{\zeta}{z}\Bigr ) \right ) \right |_{\zeta =p_i}\\=&
    A_k(u,{\bf t})-A_k(u+1, {\bf t})z^{-1}.
  \end{aligned}}
{\left. A_{k}(u,{\bf t}-[z^{-1}])=\sum_{m=0}^{M_k}a_{km}
\p_{\zeta}^{m}
\left (\zeta^{u}e^{\xi ({\bf t},\zeta )}\Bigl (1-\frac{\zeta}{z}\Bigr )
\right ) \right |_{\zeta =p_i}=
A_k(u,{\bf t})-A_k(u+1, {\bf t})z^{-1}.}
\end{equation}
Using this, it is straightforward to verify
that equation (\ref{rat5}) agrees with the general relation (\ref{BA1}),
with the \(\tau\)-function being
given by the determinant in the denominator:
\begin{equation}\label{rat6}
\tau_u ({\bf t})=\det_{i,j=1, \ldots , N}A_{i}(u\! -\! j,\, {\bf t})
=\left |\begin{array}{ccc}
A_{1}(u-1,{\bf t})&\ldots & A_{1}(u-N,{\bf t})\\
\vdots &\ddots & \vdots \\
A_{N}(u-1,{\bf t})& \ldots & A_{N}(u-N,{\bf t})
\end{array}\right |.
\end{equation}
It is a polynomial in \(u\) of degree
\(\sum\limits_{j=1}^{N}M_j\) multiplied by
\(\prod\limits_{i=1}^{N}p_{i}^{u}e^{\xi ({\bf t}, p_i)}\).

It follows from (\ref{rat5}) that the last coefficient
in (\ref{rat1}), \(w_N\),
is given by
\begin{equation}\label{rat7}
w_{N}(u,{\bf t})=(-1)^N
\frac{\tau_{u+1}({\bf t})}{\tau_u({\bf t})}.
\end{equation}
We also note the property
\begin{equation}\label{rat4b}
\p_{t_1}A_{i}(u,{\bf t})=A_{i}(u+1,{\bf t})
\end{equation}
from which one can see that the first coefficient
in (\ref{rat1}), \(w_1\),
is given by
\begin{equation}\label{rat7a}
w_{1}(u,{\bf t})=-\p_{t_1} \log \tau_u({\bf t}).
\end{equation}

\subsubsection{The adjoint BA function}

Similarly to (\ref{1002}), we have:
\begin{equation}\label{rat8}
A_{k}(u,{\bf t}+[z^{-1}])=
\left. \sum_{m=0}^{M_k}a_{km}
\p_{\zeta}^{m}
\left (\frac{\zeta^{u}e^{\xi ({\bf t},\zeta )}}{1-\zeta/z}
\right ) \right |_{\zeta =p_k},
\end{equation}
Note that this function regarded as a function of \(z\)
has a pole of order \(M_k+1\) at \(z=p_k\). The principal term is
\begin{equation}\label{rat9}
A_{k}(u,{\bf t}+[z^{-1}])=
\frac{M_k! \, a_{kM_k}p_{k}^{u+1}
e^{\xi ({\bf t}, p_k)}}{(z-p_k)^{M_k+1}}\, +
O\left ((z-p_k)^{-M_k}\right ).
\end{equation}
We also see from (\ref{rat8}) that the function
\(A_{k}(u,{\bf t}+[z^{-1}])\) has a simple zero at \(z=0\).

In order to obtain a more detailed information about singularities
of this function, let us add and subtract the term
\(z^{u} e^{\xi ({\bf t},z)}\)
in the numerator in (\ref{rat8}) and separate the nonsingular part as \(z\to p_k\)
from the singular one:
{\newcommand{\cnt}{\label{rat8b}
A_{k}(u,{\bf t}+[z^{-1}])= z^{u+1}e^{\xi ({\bf t},z)} \sum_{m=0}^{M_k}
\frac{m! \, a_{km}}{(z-p_k )^{m+1}}\CMPorArticle{\\\left.}{}-
z \Bigl (\sum_{m=0}^{M_k} a_{km}\p_{\zeta}^{m} \Bigr )\,
\frac{z^u e^{\xi ({\bf t},z)}\! -\!
\zeta^u e^{\xi ({\bf t},\zeta )}}{z-\zeta}\, \right
|_{\zeta =p_k}.}\CMPorArticle{\begin{multline}
\cnt
\end{multline}
}{\begin{equation}
\left. \cnt
\end{equation}}}
The first sum gives the multiple pole structure at the point \(p_k\)
while the second term
is obviously regular at \(p_k\) and has possible (essential)
singularities and branching only at \(0\) and \(\infty\).
Note that \(A_{k}(u,{\bf t}+[z^{-1}])\) is, by construction, a
rational function of \(z\) with simple zero at \(z=0\).
This is obvious from (\ref{rat8}) but not from (\ref{rat8b}).
On the other hand, the representation (\ref{rat8b})
explicitly shows the multi-pole structure of this function
which is rather obscure in (\ref{rat8}).

Rewriting (\ref{1002}) in the form
\[
A_k(u, {\bf t}+[z^{-1}])=A_k(u,{\bf t})+z^{-1}A_k(u+1, {\bf t}+[z^{-1}]),
\]
it is straightforward to check that
\begin{equation}\label{rat10}
\tau_u ({\bf t}+[z^{-1}])=\left |\begin{array}{cccc}
A_{1}\left (u\! -\! 1,{\bf t}\! +\! [z^{-1}]\right )&
A_{1}(u\! -\! 2,{\bf t})&
\ldots & A_{1}(u\! -\! N,{\bf t})\\
A_{2}\left (u\! -\! 1,{\bf t}\! +\! [z^{-1}]\right )&
A_{2}(u\! -\! 2,{\bf t})&
\ldots & A_{2}(u\! -\! N,{\bf t})\\
\vdots & \vdots &\ddots & \vdots \\
A_{N}\left (u\! -\! 1,{\bf t}\! +\! [z^{-1}]\right )&
A_{N}(u\! -\! 2,{\bf t})&
\ldots &A_{N}(u\! -\! N,{\bf t}).
\end{array}\right |
\end{equation}
Expanding (\ref{rat8}) in powers of \(z\), we get
\[
A_{k}(u,{\bf t}+[z^{-1}]) =A_k(u,{\bf t})+
A_k(u+1,{\bf t})z^{-1}+A_k(u+2,{\bf t})z^{-2}+\ldots
\]
and so the expansion of \(\tau_u ({\bf t}+[z^{-1}])\) around
\(\infty\) reads
\begin{equation}\label{rat11}
\tau_u ({\bf t}+[z^{-1}])=\sum_{s=0}^{\infty}z^{-s}
\left |\begin{array}{cccc}
A_{1}\left (u\! +\! s\! -\! 1,{\bf t}\right )&
A_{1}(u\! -\! 2,{\bf t})&
\ldots & A_{1}(u\! -\! N,{\bf t})\\
A_{2}\left (u\! +\! s\! -\! 1,{\bf t}\right )& A_{2}(u\! -\! 2,{\bf t})&
\ldots & A_{2}(u\! -\! N,{\bf t})\\
\vdots & \vdots &\ddots & \vdots \\
A_{N}\left (u\! +\! s\! -\! 1,{\bf t}\right )&A_{N}(u\! -\! 2,{\bf t})&
\ldots &A_{N}(u\! -\! N,{\bf t})
\end{array}\right |.
\end{equation}

We thus see that the adjoint BA function has the
determinant representation
\begin{equation}\label{rat12}
\psi^{*}_u({\bf t},z)=z^{-u}e^{-\xi ({\bf t},z)}\,
\frac{\left |\begin{array}{cccc}
A_{1}\left (u\! -\! 1,{\bf t}\! +\! [z^{-1}]\right )&
A_{1}(u\! -\! 2,{\bf t})&
\ldots & A_{1}(u\! -\! N,{\bf t})\\
A_{2}\left (u\! -\! 1,{\bf t}\! +\! [z^{-1}]\right )&
A_{2}(u\! -\! 2,{\bf t})&
\ldots & A_{2}(u\! -\! N,{\bf t})\\
\vdots & \vdots &\ddots & \vdots \\
A_{N}\left (u\! -\! 1,{\bf t}\! +\! [z^{-1}]\right )&
A_{N}(u\! -\! 2,{\bf t})&
\ldots &A_{N}(u\! -\! N,{\bf t})
\end{array}\right |}{\left |\begin{array}{cccc}
A_{1}\left (u\! -\! 1,{\bf t}\right )& A_{1}(u\! -\! 2,{\bf t})&
\ldots & A_{1}(u\! -\! N,{\bf t})\\
A_{2}\left (u\! -\! 1,{\bf t}\right )& A_{2}(u\! -\! 2,{\bf t})&
\ldots & A_{2}(u\! -\! N,{\bf t})\\
\vdots & \vdots &\ddots & \vdots \\
A_{N}\left (u\! -\! 1,{\bf t}\right )&A_{N}(u\! -\! 2,{\bf t})&
\ldots &A_{N}(u\! -\! N,{\bf t})
\end{array}\right |.}
\end{equation}
It has multiple poles at the singular points \(p_i\)
of the algebraic curve which is the Riemann sphere with
complicated cusp-like singularities. The principal parts
at the poles are easily extracted from the determinant
representation and equation (\ref{rat8b}).

Since the function
\(A_{k}(u,{\bf t}+[z^{-1}])\) has a simple zero at \(z=0\),
it is clear from
\eqref{rat6} that
the function \(\tau _u ({\bf t} +[z^{-1}])\) and thus the function
\(z^u e^{\xi ({\bf t},z)}
\psi^{*}_{u}({\bf t},z)\) have zero of order \(N\) at \(z=0\),
whence the coefficients \(\tilde a_{im}(u,{\bf t})\)
in (\ref{rat101}) appear to be
constrained by \(N\) relations
\begin{equation}\label{1001}
\sum_{i=1}^{N}\sum_{m=0}^{M_i}(-1)^m (m+n)! \, \tilde 
a_{im}(u,{\bf t})
p_{i}^{-m-n-1}=\delta_{n, 0}\,,
\quad \quad n=0,1, \ldots , N-1.
\end{equation}

\subsubsection{The multi-pole structure of the
adjoint BA function}

Let us introduce the notation
\begin{equation}\label{rat13a}
\bar A_{k}(u,{\bf t}):= \det_{
{i=1, \,\ldots , \not k , \ldots , N\atop j=1,\, \ldots \, , \, N-1}}
\, A_{i}(u+1-j, \, {\bf t})
\end{equation}
for the minor \(M_{k,N}\) of the 
\(N\! \times \! N\) matrix \(A_{i}(u+1-j)\), \(1\leq i,j\leq N\).
Then, expanding the determinant in the numerator of
(\ref{rat12}) in the first column, we obtain:
\[
\psi^{*}_u({\bf t},z)=\frac{z^{-u}e^{-\xi ({\bf t},z)}}{\tau_u ({\bf t})}
\sum_{k=1}^{N} (-1)^{k-1}
\bar A_{k}(u-2, {\bf t}) \,A_k (u-1, {\bf t}+[z^{-1}]).
\]
Using (\ref{rat8b}) we can extract the poles:
\begin{equation}\label{rat13b}
\psi^{*}_{u}({\bf t},z)=\sum_{k=1}^{N}(-1)^{k-1}
\frac{\bar A_{k}(u-2, {\bf t})}{\tau_u({\bf t})}
\sum_{m=0}^{M_k}\frac{m! \, a_{km}}{(z-p_k )^{m+1}} \,
+\, \mbox{terms regular at all \(p_k\)%
}.
\end{equation}
Therefore,
\begin{equation}\label{rat13c}
\mbox{res}_{z=p_k}\left [ (z-p_k)^n
\psi^{*}_{u}({\bf t},z)\right ]=(-1)^{k-1}n! \, a_{kn} \,
\frac{\bar A_{k}(u-2, {\bf t})}{\tau_u({\bf t})}
\end{equation}
for all \(n\geq 0\), or, in terms of the \(\tau \)-function,
\begin{equation}\label{rat13d}
\mbox{res}_{z=p_k}\left [ (z-p_k)^n z^{-u}e^{-\xi ({\bf t}, z)}
\tau _u ({\bf t}+[z^{-1}])\right ]=(-1)^{k-1}n! \, a_{kn} \,
\bar A_{k}(u-2, {\bf t}).
\eeq
Note that the expression in the r.h.s. has the same structure
as (\ref{rat6}) and, therefore, is a polynomial \(\tau \)-function.
This fact will be used below for the B\"acklund transformations.

We also note the factorization formula
\begin{equation}\label{rat15c}
\mbox{res}_{z=p_k} \!\left (z^u e^{\xi ({\bf t},z)}
\psi^{*}_{u'} ({\bf t'}, z)\right )=(-1)^{k-1}
 \frac{\bar A_{k}(u'\! -\! 2,{\bf t'})}{\tau_{u'}({\bf t'})}
\, A_k (u, \, {\bf t})
\end{equation}
which follows from (\ref{rat13b}). 
At \(u'=u=t'_j=t_j=0\) this formula gives the relation between
\(\tilde{a}_{k0} \equiv \tilde{a}_{k0}(0,0) \) and \(a_{k0}\):
\begin{equation}\label{rat15d}
\tilde a_{k0}=(-1)^{k-1}\frac{\bar A_{k}(-2,0)}{\tau_0(0)}\, a_{k0}.
\end{equation}

\subsection{Undressing B\"acklund transformations
for the rational solutions}
\label{sec:family-backl-transf}

As it will be clear in section 5, the main relations of the
Bethe ansatz method
are naturally built  into
the construction of rational solutions to the MKP hierarchy.
In particular, the functions
\(\bar A_{k}(u,{\bf t})\) and
\(A_{k}(u,{\bf t})\)
will be identified, up to some irrelevant factors, with
the (eigenvalues of) \(T\)-operators on
the first and the last levels of nesting in the nested
Bethe ansatz scheme.
The parameters \(p_i\) will be identified with
eigenvalues of the twist matrix \(g\).
We will also see that all these formulas can be
understood in the operatorial sense  (in the quantum space of
the spin chain) since they involve only commuting \(T\)-operators.

The nested Bethe ansatz scheme itself
appears to be equivalent to a chain of some
special B\"acklund transformations of the initial
rational MKP solution with Krichever's data \(p_i, a_{im}\)
that ``undress'' it to the trivial solution by reducing the
number of singular points in succession.
Here we present the idea solely in terms of the MKP
hierarchy
leaving the precise identification with objects from
quantum integrable spin chains for the next section.

Adding/removing a point \(p_i\) to/from the Krichever's data
of a rational solution is a B\"acklund transformation.
It sends a polynomial \(\tau\)-function to another one.
The key equation that allows one to implement
such a transformation is (\ref{rat13d}). 
Let us start with the \(n=0\) case.
Specifically, consider the function
\begin{equation}\label{Back1}
\tau_{u}^{[i]}({\bf t})=(-1)^{i-1}\mbox{res}_{z=p_i}\left (
z^{-u-1}e^{-\xi ({\bf t},z)}\tau_{u+1}({\bf t}+[z^{-1}])\right ).
\end{equation}
According to the \(n=0\) case of (\ref{rat13d}),
it is equal to
\begin{equation}\label{Back2}
\tau_{u}^{[i]}({\bf t})=
a_{i0}
\left |
\begin{array}{cccc}
A_{1}\left (u\! -\! 1,{\bf t}\right )& A_{1}\, (u\! -\! 2,{\bf t})&
\ldots & A_{1}(u\! -\! N\! +\! 1,{\bf t})\\
\vdots & \vdots &\ddots & \vdots \\
\,\,\,\,A_{i-1}\left (u\! -\! 1,{\bf t}
\right )& \,\,\,\,A_{i-1}(u\! -\! 2,{\bf t})&
\ldots & \,\,\,A_{i-1}(u\! -\! N\! +\! 1,{\bf t})\\
\,\,\,\,A_{i+1}\left (u\! -\! 1,{\bf t}\right )&
\,\,\,\,A_{i+1}(u\! -\! 2,{\bf t})&
\ldots & \,\,\,\,A_{i+1}(u\! -\! N\! +\! 1,{\bf t})\\
\vdots & \vdots &\ddots & \vdots \\
A_{N}\left (u\! -\! 1,{\bf t}\right )&A_{N}(u\! -\! 2,{\bf t})&
\ldots &A_{N}\, (u\! -\! N\! +\! 1,{\bf t})
\end{array}
\right |,
\end{equation}
i.e., to the minor \(M_{i,1}\) of the matrix \(A_{i}(u+1-j)\)
multiplied by \(a_{i0}\). Therefore, it is 
a \(\tau\)-function, i.e., it
satisfies the same 3-term Hirota equation as \(\tau_u({\bf t})\) does and
\(\tau \rightarrow \tau^{[i]}\) is indeed a B\"acklund
transformation. Below we assume that \(a_{i0}\neq 0\) for all
\(i=1, \ldots , N\), then the transformation is non-trivial.

Having at hand \(\tau_u^{[i]}({\bf t})\), one can construct the BA
function
\begin{equation}\label{Back3}
\psi^{[i]}_u({\bf t}, z)=z^u e^{\xi ({\bf t},z)}
\, \frac{\tau_u^{[i]}({\bf t} -[z^{-1}])}{\tau_u^{[i]}({\bf t})}
\end{equation}
which obeys the same Krichever's conditions (\ref{rat2a})
at the points \(p_1, \ldots ,p_N\) except the point \(p_i\),
where no condition is imposed. Note that the BA function
\(\psi^{[i]}\) has the expansion around the point \(z=0\)
of the same form (\ref{rat1}) but with a pole 
of order \(N-1\) rather than \(N\), so the number
of conditions again matches the number of unknown coefficients
of the singular part at \(z=0\).
The adjoint BA function is
\begin{equation}\label{Back4}
\psi^{*[i]}_u({\bf t}, z)=z^{-u} e^{-\xi ({\bf t},z)}
\, \frac{\tau_u^{[i]}({\bf t} +[z^{-1}])}{\tau_u^{[i]}({\bf t})}.
\end{equation}
It has multiple poles at the points
\(p_1, \ldots ,p_N\) except the point \(p_i\).

This process can be continued by taking
the residue of \(z^{-u-1}e^{-\xi ({\bf t},z)}
\tau_{u+1}^{[i]}({\bf t} +[z^{-1}])\)
at some other singular point \(p_{j}\), \(j\neq i\):
{\newcommand{\cnt}{\label{Back5}
\tau_{u}^{[i j]}({\bf t})=
(-1)^{j}\varepsilon_{ij}
\mbox{res}_{z=p_{j}}\left (
z^{-u-1}e^{-\xi ({\bf t},z)}\tau^{[i]}_{u+1}({\bf t}+[z^{-1}])
\right )\CMPorArticle{\nonumber
\\[.3cm]}{\\\\}
=(-1)^{i+j-1}\varepsilon_{ij}
\mbox{res}_{\substack{\\z_{i}=p_{i}\\z_{j}=p_{j}}}\Bigl [
(z_{i}z_{j})^{-u-2}e^{-\xi ({\bf t},z_{i})-\xi ({\bf t},z_{j})}
(z_{j}\! -\! z_{i})\tau_{u+2}({\bf t}+[z_{i}^{-1}]+
[z_{j}^{-1}])\Bigr ]\CMPorArticle{\nonumber
\\[.3cm]}{\\\\}
=\,\, a_{i 0}a_{j 0}\, \det\limits_{\substack{\\
1\leq r \leq N, \neq i, j\\s=1, \ldots , N-2}}
\Bigl [ A_r (u-s, {\bf t})\Bigr ].}
\CMPorArticle{\begin{gather}
\cnt
\end{gather}}{\begin{equation}
\begin{array}{c}
\cnt
\end{array}
\end{equation}}}
Here and below \(\varepsilon_{ij}=1\) if
\(i<j\) and \(-1\) if \(i>j\).

We thus obtain a chain of B\"acklund transformations
\(\tau \rightarrow \tau^{[i_1]} \rightarrow \tau^{[i_1 i_2]}
\rightarrow \ldots \).
The general recursive formula at the \(n\)-th level is
{\newcommand{\ctnt}{\tau_{u}^{[i_1 \ldots i_n]}({\bf t}) \CMPorArticle{}{&}
=
(-1)^{i_{n}-1-
\mathrm{Card} \{ k | i_{k} < i_{n}, 1 \le k \le n-1 \} 
}\CMPorArticle{\\\times~}{}
\mbox{res}_{ z_{i_{n}}=p_{i_{n}} }
\left(
z_{i_{n}}^{-u-1}e^{-\xi ({\bf t},z_{i_{n}})}
\tau_{u+1}^{[i_1\ldots i_{n-1}]}
({\bf t}+[z^{-1}_{i_{n}}])
\right),
\label{Backrec}}
\CMPorArticle{\begin{multline}
\ctnt
\end{multline}
}{\begin{align}
\ctnt
\end{align}}}
which leads to
\begin{equation}
\label{Back5gen}
\CMPorArticle{\begin{aligned}}{\begin{array}{c}\displaystyle}
\tau_{u}^{[i_1\ldots i_n]}({\bf t})=\CMPorArticle{&}{} (-1)^{d_n}
\mbox{res}_{\substack{\\z_{i_k}=p_{i_k}\\1\leq k\leq n}}
\left [ \Bigl (\prod_{\alpha =1}^{n}z_{i_{\alpha}}^{-u-n}
e^{-\xi ({\bf t}, z_{i_{\alpha}})} \Bigr )\CMPorArticle{\right.\\&\qquad\qquad\qquad\qquad\left.}{}\Delta_n(z_{i_1}, \ldots ,
z_{i_n})\tau_{u+n}\Bigl ({\bf t}+\sum_{\beta =1}^{n}
[z_{i_{\beta}}^{-1}]\Bigr )\right ]
\\
 \CMPorArticle{ =&}{\\\displaystyle=}\, \Bigl (\prod_{\alpha =1}^{n}
a_{i_{\alpha} 0}\, \Bigr )
\det\limits_{\substack{\\
1\leq r \leq N, \neq i_1,\ldots , i_n\\s=1, \ldots , N-n}}
\Bigl [ A_r (u-s, {\bf t})\Bigr ].
\CMPorArticle{\end{aligned}}{\end{array}}
\end{equation}
In the first line \(d_n = i_1 +\ldots +i_n +
\frac{1}{2}n(n+1)\) and
\beq\label{Vander}
\Delta_n(z_{i_1}, \ldots ,
z_{i_n})=\prod_{\substack{\\i>j\\i,j \in \{i_1 , \ldots , i_n\}}}.
\!\!\!\! (z_i-z_j)
\eeq
is the Vandermonde determinant.
In particular, on the second to last level we have
\begin{align}
\tau_u^{[12\ldots \not j \ldots N]}({\bf t})
=\Bigl (\prod_{i=1, \neq j}^{N}
a_{i0}\Bigr ) A_j(u-1, {\bf t}) 
 \label{basictau}
\end{align}
and on the last level
\(
\tau_u^{[12 \ldots N]}({\bf t})
=\prod_{i=1}^{N} a_{i0}=\mbox{const}\). Note that
\( a_{i0}=A_i(0,0)\).

By taking residues of
the Hirota equations in the form (\ref{bi2}),
(\ref{bi3}), one can obtain
bilinear  relations for the \(\tau\)-functions on the neighboring
levels of this chain. In particular, we have:
\begin{equation}\label{Back6}
\varepsilon_{ij}\tau_u^{[ij]}({\bf t})
\tau_{u+1}^{[k]}({\bf t}) +
\varepsilon_{jk}\tau_u^{[jk]}({\bf t})
\tau_{u+1}^{[i]}({\bf t})
+\varepsilon_{ki}\tau_u^{[ki]}({\bf t})
\tau_{u+1}^{[j]}({\bf t})=0,
\end{equation}
\begin{equation}\label{Back7}
\varepsilon_{ij}\tau_u^{[ij]}({\bf t})
\tau_{u+1}({\bf t})=
\tau_{u}^{[i]}({\bf t})\tau_{u+1}^{[j]}({\bf t})
-\tau_{u}^{[j]}({\bf t})\tau_{u+1}^{[i]}({\bf t}).
\end{equation}
We can also rewrite the determinant \eqref{Back5gen} in terms of \eqref{basictau}:
\begin{align}
\tau_{u}^{[i_1\ldots i_n]}({\bf t})
\displaystyle{=\, \Bigl (\prod_{\alpha =1}^{N}
a_{\alpha 0}\, \Bigr )^{-N+n+1}
\det\limits_{\substack{\\
1\leq r \leq N, \neq i_1,\ldots , i_n\\s=1, \ldots , N-n}}
\Bigl [ \tau_{u-s+1}^{[1,2,\dots, \not r, \dots, N]} ({\bf t})\Bigr ].}
 \label{Back5gen-2}
\end{align}
Note that \eqref{Back6} and \eqref{Back7}
are respectively the Pl\"{u}cker and Jacobi
identities for \eqref{Back5gen-2}.

Equation (\ref{rat13d}) allows one to make 
the same B\"acklund transformations by picking the coefficients
in front of higher order poles of the function 
\(\tau_{u}({\bf t}+[z^{-1}])\) at \(z=p_1, \ldots , p_N\). 
The results differ by normalization factors independent 
of \(u, {\bf t}\). For example, instead of taking residues
one may pick up the highest singularity (the pole 
of order \(M_i +1\) at \(p_i\)) and define
\begin{equation}\label{Back8}
\begin{array}{lll}
\tau_{u}^{(i)}({\bf t})&=&(-1)^{i-1}\mbox{res}_{z=p_i}\Bigl (
(z\! -\! p_i)^{M_i}
z^{-u-1}e^{-\xi ({\bf t},z)}\tau_{u+1}({\bf t}+[z^{-1}])\Bigr )
\\ && \\
&=&(-1)^{i-1} p_{i}^{-u-1}e^{-\xi ({\bf t},p_i)}
\, \mbox{res}_{z=p_i}\Bigl (
(z\! -\! p_i)^{M_i}
\tau_{u+1}({\bf t}+[z^{-1}])\Bigr )
\\ && \\
&=&M_{i}! \, a_{iM_i} \bar A_i(u-1, {\bf t}).
\end{array}
\end{equation}
In a similar way, one may define the ``undressing 
chain'' of such transformations
as in \eqref{Backrec} and (\ref{Back5gen}): 
\begin{multline}
\tau_{u}^{(i_1 \ldots i_n)}({\bf t})= 
(-1)^{i_{n}-1-
\mathrm{Card} \{ k | i_{k} < i_{n}, 1 \le k \le n-1 \} 
}
\\
\times 
\mbox{res}_{ z_{i_{n}}=p_{i_{n}} }
\left( (z_{i_{n}}-p_{i_{n}} )^{M_{i_{n}}}
z_{i_{n}}^{-u-1}e^{-\xi ({\bf t},z_{i_{n}})}
\tau_{u+1}^{(i_1\ldots i_{n-1})}
({\bf t}+[z^{-1}_{i_{n}}])
\right), 
 \label{Back9-def}
\end{multline}
which leads to
\beq\label{Back9}
\tau _{u}^{(i_1 \ldots i_n)}({\bf t})=
\Bigl (\prod_{\alpha =1}^{n}M_{i_{\alpha}}\! !\, 
a_{i_{\alpha} M_{i_{\alpha}} }\, \Bigr )
\det\limits_{\substack{\\
1\leq r \leq N, \neq i_1,\ldots , i_n\\s=1, \ldots , N-n}}
\Bigl [ A_r (u-s, {\bf t})\Bigr ].
\eeq
The construction of the undressing based on the highest poles
is somewhat more complicated because it explicitly 
uses the multiplicities
of the highest poles but the advantage is that it 
always gives the non-vanishing result
(because the coefficients in front of the highest poles cannot
vanish by their definition).

\subsection{Fermionic realization of rational solutions}

Here we suggest a realization of the rational solutions
as vacuum expectation values of some fermionic operators.

As before, we fix \(N\) points \(p_i\) and
the coefficients \(a_{im}\).
Let us introduce the following fermionic
operators:
\begin{equation}\label{fermi1}
\Psi _i (z):= \sum_{m=0}^{M_i}a_{im}\, \p_{z}^{m}\psi (z)
\end{equation}
The operators \(\Psi_i(p_i)\) are particular
cases of \(\Psi_{I_i}\) (\ref{I1})
when each set \(I_i\) contains just one element \(N+1-i\). 
Therefore, we already know that
the matrix element
\begin{equation}\label{fermi2}
\tau_n({\bf t})=\lvacn e^{J_+({\bf t})}\Psi_1(p_1)\ldots
\Psi_N(p_N) \! \left |\, n\! -\! N\right >
\end{equation}
is a \(\tau\)-function.

On the other hand, we can calculate it
using the Wick's theorem in the form (\ref{Wick3aa}).
We get:
\begin{align}
\begin{array}{lll}
\tau_n({\bf t})&=\displaystyle{\det_{i,j=1, \ldots , N}
\left < n-j+1\right |
e^{J_+({\bf t})}\Psi_i(p_i)\left | n-j\right >}
\\ & \\
&=\displaystyle{\left. \det_{i,j=1, \ldots , N}\sum_{m=0}^{M_{i}}
a_{im}\p_{z}^{m}\left [e^{\xi ({\bf t},z)}
\left < n\! -\! j\! +\! 1\right |
\psi(z)\psistar_{n-j}\left | n\! -\! j\! +\! 1\right >\right ]
\right |_{z=p_i}}
\\ & \\
&=\displaystyle{\left. \det_{i,j=1, \ldots , N}\sum_{m=0}^{M_{i}}
\sum_{k\in \z}a_{im}\p_{z}^{m} \left [z^k e^{\xi ({\bf t},z)}\left <
n\! -\! j\! +\! 1\right |  \psi_k
\psistar_{n-j}\left | n\! -\! j\! +\! 1\right >\right ] \right |_{z=p_i}}
\\ & \\
&=\displaystyle{\left. \det_{i,j=1, \ldots , N}\sum_{m=0}^{M_{i}}
a_{im}\p_{z}^{m} \Bigl [z^{n-j}e^{\xi ({\bf t},z)}\Bigr ] \right |_{z=p_i}}
\\ & \\
&=\displaystyle{\det_{i,j=1, \ldots , N}A_i(n\! -\! j, {\bf t})}
\end{array} 
 \label{tau-fermi0}
\end{align}
which coincides with the \(\tau\)-function (\ref{rat6})
after the change \(n\to u\). Let us also note the formulas
\begin{equation}\label{fermi3}
\tau_n \left ({\bf t}\! -\! [z^{-1}]\right )=z^{-n}e^{-\xi ({\bf t}, z)}
\left <n+1\right | e^{J_+({\bf t})}\psi (z)\Psi_1(p_1)\ldots
\Psi_N(p_N) \! \left |\, n\! -\! N\right >
\end{equation}
\begin{equation}\label{fermi4}
\tau_n \left ({\bf t}\! +\! [z^{-1}]\right )=z^{n-1}e^{\xi ({\bf t}, z)}
\left <n-1\right | e^{J_+({\bf t})}\psistar (z)\Psi_1(p_1)\ldots
\Psi_N(p_N) \! \left |\, n\! -\! N\right >
\end{equation}
which directly follow from the bosonization rules (\ref{bosonization}).

The simplest example is \(M_i=0\) when
\(\Psi_i(p_i)=\psi (p_i)\)
(we take \(a_{i0}=1\))
and
\[
\tau^{(M_i=0)}_n({\bf t})=\lvacn e^{J_+({\bf t})}\psi (p_1)\ldots
\psi(p_N) \! \left |\, n\! -\! N\right >
\]
which is easily calculated to be
\(\displaystyle{
\tau^{(M_i=0)}_n({\bf t})=\prod_{i>j}^{N}(p_{i}^{-1}-p_{j}^{-1})\,
\prod_{k=1}^{N}p_{k}^{n-1} e^{\xi ({\bf t}, p_k)}}
\).

In the same way as \eqref{fermi2}, we have a fermionic realization 
of the intermediate \(\tau\)-functions for the B\"{a}cklund transformations. The \(\tau\)-functions corresponding to 
 \eqref{Back5gen} have 
the form
\begin{align}
\tau_{n}^{[i_1\ldots i_m]}({\bf t})&=
  \, \Bigl (\prod_{\alpha =1}^{m}
a_{i_{\alpha} 0}\, \Bigr ) 
\lvacn e^{J_+({\bf t})} 
\prod_{j=1, \ne i_{1},\dots, i_{m}}^{\overrightarrow{N}}
\Psi_j(p_j)
 \! \left |\, n\! -\! N+m \right >, 
 \label{tau-bac1}
\end{align}
where \(m\) is a non-negative integer (\(0\le  m \le N\)). 
The \(\tau\)-functions corresponding to \eqref{Back9-def} 
have the same form except for the prefactor \((\cdots)\). 
These \(\tau\)-functions reduce to \eqref{fermi2} for \(m=0\). 
The determinant formulas \eqref{Back5gen} 
for these \(\tau\)-functions 
 follow from the Wick's theorem \eqref{Wick3aa} in 
the same way as \(m=0\) case \eqref{tau-fermi0}. 
The BA functions for \eqref{tau-bac1} are defined by
\begin{equation}\label{BA-bac}
\psi^{[i_1\ldots i_m]}_n({\bf t}, z)=z^n e^{\xi ({\bf t},z)}
\, \frac{\tau_n^{[i_1\ldots i_m]}
({\bf t} -[z^{-1}])}{\tau_n^{[i_1\ldots i_m]}({\bf t})}.
\end{equation}
The BA functions for \eqref{Back9-def} are defined exactly in the same way. In addition, both of them coincide under the condition that the  
prefactor of the \(\tau\)-functions
are not zero. 
By using a formula similar to \eqref{fermi3}, we obtain
\begin{align}
\psi^{[i_1\ldots i_m]}_n({\bf t}, z)&=
\frac{
\left <n+1\right | e^{J_+({\bf t})} 
\psi(z) 
\prod_{j=1, \ne i_{1},\dots, i_{m}}^{\overrightarrow{N}}
\Psi_j(p_j)
 \! \left |\, n\! -\! N+m \right >
}{
\lvacn e^{J_+({\bf t})} 
\prod_{j=1, \ne i_{1},\dots, i_{m}}^{\overrightarrow{N}}
\Psi_j(p_j)
 \! \left |\, n\! -\! N+m \right >
}
.
\label{tau-bac3}
\end{align}
Then the Krichever's conditions for the B\"{a}cklund transformations
\begin{align}
\sum_{j=0}^{M_k} a_{kj}\,
\p_{z}^{j} \psi_n^{[i_1\ldots i_m]}
 ({\bf t},z)\Bigr |_{z=p_k}\!\! =0
\quad \text{for} \quad k \in \{1,2,\dots, N\} \setminus \{i_{1},\dots, i_{m}\}
\end{align}
follow straightforwardly from \eqref{tau-bac3} and the relation \((\Psi_k(p_k))^{2}=0\).

\section{The \texorpdfstring{$\tau$}{tau}-function in quantum 
integrable models}
\label{sec:quantum}

\subsection{\texorpdfstring{$T$}{T}-operators}
\label{sec:t-operators}

We consider generalized quantum integrable spin chains with
rational \(GL(N)\)-invariant \(R\)-matrix. The \(R\)-matrix acting
in the tensor product of the \(N\)-dimensional vector representation
of \(GL(N)\) and an arbitrary finite dimensional representation
\(\pi_{\lambda}\) (labeled by a Young diagram \(\lambda\)) has the form
\begin{equation}\label{QT1}
R^{\lambda}(u)=u\,
\I \otimes \I 
+\sum_{i,j=1}^{N}e_{ji}\otimes \pi_{\lambda}(e_{ij}),
\end{equation}
where \(u\in \CC\) is the spectral parameter\footnote{Our definition of the
spectral parameter differs from that in
\cite{KSZ08,Kazakov:2007na} by a factor of 2}
and \(e_{ij}\) are standard generators
of the algebra \(gl(N)\) with matrix elements 
\( (e_{ij})_{\alpha \beta}=\delta_{i\alpha}\delta_{j\beta}\).  
In the second term, the vector representation (the one for the Young  diagram with one box) \(\pi_{\square }(e_{ji})\) 
is abbreviated to \(e_{ji}\). 

A large family of commuting operators
(quantum transfer matrices or simply \(T\)-operators) can be constructed as
\begin{equation}\label{QT2}
T^{\lambda }(u)=\mbox{tr}_{\pi_\lambda}\left (
R^{\lambda}(u-\xi_L)\otimes \ldots \otimes R^{\lambda}(u-\xi_1)
(\I^{\otimes L} \otimes \pi_{\lambda}(g) ) \right ),
\end{equation}
where \(\xi_i\) are arbitrary parameters (inhomogeneities at the lattice sites)
and \(g\in GL(N)\) is called the twist matrix.
For technical simplicity, we assume below that the twist matrix
has distinct, non-zero eigenvalues. 
The trace is taken in the
auxiliary space where the representation \(\pi_{\lambda}\) acts.
The \(T\)-operators act in the physical Hilbert space of the model
\({\cal H}=(\CC ^N)^{\otimes L}\).
The Yang-Baxter equation implies that the \(T\)-operators with the same \(g\)
commute for all \(u\) and \(\lambda\) and can be diagonalized simultaneously.
It follows then from (\ref{QT1}) and (\ref{QT2})
that all their eigenvalues are polynomials
in \(u\) of degree \(\leq L\).

It is natural to define the \(T\)-operator
for the trivial representation (empty Young diagram) as
\begin{equation}\label{phia}
T^{\emptyset }(u) =\prod_{i=1}^{L}(u-\xi_i),
\end{equation}
where the multiplication by the identity operator in the r.h.s. is
implicit. We introduce a special notation
\begin{equation}\label{phi}
\phi (u):=\prod_{i=1}^{L}(u-\xi_i)
\end{equation}
for the polynomial in the r.h.s. This polynomial characterizes the
model through its zeros \(\xi_i\). Another normalization,  sometimes
more convenient, is to set
\begin{equation}\label{QT3}
{\sf T}^{\lambda }(u)=\frac{T^{\lambda }(u)}{T^{\emptyset }(u)}.
\end{equation}
In this normalization all \(T\)-operators are rational functions
which may have poles only at the points \(\xi_i\) and
\({\sf T}^{\emptyset}(u)=1\).

\subsection{Determinant formulas for \texorpdfstring{$T$}{T}-operators}

The \(T\)-operators for different representations in the auxiliary space
are known to obey an infinite number of functional relations which
follow from the fusion procedure and, eventually, from the Yang-Baxter
equation.

First, there are determinant formulas which express \(T^{\lambda }(u)\)
for arbitrary \(\lambda\) through \(T\)-operators for symmetric representations
\(T_s(u):=T^{(s)}(u)\) corresponding to one-row diagrams \((s)\)
\cite{Chered,BR90}.
They are called Cherednik-Bazhanov-Reshe\-tikhin (CBR) formulas
or quantum Jacobi-Trudi identities. In our
normalization they have the form 
\begin{equation}\label{det1}
T^{\lambda }(u)=\left (\prod_{k=1}^{\lambda_{1}'-1}
\phi (u-k)\right )^{-1}\det_{i,j=1, \ldots , \lambda_{1}'}
T_{\lambda_i-i+j}(u-j+1).
\end{equation}
A direct proof ``from the first principles'' for the models with the rational
\(GL(N|M)\)-invariant \(R\)-matrix was given in \cite{Kazakov:2007na}.
The proof uses the operation of co-derivative with respect to the twist matrix
\(g\) (see below).
When the dependence on the spectral parameter disappears (say, in the limit
\(u\to \infty\)) one obtains the standard Jacobi-Trudi formulas
(\ref{schur2}) for characters.
We note in passing that the determinant formula
for the  generalized Schur functions 
(constructed in terms of the Gauss decomposition 
of a matrix 
(resp.\ orthogonal polynomials)) obtained in 
\cite{NNSY00} (resp.\ 
\cite{SV09}) has a structure similar to (\ref{det1}).

Second, there are determinant formulas of the Giambelli type
which express \(T^{\lambda }(u)\)
for arbitrary \(\lambda\) through  the \(T\)-operators
\(T_{l,k}(u):=T^{(l+1, 1^k)}(u)\) corresponding to the hook diagrams
\((l+1, 1^k)\) \cite{KOS96}:
\begin{equation}\label{det2}
T^{\lambda }(u)=\left (\phi (u)\right )^{-d(\lambda )+1}
\det_{i,j=1, \ldots , d(\lambda )}T_{\lambda_i -i, \lambda_{j}'-j}(u).
\end{equation}
Note that there are no shifts of the spectral parameter here.
In Appendix C we prove that the determinant representation
of the Giambelli type (\ref{det2}) follows from
the quantum Jacobi-Trudi identities (\ref{det1}).

In the normalization (\ref{QT3}) the pre-factors
cancel and the determinant formulas look simpler:
\begin{equation}\label{det1a}
{\sf T}^{\lambda }(u)=\det_{i,j=1, \ldots , \lambda_{1}'}
{\sf T}_{\lambda_i-i+j}(u-j+1)=
\det_{i,j=1, \ldots , d(\lambda )}{\sf T}_{\lambda_i -i, \lambda_{j}'-j}(u).
\end{equation}

Below we will argue that these formulas allow one to
establish a close connection
with classical MKP integrable hierarchy and to recover a hidden
free fermionic structure in the auxiliary space.

\subsection{The master \texorpdfstring{$T$}{T}-operator}
\label{sec:master-t-operator}

Let us introduce a generating function of the \(T\)-operators
(the master \(T\)-operator) depending on an infinite number of
parameters \({\bf t}=\{t_1, t_2, \ldots \}\):
\begin{equation}\label{master}
T(u,{\bf t})=\sum_{\lambda}s_{\lambda}({\bf t})T^{\lambda }(u).
\end{equation}
These operators commute for different values of the
parameters: \CMPorArticle{\linebreak}{} \([T(u,{\bf t}),\, T(u',{\bf t'})]=0\).
The \(T\)-operators \(T^{\lambda}(u)\) can be restored from the
master \(T\)-operator according to the formula
\begin{equation}\label{master33}
\left. \phantom{\int}
T^{\lambda}(u)=s_{\lambda}(\tilde \p )T(u,{\bf t})\right |_{{\bf t}=0},
\end{equation}
where \(\tilde \p =\{\p_{t_1} , \frac{1}{2}\p_{t_2}, \frac{1}{3}\p_{t_3},
\ldots \, \}\). In particular, \(T(u, 0+[z^{-1}])\) is the
generating series for \(T\)-operators corresponding to the 
symmetric representations:
\beq\label{master44}
T(u, [z^{-1}])=\sum_{s\geq 0}z^{-s}T_s(u).
\eeq

\subsubsection{The master \texorpdfstring{$T$}{T}-operator as a \texorpdfstring{$\tau$}{tau}-function}

Comparing formulas (\ref{det1}) with (\ref{tau4a}) and
(\ref{det2}) with (\ref{tau3}), one can see that they coincide
after the identification \(u\leftrightarrow n\) and
\(T^{\lambda }(u)\leftrightarrow c_{\lambda}(n)\).
According to section \ref{sec:Jacobi-Trudi},
this implies that
there exists a fermionic group-like element \(\hat G\)
(an operator in the
quantum space of the model) and a \(u\)-dependent operator
\(\hat C(u)\) such that
\begin{equation}\label{master1}
T^{\lambda }(u)=(-1)^{b(\lambda )}\hat C(u)
\lbr \lambda , u\right |\hat G \left |u\rbr
\end{equation}
(which makes sense at least for integer values of \(u\)), and thus
the master \(T\)-operator is the \(\tau\)-function of the MKP hierarchy:
\begin{equation}\label{master2}
T(u,{\bf t})=\hat C(u)\lbr u \right |e^{J_{+}({\bf t})}\hat G\left |u\rbr .
\end{equation}
Setting \(t_k=0\), we get \(T(u,0)=T^{\emptyset }(u)\) whence
\begin{equation}\label{master2a}
\hat C(u)\lbr u\right |\hat G \left |u\rbr =\phi (u)
\end{equation}
(multiplication by the identity operator in the quantum space of
the model is implied in the r.h.s.)

A weaker statement, that
the master \(T\)-operator is the \(\tau\)-function of the KP-hierarchy if
\(T^{\lambda }(u) \) in \eqref{master} has the CBR-determinant
representation,
can be proved independently in the following way.
Suppose \(T^{\lambda }(u) \) has the CBR determinant
representation \eqref{det1}.
Then it can be expressed  as the
Giambelli determinant \eqref{det2} in the Frobenius
notation for the Young diagram
\(\lambda =(\alpha_{1},\dots, \alpha_{d} | \beta_{1}, \dots, \beta_{d})\)
(see Appendix C).
For any \(k,l \in \{1,2,\dots, d(\lambda )\}\), \(k<l\), the Jacobi identity for this Giambelli determinant reads:
\begin{multline}
T^{(\alpha_{1},\dots, \hat{\alpha}_{k}, \dots, \hat{\alpha}_{l}, \dots , \alpha_{d} |
\beta_{1}, \dots, \hat{\beta}_{k}, \dots, \hat{\beta}_{l}, \dots ,
\beta_{d})}(u)
T^{(\alpha_{1},\dots, \alpha_{d} |
\beta_{1}, \dots, \beta_{d})}(u)
\\
-
T^{(\alpha_{1},\dots, \hat{\alpha}_{k}, \dots , \alpha_{d} |
\beta_{1}, \dots, \hat{\beta}_{k}, \dots, \beta_{d})}(u)
T^{(\alpha_{1},\dots, \hat{\alpha}_{l}, \dots , \alpha_{d} | \beta_{1},
\dots, \hat{\beta}_{l}, \dots, \beta_{d})}(u)
\\
+
T^{(\alpha_{1},\dots, \hat{\alpha}_{k}, \dots , \alpha_{d} | \beta_{1},
\dots, \hat{\beta}_{l}, \dots, \beta_{d})}(u)
T^{(\alpha_{1},\dots, \hat{\alpha}_{l}, \dots , \alpha_{d} | \beta_{1},
\dots, \hat{\beta}_{k}, \dots, \beta_{d})}(u)
=0,
\end{multline}
where \( \hat{\alpha}_{k} \) (resp.\  \( \hat{\beta}_{l} \))
means that \(\alpha_{k} \)
(resp.\  \(\beta_{l} \))   is removed from the original diagram
\((\alpha_{1},\dots, \alpha_{d} | \beta_{1}, \dots, \beta_{d})\).
This is nothing but the 3-term relation given
in \cite{Sato81}, which is known to be the necessary and sufficient
condition that \(T(u,{\bf t})\) solves the KP hierarchy\footnote{\(\alpha_{k}\)
and \(\beta_{k}\) here
are related to the parameters
in page 959 of \cite{JM83} as \(i_k=\alpha_k, j_k= -\beta_k -1, r=d\).
}.
Therefore, \(T(u,{\bf t})\) is a \(\tau\)-function.
In other words, the functional relations (\ref{det1}), (\ref{det2}) mean
that the \(T\)-operators \(T^{\lambda }(u)\)
are Pl\"ucker coordinates for the \(\tau\)-function \(T(u,{\bf t})\).

We emphasize that the notion of the master \(T\)-operator is independent of the
specific form of the \(R\)-matrix\footnote{
For  the trigonometric case  (or the rational case), we can define the
``universal master \(T\)-operator''.
Let \(R\) be the universal R-matrix associated with
 \(U_{q}(\hat{gl}(N))\). \(R\) is considered to be an element of
\(B_{+} \otimes B_{-}\), where \(B_{\pm}\) are Borel subalgebras.
Let \(\pi_{\lambda }(u) \) be an evaluation representation of \(U_{q}(\hat{gl}(N))\)
based on a tensor representation of \(U_{q}(gl(N))\)
 labeled by the Young diagram \(\lambda \).
This representation depends on the spectral parameter \(u\) 
since the evaluation map does.
We can define the universal \(T\)-operator (cf.\ \cite{Bazhanov:1998dq}) as
\(T^{\lambda }(u)=(\mathrm{Tr} \otimes \mathrm{id}) (\pi_{\lambda }(u) \otimes \mathrm{id})((g \otimes 1)R)\),
where \(g\) is a twist element (a function of the Cartan elements).
Then our universal master T-operator is obtained by substituting this into \eqref{master}.
Now \eqref{master} is an element of  \(B_{-}\).
One can obtain a master \(T\)-operator
on a particular quantum space \(V \) if
we consider a representation of \(B_{-}\).
For example, if we consider \(V=\pi_{(1)}(\xi_{1}) \otimes \pi_{(1)}(\xi_{2}) \otimes \cdots \otimes \pi_{(1)}(\xi_{L})\),
where \(\{ \pi_{(1)}(\xi_{j})\}_{j=1}^{L}\) are fundamental representations
we obtain the master \(T\)-operator for a lattice model
in section \ref{sec:master-t-operator-1} (as a limit \(q \to 1\)).
If we consider a space of the vertex operators of CFT for
\(V\), we will obtain a master \(T\)-operator for CFT.
It is tempting to rewrite the universal master \(T\)-operator as
a kind of determinant over some function of \((\pi_{(1)}(u) \otimes \mathrm{id})(R)\),
where \(\pi_{(1)}(u)\) is the fundamental representation.
This should be a generalization of a generating function of transfer matrices (cf.\ \cite{Talalaev04}).}.
The key point is that \(T^{\lambda }(u)\) in
\eqref{master}
is the transfer matrix obtained by the standard fusion procedure
in the auxiliary space (the horizontal leg)
labeled by the Young diagram
\(\lambda \). The master \(T\)-operator is thus well-defined
for spin chains with trigonometric and elliptic \(R\)-matrices
as well and the
statement that it is a \(\tau\)-function still holds.
The statement is also applicable to
the \(T\)-operator corresponding to any higher representation
in the quantum space of the model.
It can be also applicable to the \(q\)-characters of
representations of quantum affine algebras
\cite{FR99}.

\subsubsection{Hirota equations for the master \texorpdfstring{$T$}{T}-operator}
\label{sec:HirMasterT}

As any \(\tau\)-function, the master \(T\)-operator satisfies the
bilinear identity
\begin{equation}\label{bi2a101}
\oint_{C} e^{\xi ({\bf t}-{\bf t'}, z)}
z^{u-u'} T(u, {\bf t} -[z^{-1}])\, T(u', {\bf t'} +[z^{-1}]) \, dz=0
\end{equation}
for all \(u,u', {\bf t}, {\bf t'}\). 
Here the integrand should be interpreted as a formal power series 
on \(z\) and the contour \(C\) should encircle \(z=\infty\). 
By standard manipulations, one can derive from (\ref{bi2a101}) the
infinite KP and MKP hierarchies of differential equations.
Making the shift of variables \({\bf t} \to {\bf t}+
{\bf a}\), \({\bf t}^{\prime} \to {\bf t}-{\bf a}\)
in \eqref{bi2a101}, where
\({\bf a}=(a_{1},a_{2},\dots )\) and setting
\(u-u^{\prime} \in {\mathbb Z}_{\ge 0}\),
one obtains, in the usual way, the following Hirota-type bilinear
differential equation for the MKP hierarchy:
\begin{align}
\sum_{m=0}^{\infty} h_{m}(2{\bf a}) h_{m+u-u^{\prime} +1}
(-\tilde{{\mathbf D}}) e^{\sum_{k=1}^{\infty}
a_{k}D_{t_{k}} } T(u,{\bf t}) \cdot T(u^{\prime},{\bf t})=0.
\end{align}
Here \( \tilde{{\mathbf D}}_{t} =(D_{t_{1}},
\frac{1}{2}D_{t_{2}},\frac{1}{3}D_{t_{3}},\ldots )\), where \(D_{t_{k}}\)
is the Hirota derivative with respect to \(t_{k}\).
For example, the first equation of the MKP hierarchy reads
\begin{equation}\label{MKP1}
\p_{t_2}\log \frac{T(u+1, {\bf t})}{T(u, {\bf t})}=
\p_{t_1}^{2}\log \Bigl (T(u+1, {\bf t})T(u, {\bf t})\Bigr )
+\Bigl ( \p_{t_1}\log \frac{T(u+1, {\bf t})}{T(u, {\bf t})}\Bigr ).
\end{equation}

The bilinear relations (\ref{bi2}), (\ref{bi3}) for
\(T(u,{\bf t})\) can be written as
\begin{equation}\label{bi2a}
(z_2-z_3)T\left (u, {\bf t}+[z_{1}^{-1}]\right )T\left (u,
{\bf t} +[z_{2}^{-1}]+[z_{3}^{-1}]\right ) \, + (231) +(312)\, =0
\end{equation}

\begin{equation}\label{bi3a}
\begin{array}{c}
z_2 T\left (u+1, {\bf t}+[z_{1}^{-1}]\right )
T\left (u, {\bf t}+[z_{2}^{-1}]\right )
-z_1 T\left (u+1, {\bf t}+[z_{2}^{-1}]\right )
T\left (u, {\bf t}+[z_{1}^{-1}]\right )
\\ \\
+\,\, (z_1 -z_2)T\left (u+1, {\bf t}+[z_{1}^{-1}] +[z_{2}^{-1}]\right )
T\Bigl (u,{\bf t}\Bigr )=0
\end{array}
\end{equation}
(comparing to (\ref{bi2}), (\ref{bi3}), we have made the
overall shift of the time variables \({\bf t} \rightarrow
{\bf t}+[z_{1}^{-1}]+[z_{2}^{-1}]+[z_{3}^{-1}]\) in (\ref{bi2a}) and
\({\bf t} \rightarrow {\bf t}+[z_{1}^{-1}]+[z_{2}^{-1}]\) in (\ref{bi3a})).

For any positive integer \(K\) and any subset 
\( \{ i_{1},i_{2},\dots, i_{n} \} \subset \{1,2,\dots, K \}\) where all  \(i_k\) are different,
we introduce the operator
\begin{align}
T^{ \{ i_{1},i_{2},\dots, i_{n} \}  }(u, {\bf t}) = T(u, {\bf t} +
\sum_{k=1}^{n} [z_{i_{k}}^{-1}]).
 \label{mastershift}
\end{align}
Solving \eqref{bi3a} for \eqref{mastershift} recursively, we obtain the determinant formulas
\begin{align}
T^{ \{ i_{1},i_{2},\dots, i_{n} \}  }(u, {\bf t}) &=
\frac{
\mathrm{det}_{1 \le k,j \le n} \left( z_{i_{k}}^{j-n} 
T^{ \{ i_{k} \}  }(u-j+1, {\bf t}) \right)
}
{
\mathrm{det}_{1 \le k,j \le n} \left( z_{i_{k}}^{j-n} \right)
\prod_{k=1}^{n-1}  T(u-k, {\bf t})
} ,
\label{masterdet1}
\\[6pt]
T^{\{1,2,\dots, K \} \setminus \{ i_{1},i_{2},
\dots, i_{n} \}  }(u, {\bf t}) &=
\frac{
\mathrm{det}_{1 \le k,j \le n} \left( z_{i_{k}}^{j-n}
T^{ \{1,2,\dots, K \} \setminus  \{ i_{k} \}  }(u+j-1, {\bf t}) \right)
}
{
\mathrm{det}_{1 \le k,j \le n} \left( z_{i_{k}}^{j-n} \right)
\prod_{k=1}^{n-1}  T^{\{1,2,\dots, N \} }(u+k, {\bf t})
} ,
\label{masterdet2}
\end{align}
Note that \eqref{bi2a} is a Pl\"{u}cker identity for \eqref{masterdet1} and
\eqref{bi3a} is the Jacobi identity for \eqref{masterdet1}.

\subsubsection{The master \texorpdfstring{$T$}{T}-operator in models with rational \texorpdfstring{$R$}{R}-matrices}
\label{sec:master-t-operator-1}

For models with rational \(R\)-matrices, the \(T\)-operators
\(T^{\lambda}(u)\) (\ref{QT2})
admit a more explicit
realization in terms of the co-derivative operators
acting on functions of the
twist matrix \(g\in GL(N)\) \cite{Kazakov:2007na}:
\begin{equation}\label{QT4}
\begin{array}{lll}
T^{\lambda }(u)&=&(u-\xi_1 +\hat D)\otimes \ldots \otimes
(u-\xi_L +\hat D)\chi _{\lambda }(g)
\\ &&\\
&= &\displaystyle{
\bigotimes_{i=1}^{L}\, (u-\xi_i +\hat D)\chi _{\lambda }(g)},
\end{array}
\end{equation}
where \(\chi_{\lambda}(g)=\mbox{tr}\,
\pi_{\lambda}(g)\) is the character of \(g\) in the representation
\(\pi_{\lambda}\). All necessary information about the co-derivative \(\hD\) is given in Appendix D.

Set
\begin{equation}\label{F}
F_g({\bf t})=\exp \left (\sum_{k\geq 1}t_k \, \mbox{tr}\, g^k\right ).
\end{equation}
As immediately follows from its definition, the master \(T\)-operator (\ref{master})
can be written in terms of the co-derivative  as follows
\begin{equation}\label{master3}
T(u, {\bf t})=\bigotimes_{i=1}^{L}\, (u-\xi_i +\hat D)F_g({\bf t}).
\end{equation}
Note that each term in the direct product introduces a spin into the underlying quantum spin chain.

 Alternatively, using the fact that \((\det g)^{-u} \hat D (\det g)^{u}=\hat D
+u\), we can also write it as
\begin{equation}\label{master301}
T(u, {\bf t})=(\det g)^{-u} \bigotimes_{i=1}^{L}\,
(\hat D -\xi_i)F_{g}(u,{\bf t}),
\end{equation}
hiding the spectral parameter into the trivial \(\tau\)-function (with no spins) 
\[
F_{g}(u,{\bf t})=(\det g)^{u} F_g({\bf t})=(\det g)^{u} \exp
\Bigl (\sum\limits_{k\geq 1}t_k \, \mbox{tr}\, g^k\Bigr ).
\]
In this form it is clear that the spectral parameter is naturally
combined with \(t_i\) into a hierarchy of times \(u=t_0, t_1, t_2,
\ldots\):
\begin{align}
 F_{g}(u,{\bf t})
=\exp
\Bigl (\sum\limits_{k\geq 0}t_k \, \mbox{tr}\, g^{(k)}\Bigr ),
\quad \textrm{where~} g^{(k)}=\left\{
  \begin{array}{cc}
    g^{k}&\textrm{if~} k>0\\
    \log g&\textrm{if~} k=0.
  \end{array}
\right.
\end{align}
The explicit form of the simplest \(T\)-operators for a chain of length \(L=1,2\)  is:
\begin{align}
  T_{(L=1)}(u,{\bf t})=&\left ((u-\xi_1){\I} +\sum_{k\geq
      1}kt_kg^k\right ) \exp \Bigl ( \sum_{l\geq 1}t_l \, \mbox{tr}\,
  g^l\Bigr ),
\label{eq:Texplicit1}
\end{align}
{\newcommand{\cntnt}{ \label{eq:Texplicit2}
T_{(L=2)}(u,{\bf t})=\CMPorArticle{}{&}\left ((u-\xi_1)(u-\xi_2)\, {\I}\otimes {\I} +
(u-\xi_1)\sum_{k\geq 1}kt_k \, {\I}\otimes g^k\CMPorArticle{\right.\\\qquad\qquad}{}
+(u-\xi_2)\sum_{k\geq 1}kt_k \, g^k \otimes {\I} 
  \CMPorArticle{}{\right.\nonumber\\&} +\, \sum_{k\geq 1}\sum_{k'\geq 1}
kk' t_k t_{k'}\, g^k \otimes g^{k'}
\CMPorArticle{\\}{}\left.
+\sum_{k\geq 1}\sum_{\alpha =0}^{k-1}kt_k\, {\cal P}\,
(g^{\alpha}\otimes g^{k-\alpha})\right )
\exp \Bigl ( \sum_{l\geq 1}t_l \, \mbox{tr}\, g^l\Bigr ),} \CMPorArticle{\begin{multline}
\cntnt
\end{multline}
}{\begin{align}
\cntnt
\end{align}}}
where \({\cal P}\) is the permutation operator (see Appendix D for  definitions).
For clarity, the unit matrix \(\I\) is everywhere written explicitly.

Using the method of \cite{KLT10}, it is possible to show that
the master \(T\)-operator contains the whole family of Baxter
\(Q\)-operators as well as the \(T\)-operators that emerge
on intermediate levels of the nested Bethe ansatz.
This will be explained
in section \ref{sec:backlund-flow-baxter}.

It follows from (\ref{master3}) that
\begin{equation}\label{master3a}
\begin{array}{l}\displaystyle{
T(u, {\bf t} - [z^{-1}])=\bigotimes_{i=1}^{L}\,
(u \! -\! \xi_i \! +\! \hat D)
\left [(\det(\I -z^{-1}g)) F_g({\bf t})\right ],}
\\ \\\displaystyle{
T(u, {\bf t} + [z^{-1}])=\bigotimes_{i=1}^{L}\,
(u \! -\! \xi_i \! +\hD)
\left [(\det(\I -z^{-1}g))^{-1}F_g({\bf t})\right ]}.
\end{array}
\end{equation}
Using these formulas, one can explicitly verify that
\beq \label{adddef2}
\lim_{z\to 0}\left [(-z)^{\pm N} T(u, {\bf t} \mp [z^{-1}])\right ]
=(\det g)^{\pm 1}T(u\pm 1, {\bf t}).
\eeq
Comparing with the prescription
\eqref{adddef1}, we see that there is an extra factor
\((\det g)^{\pm 1}\) which is due to a slightly different definition
of the \(\tau \)-function: in our case it is a pure polynomial in
\(u\) while in \eqref{adddef1} it is a polynomial multiplied by
an exponential function.

Denoting \(\hcD_L(u):=\bigotimes_{i=1}^{L}(u \! -
\! \xi_i \! +\hD)\)
and \(w(z)=(\det(\I -z g))^{-1}\) for brevity,
we can rewrite (\ref{bi2a}),
(\ref{bi3a}) in the form
\begin{equation}\label{bi2b}
(z_2-z_3)\left (\hcD_L(u)w(z_1^{-1})F_g\right )
\left (\hat {\cal D}_L(u)w(z_2^{-1})w(z_3^{-1})F_g\right )
\, + (231) +(312)\, =0,
\end{equation}
\CMPorArticle{\begin{multline}\label{bi3b}
z_2 \left (\hat {\cal D}_L(u+1)w(z_1^{-1})F_g \right )
\left (\hat {\cal D}_L(u)w(z_2^{-1})F_g \right ) \\-
z_1 \left (\hat {\cal D}_L(u+1)w(z_2^{-1})F_g \right )
\left (\hat {\cal D}_L(u)w(z_1^{-1})F_g  \right )
\\
+(z_1-z_2)\left (\hat {\cal D}_L(u+1)w(z_1^{-1})w(z_2^{-1})F_g\right )
\left (\hat {\cal D}_L(u)F_g\right ) =0.
\end{multline}}{\begin{equation}\label{bi3b}
\begin{array}{c}
z_2 \!\left (\hat {\cal D}_L(u+1)w(z_1^{-1})F_g \!\right )\!
\left (\hat {\cal D}_L(u)w(z_2^{-1})F_g\! \right )\! -\!
z_1 \!\left (\hat {\cal D}_L(u+1)w(z_2^{-1})F_g\! \right )\!
\left (\hat {\cal D}_L(u)w(z_1^{-1})F_g \! \right )
\\[.3cm]
+(z_1-z_2)\left (\hat {\cal D}_L(u+1)w(z_1^{-1})w(z_2^{-1})F_g\right )
\left (\hat {\cal D}_L(u)F_g\right ) =0.\end{array}
\end{equation}}
In the second equation, one can recognize the ``master identity'' from
\cite{KLT10} (to identify them, the redefinition \(z_i\to 1/z_i\)
is required). The relation (\ref{adddef2}) implies that 
equation (\ref{bi3b}), which is the Hirota equation
for  MKP hierarchy,
follows from (\ref{bi2b}), which is the Hirota equation for
the KP hierarchy,
in the limit \(z_3\to 0\). Note that at \(t_i=0\) (i.e., \(F_g=1\))
and \(u=\xi_1=\dots=\xi_L=0\), the third
term in equation (\ref{bi3b}) vanishes since
\(\hcD_L\cdot1=0\) and the equation becomes
the ``basic identity'' (5.9)   from \cite{Kazakov:2007na}
\begin{equation}\label{bi3c}
z \left (\left(1+\hD\right)^{\otimes L}w(z)\right )
\left (\hD^{\otimes L}w(z')\right )=
z' \left (\left(1+\hD\right)^{\otimes L}w(z')\right )
\left (\hD^{\otimes L}w(z)\right ),
\end{equation}
where we put \(z=z_{1}^{-1}, z^{\prime}=z_{2}^{-1}\).

When \(z\to 0\), \eqref{bi3c} reduces to the following identity
\cite{Kazakov:2007na} linear in \(w\),
\begin{equation}\label{eq:linearId-1}
\hD^{\otimes L} w(z')=z'\left(g_1\cP_{1\dots L}\right)\left(\left(1+\hD\right)^{\otimes L}w(z')\right),
\end{equation}
where we used \(w(z)=1+z\,\tr g+O(z^{2})  \) and
\begin{equation}
\hD^{\otimes L}\,\tr g= g_1\cP_{1\dots L}.
\end{equation}
Here \(\cP_{1\dots L}=\cP_{12}\cP_{23}\cdots\cP_{L-1,L}\)
is the full cyclic permutation in the quantum space  stemming from
\eqref{co5} and from the repeated use of  \eqref{Dg}
(see Appendix D).
The subscript in \(g_1\) means that it acts, as a matrix, in the space number 1 of the spin chain.    Note that the order of operators in \eqref{bi3c} is irrelevant due to the commutation of transfer-matrices with arbitrary auxiliary representations.
Similarly, for \(z'\to 0\), in the first order in \(z'\) we obtain:
\begin{equation}\label{eq:linearId-2}
z\left(1+\hD\right)^{\otimes L}w(z)=\left(\hD^{\otimes L}w(z)\right)\left(g_1\cP_{1\dots L}\right)^{-1}.
\end{equation}
It is a direct consequence of \eqref{eq:linearId-1}
if we recall that the two factors in \eqref{eq:linearId-2} commute: both are particular cases of the same family of commuting transfer-matrices. Eqs.\eqref{eq:linearId-1} and \eqref{eq:linearId-2} were proven in a combinatorial way in \cite{Kazakov:2007na}.
Multiplying the left hand sides and the right hand sides of
 \eqref{eq:linearId-1} and \eqref{eq:linearId-2} we restore again the relation \eqref{bi3c} which served there as the main tool for the direct proof of the CBR
formulas.

As is explained above, equation \eqref{eq:linearId-2} was proven in
\cite{Kazakov:2007na} in a combinatorial way. 
Let us now show how the case \(F_g=1\) 
of \eqref{bi3b}, i.e.,
\begin{multline}
  \label{eq:simplestMID}
z_2 \!\left (\hat {\cal D}_L(u+1)w(z_1^{-1}) \!\right )\!
\left (\hat {\cal D}_L(u)w(z_2^{-1})\! \right )\! -\!
z_1 \!\left (\hat {\cal D}_L(u+1)w(z_2^{-1})\! \right )\!
\left (\hat {\cal D}_L(u)w(z_1^{-1}) \! \right )
\\
+(z_1-z_2)\left (\hat {\cal D}_L(u+1)w(z_1^{-1})w(z_2^{-1})\right )
\left (\hat {\cal D}_L(u)1\right ) =0.
\end{multline}
can be deduced from it.
First we notice that when \(L\geq
1\), equation \eqref{bi3c} is the particular case of 
\eqref{eq:simplestMID}
when \(\xi_i=u\) for all \(i=1,\ldots L\). 
Next we notice that although \eqref{bi3c} only holds if \(L\geq 1\), the
third term in \eqref{eq:simplestMID} is such that
\eqref{eq:simplestMID} 
also  holds (trivially) when \(L=0\). Since the left hand side
of \eqref{eq:simplestMID} is polynomial in the variables 
\(u_i:=u - \xi_i \), this allows for a 
recurrence procedure which proves that the
term with an arbitrary degree in each \(u_i=(u - \xi_i)\) 
in \eqref{eq:simplestMID} is equal to zero. 
See Appendix E for 
a detailed proof on this.

\subsubsection{Proof of the determinant representation and the Hirota equation for the master \texorpdfstring{$T$}{T}-operator }

A chain of arguments in \cite{Kazakov:2007na,KLT10} allows one
to give a self-contained proof ``from the first principles'' of
the central claim of this paper: the master \(T\)-operator
for rational spin chains is a tau-function of the MKP hierarchy. 
Here we give a direct proof of this statement using the representation
of the master \(T\)-operator through the co-derivatives.

The master identity in the form \eqref{bi3a} or \eqref{bi3b} 
directly follows from the determinant formula 
\eqref{masterdet1} by applying the
Jacobi identity. 
So in order to prove the master identity it is enough to prove \eqref{masterdet1}. In fact, for this proof we can assume, without loss of generality, that \({\bf t}=0\). Indeed, 
the product\footnote{Here we change all \(z_k\to \frac{1}{z_k}\) 
for convenience.}
\begin{equation} \left.F_g({\bf t}+\sum_{k=1}^K[z_k])\right|_{{\bf t}=0}=\prod_{k=1}^K w(z_k) \end{equation}       
can approximate \(F_g({\bf t})\) for any set of \({\bf t}\) with any needed accuracy, if \(K\) is sufficiently large, by an appropriate choice of 
the parameters \(z_1,z_2,\dots,z_K\). In other words, we change 
the variables from  \({\bf t}\) to \(z_1,z_2,\dots,z_K\)
(which is the Miwa change of variables):
\begin{equation}
 T^{ \{1,2,\dots,K \}  }(u):= T\left(u, \sum_{k=1}^K[z_k]\right)
=
\hcD_L(u)\,\prod_{k=1}^K w(z_{k})
\end{equation} 
or, for a subset \( \{ i_{1},i_{2},\dots, i_{n} \} 
\subset \{1,2,\dots, K \}\),
\begin{equation}\label{Tzdef}
 T^{ \{ i_{1},i_{2},\dots, i_{n} \}  }(u)
:= T\left(u, \sum_{k=1}^n [z_{t_k}]\right)
=\hcD_L(u)\,\prod_{k=1}^n w(z_{i_k})\,.
\end{equation} 
Then, in order to prove \eqref{bi3b} for arbitrary \({\bf t}\), it is
sufficient to prove \eqref{bi3b} for arbitrary \({\bf
  t}=0+\sum_{k=1}^K[z_k]\), which is equivalent to proving
\eqref{masterdet1} when \({\bf   t}=0\). Therefore, we only have to prove the identity
\begin{equation}
T^{ \{ i_{1},i_{2},\dots, i_{n} \}  }(u) =
\frac{
\mathrm{det}_{1 \le k,j \le n} \left( z_{i_{k}}^{n-j} 
\hcD_L(u-j+1)\,w(z_{i_k}) \right)
}
{
\mathrm{det}_{1 \le k,j \le n} \left( z_{i_{k}}^{n-j} \right)
\prod_{k=1}^{n-1}\prod_{l=1}^{L}  (u-\xi_l-k)
} ,
\label{masterdet1w}
\end{equation}
This formula expresses the general 
\(T^{ \{ i_{1},i_{2},\dots, i_{n} \}  }(u)\) through  \(T^{ \{ i\}  }(u)\) with only one index (the generating operator for transfer matrices in symmetric representations). 
The proof goes along the same lines as the proof of the CBR formula in  \cite{Kazakov:2007na}. It consists of two steps:
\begin{enumerate}
\item We should prove that \(T^{ \{ i_{1},i_{2},\dots, i_{n} \}  }(u)\) is a polynomial operator of the power \(L\), as it is clear from the definition; 
\item 
We should prove that the coefficients of polynomials 
in the r.h.s. and the l.h.s. of \eqref{masterdet1w} are equal.  
\end{enumerate}
In order to prove the first statement, it is enough 
to show that the determinant in the numerator vanishes at all zeros of the denominator, namely at \(u=\xi_l+j,\quad l=1,\dots,L,\,\, j=1,\dots,n-1\). 
For that, let us compare two consecutive columns, the \(j\)'th and the \((j+1)\)'th,    of the matrix in the numerator. If the \(2\times 2\) minor determinant
\begin{equation}
\mathrm{det}
\begin{pmatrix}  z_{i_{k}}^{n-j} 
\hcD_L(u-j+1)\,w(z_{i_k}) &  z_{i_{k}}^{n-j-1} 
\hcD_L(u-j)\,w(z_{i_k}) \\
z_{i_{k'}}^{n-j} 
\hcD_L(u-j+1)\,w(z_{i_{k'}}) &  z_{i_{k'}}^{n-j-1} 
\hcD_L(u-j)\,w(z_{i_{k'}}) \\
\end{pmatrix}
\end{equation}
built out of 
any two arbitrary rows \(k\) and \(k'\) of these two columns
are zero, then the whole determinant vanishes\footnote{Of course these
  minors vanish  only
  under the constraint that
\({u=\xi_l+j}\), where \({l=1,\dots,L}\), and \({j=1,\dots,n-1}\).
}. Indeed, the corresponding operatorial identity 
\begin{equation}
\label{bi45}
z \left (\hcD_L(\xi_l+1)w(z)\right )
\left (\hcD_L(\xi_l)w(z')\right ) -
z' \left (\hcD_L(\xi_l+1)w(z')\right )
\left (\hcD_L(\xi_l)w(z)\right ) = 0
\end{equation} 
appeared in \cite{Kazakov:2007na} in relation to the proof of the
CBR formula. This identity can be deduced from
\eqref{eq:simplestMID}, where the third term is equal to zero at
position \(u=\xi_l\). The identity \eqref{eq:simplestMID} itself is, as
shown above, deduced from the relation \eqref{bi3c}
(which is proven diagrammatically in \cite{Kazakov:2007na}) by
recurrence over the size \(L\) of the spin chain.

The proof of the second statement, about the equality of coefficients in polynomials on both sides of \eqref{masterdet1w}, can be achieved by comparing their  large \(u_i=u-\xi_i\) asymptotics.  
First, we can rewrite \eqref{masterdet1w} as follows:  
\begin{align}\label{uass1}
T^{ \{ i_{1},i_{2},\dots, i_{n} \}  }(u) = 
\frac{u_1 u_2
\cdots u_L \det_{1\le k,j\le n}  
\left(
 z_{i_{k}}^{n-j} 
\otimes_{l=1}^L\left(1+\frac{1}{u_l+1-j}\hD\right)w(z_{i_k})
\right)
}
{
\mathrm{det}_{1 \le k,j \le n} \left( z_{i_{k}}^{n-j} \right)
} .
\end{align} 
The r.h.s. must be a linear polynomial in each of the variables \(u_l\). We can then reconstruct this polynomial from the large \(u_l\) asymptotics. For example, for large \(u_1\) we find
\begin{align}\label{uass2}
T^{ \{ i_{1},i_{2},\dots, i_{n} \}  }(u) \simeq 
\frac{u_2 u_3
\cdots  u_L(u_1+\hD)
\CMPorArticle{ {\displaystyle \det_{~\quad\mathclap{1\le k,j\le n}\quad~}  }}{
 \det_{1\le k,j\le n}  }
\left(
 z_{i_{k}}^{n-j} 
\otimes_{l=2}^L\left(1\CMPorArticle{\!+\!}{+}\frac{1}{u_l+1-j}\hD\right)w(z_{i_k})
\right)
}
{
\mathrm{det}_{1 \le k,j \le n} \left( z_{i_{k}}^{n-j} \right)
} ,
\end{align} 
where we used the fact that at large \(u_1\) the co-derivative in  the bracket \CMPorArticle{\linebreak}{} \(\left(1+\frac{1}{u_1+1-j}\hD\right) 
\simeq \left(1+\frac{1}{u_1}\hD\right)  \) can occur only once in each of \(L!\) terms of determinant 
and the Leibniz rule to give the right overall asymptotics \(u_1\times1/u_1=1\). 
Expanding in this way for each of the remaining \(u_l\)'s we  recover
in the r.h.s. the original definition \eqref{Tzdef}.  

In this way we have proved the determinant formula \eqref{masterdet1}. It amounts not only to the proof of the master identity \eqref{bi3a} but also to the fact that the master \(T\)-operator satisfies the whole general Hirota bilinear identity \eqref{bi2a101} for the MKP hierarchy. 

Let us note that the commutativity of two general master operators \eqref{commT} follows, in virtue of the determinant representation \eqref{masterdet1}, from the commutativity of generating functions of \(T\)-operators for the symmetric tensor representations
\begin{equation}\label{commw}
 \left (\hcD_L(u)w(z)\right )
\left (\hcD_L(u')w(z')\right )=
 \left (\hcD_L(u')w(z')\right )
\left (\hcD_L(u)w(z)\right ).
\end{equation}  
A proof of this last relation  can be also found in   \cite{Kazakov:2007na}.

\subsubsection{Analyticity properties of the BA functions related to the master \texorpdfstring{$T$}{T}-operator}

Equations (\ref{master3a}) also allow us to characterize the class of
solutions of the classical Hirota equations
relevant to the quantum spin chain. First of all,
it is obvious that the master
\(T\)-operator is a polynomial in \(u\) of degree \(L\). Second,
consider the (operator-valued) BA function and its adjoint:
\begin{equation}\label{bi3d}
\psi_{u}({\bf t}, z)=z^u e^{\xi ({\bf t}, z)}
\frac{T(u, {\bf t} - [z^{-1}])}{T(u, {\bf t})}\,,
\quad
\psi^*_{u}({\bf t}, z)=z^{-u} e^{-\xi ({\bf t}, z)}
\frac{T(u, {\bf t} + [z^{-1}])}{T(u, {\bf t})}.
\end{equation}
From (\ref{master3a}) it is clear that  the function
\(\psi\) is of the form (\ref{rat1})
while the function \(\psi^*\) is of the form (\ref{rat101})
with the poles at the points \(p_i\) which are the {\it eigenvalues
of the twist matrix \(g\)} (assumed to be all distinct and nonzero).
This means that in the appropriate basis \(g\) can be diagonalized:
\begin{align}
\label{eq:g=diagp}
  g=&\mathrm{diag}\left(p_1,p_2, \ldots ,p_N\right)\,.
\end{align}
The maximal possible order of each pole is \(L+1\).
Indeed, the
\(T\)-operator {\bf }
for the empty spin chain
 (at \(L=0\)) gives first
order poles at \(p_i\):
\begin{equation}\label{bi3e}
T_{(L=0)}(u, {\bf t}+[z^{-1}])= \Bigl (\prod_{i=1}^{N}
\frac{z}{z-p_i}\Bigr )
\exp \Bigl ( \sum_{i=1}^{N}\sum_{k\geq 1}t_k p_i^k \Bigr ),
\end{equation}
and each co-derivative increases the order of each pole by 1.
Note also that (\ref{bi3e}) (and thus
the adjoint BA function) has zero of order \(N\)
at \(z=0\).

\subsection{The wave operators as non-commutative generating series}

The wave operators (\ref{BA6}), (\ref{BA6a}) for the master
\(T\)-operator are of the form
\begin{equation}\label{master6}
W(u,{\bf t})=\sum_{a\geq 0}\frac{h_a(-\tilde \p )T(u,{\bf t})}{T(u,{\bf t})}
\, e^{-a\p_u},
\end{equation}
\begin{equation}\label{master6a}
W^{-1}(u,{\bf t})=\sum_{s\geq 0}e^{-s\p_u} \,
\frac{h_s(\tilde \p )T(u+1,{\bf t})}{T(u+1,{\bf t})}.
\end{equation}
Expanding both sides of (\ref{master3a}) in powers of \(z\), we get
\begin{equation}\label{master4}
h_k(\tilde \p )T(u,{\bf t})=\bigotimes_{i=1}^{L}\,
(u \! -\! \xi_i \! +\! \hat D)
\left [\chi_k(g)F_g({\bf t})\right ],
\end{equation}
\begin{equation}\label{master5}
h_k(-\tilde \p )T(u,{\bf t})=\bigotimes_{i=1}^{L}\,
(u \! -\! \xi_i \! +\! \hat D)
\left [(-1)^k \chi^k(g)F_g({\bf t})\right ],
\end{equation}
where \(\chi_k(g)\), \(\chi^k(g)\) are characters of symmetric 
and antisymmetric
representations respectively.
From (\ref{master5}) it is clear that \(h_a(-\tilde \p )T(u,{\bf t})=0\)
for \(a>N\) because the antisymmetric \(GL(N)\) character \(\chi^a(g)\)
vanishes. Therefore, the series (\ref{master6}) is truncated:
\begin{equation}\label{master6b}
W(u,{\bf t})=\sum_{a=0}^{N}\frac{h_a(-\tilde \p )T(u,{\bf t})}{T(u,{\bf t})}
\, e^{-a\p_u}.
\end{equation}
Moreover, since for \(GL(N)\) \(\chi^N(g)=\det g\),
we have from (\ref{master5}):
\[
\begin{array}{lll}
h_N(-\tilde \p )T(u,{\bf t})&=&\displaystyle{
\bigotimes_{i=1}^{L}\, (u \! -\! \xi_i \! +\! \hat D)
\left [(-1)^N \det g F_g({\bf t})\right ]}
\\ &&\\
&=&\displaystyle{
(-1)^N \det g \bigotimes_{i=1}^{L}\,
(u +1 \! -\! \xi_i \! +\! \hat D)F_g({\bf t})}
\\ &&\\
&=&(-1)^N \det g \, T(u+1,{\bf t}),
\end{array}
\]
and so the coefficient in front of the
last term in (\ref{master6b}) is equal to  \CMPorArticle{\linebreak}{}
\(\displaystyle{
(-1)^N \det g \, \frac{T(u+1,{\bf t})}{T(u,{\bf t})}}
\)
as it should, as soon as the points \(p_i\)
are identified with eigenvalues of the matrix \(g\) (compare
with (\ref{rat7}) and recall that the \(\tau \)-function
from section 4 and the master \(T\)-operator from this section should
be identified as \( \tau_u ({\bf t})=(\det g)^u T(u, {\bf t})\)).

We conclude that the \(\tau\)-function (\ref{master})
corresponding to the \(GL(N)\)-invariant
spin chain on \(L\) sites must obey the
following two conditions:
\begin{itemize}
\item
It is a polynomial in \(u\) of degree \(L\)
multiplied by exponential function of a linear
function of times,
\item
The series for the wave operator (or, equivalently, the expansion of the
Baker-Akhiezer function) truncates at the \(N\)-th term.
\end{itemize}

At last, we remark that the wave operator (\ref{master6})
at \(t_i=0\) is nothing else than the ``non-commutative generating
series'' of \(T\)-operators for antisymmetric (fundamental)
representations \(T^a(u):=T^{(1^a)}(u)\):
\begin{equation}\label{master7}
W(u,0):=W(u)=\sum_{a=0}^{N}(-1)^a \frac{T^a(u)}{\phi (u)}
\, e^{-a\p_u}.
\end{equation}
Similarly, \(W^{-1}(u)\) is the ``non-commutative generating
series''\footnote{
It is known that one can generate the
\(T\)-operators for infinite-dimensional representations
just by changing
the expansion point of the non-commutative generating series \cite{Beisert:2005di,Gromov:2010vb}.
It will be very interesting to see how this phenomenon
shows up for the master \(T\)-operator and 
to realize formulas for T-functions in 
\cite{Tsuboi:2011iz} as operators.
}
 of \(T\)-operators for symmetric
representations \(T_s(u):=T^{(s)}(u)\)
(cf. \cite{KLWZ97}):
\begin{equation}\label{master7a}
W^{-1}(u)=\sum_{s=0}^{\infty} e^{-s\p_u}\, \frac{T_s(u+1)}{\phi (u+1)}=
\sum_{s=0}^{\infty}\frac{T_s(u\! -\! s\! +\! 1)}{\phi (u\! -\! s\! +\! 1)}\,
e^{-s\p_u}.
\end{equation}
The operator \(T^N(u)=\det g \, T^{\emptyset }(u+1)=
\det g \, \phi (u+1)\) is known as the quantum determinant
of the quantum monodromy matrix. These formulas become
more simply looking in the normalization (\ref{QT3}):
\begin{equation}\label{master7b}
W(u)=\sum_{a=0}^{N}(-1)^a {\sf T}^a(u)
\, e^{-a\p_u}, \quad \quad
W^{-1}(u)=\sum_{s=0}^{\infty} e^{-s\p_u}\, {\sf T}_s(u+1).
\end{equation}
It would be interesting to clarify 
the role of the classical Lax operator (\ref{Lax}) in the
quantum spin chain.

\subsection{Undressing B\"acklund flow and Baxter \texorpdfstring{$Q$}{Q}-operators}
\label{sec:backlund-flow-baxter}
The construction of \(Q\)-operators originally 
introduced by Baxter \cite{Bax72}
in his solution to the 8-vertex model
is now generalized and developed in the literature
in various directions (see e.g. \cite{Bazhanov:1998dq,osc-app,otherQ}).
In this section, we explain how
the Baxter \(Q\)-operators emerge from the master \(T\)-operator
using the approach of \cite{KLT10}.

\subsubsection{Preliminary remarks}

We will construct a set of 
\(Q\)-operators
which can be identified with 
the Baxter \(Q\)-operators 
in the sense that:
\begin{enumerate}

\item The \(Q\)-operators 
are building blocks for the \(T\)-operators meaning that
the latter are written in terms of the former 
in the Wronskian form. The Wronskian formulas resolve
functional relations 
for the \(T\)- and \(Q\)-operators known as \(TQ\)-relations. 

\item 
The \(Q\)-operators satisfy bilinear functional relations 
(the \(QQ\)-relations). With the additional input of analytical 
properties of the \(Q\)-operators (which are polynomials
in \(u\) in our case), they imply the system 
of Bethe equations for their roots.

\end{enumerate}

\noindent
Our construction of the \(Q\)-operators is directly related
to the undressing chain of
B\"acklund transformations
from section \ref{sec:family-backl-transf}.
In our approach, the nested
Bethe ansatz is represented by this undressing chain, where
the size of the matrices involved in the Wronskian-like solution is
successively reduced, as in
equation \eqref{Back5gen}.
As it was shown in section 4.2, the undressing procedure 
basically consists in picking the singular terms of the 
function \(T(u, {\bf t}+[z^{-1}])\) (in this section it is
an operator) at the poles at \(z=p_1, p_2, \ldots , p_N\). Since the poles are 
in general of a high order, there are different versions of this
procedure according to the possibility to consider different terms
in the singular part of the Laurent expansion. As it was argued in
section 4.1.5, they are essentially equivalent and differ by 
certain normalization factors
only. Below in this section we consider in some detail two 
distinguished choices: picking the coefficients in front of 
the highest and the lowest poles.

As it was already mentioned, the master
\(T\)-operator introduced in
section \ref{sec:master-t-operator-1} 
is a polynomial \(\tau\)-function in the variable
\(u\) while the \(\tau\)-functions from section 4 
are polynomials multiplied by the
overall exponential factor \(\prod_{i=1}^N p_i^u\),
so that the
definition of the
master \(T\)-operator differs from the \(\tau\)-function \eqref{rat6}
by the factor \(\left(\det g\right)^{-u}\). More precisely,
let \(\left |\omega \right\>\) be a
basis of eigenvectors of the 
master \(T\)-operator and let 
\(\left <\omega\right |\) be the
dual basis such that \( \left <\omega \middle |
\omega' \right >=\delta_{\omega , \omega '}\)). 
Then 
\beq\label{ruleop0}
\left < \omega \right |
T(u,{\bf t})\left |\omega \right\>=(\det g)^{-u}
\tau_u ({\bf t}; \left |\omega \right\>),
\eeq
where the \(\tau \)-function in the r.h.s. depends on the 
state \(\left |\omega \right \>\).
According to this,
the undressing procedure should be slightly modified in order 
to preserve
polynomiality of the \(\tau \)-functions at each level.

In this section we will work in the basis 
\({\sf e}_{\alpha}\) where the twist 
matrix \(g \) is diagonal: \( g {\sf e}_{\alpha}=p_{\alpha}
{\sf e}_{\alpha}\), \( \alpha =1, 2, \ldots , N\):
\[
{\mathsf e}_{i}= ^{\, {\sf t}}(\underbrace{0,0,\cdots,0}_{i-1},1,
 \underbrace{0,\cdots,0}_{N-i} ).
\]
A basis of eigenvectors
of \(g^{\otimes L}=\underbrace{g\otimes g\otimes \ldots \otimes g}_{L}\)
is then given by
\[ \left | 
\alpha_1 \alpha_2 \ldots \alpha_L\right >=
{\sf e}_{\alpha_1}\otimes {\sf e}_{\alpha_2 }\otimes
\ldots \otimes {\sf e}_{\alpha_L}
\]
with eigenvalues \(\displaystyle{p_{\alpha_1}p_{\alpha_2}
\ldots p_{\alpha_L}=
\prod_{j=1}^{N}p_{i}^{M_i}}\), where \(M_i\) is the number of
indices \(\alpha_{l}\) in the vector 
\(\left | 
\alpha_1 \alpha_2 \ldots \alpha_L\right >\)
equal to \(i\). These numbers can be regarded as eigenvalues 
of the operator
\footnote{Note that \({\mathsf m}_{i}\) is a Cartan element
 \(e_{ii}\) of \(gl(N)\) in the fundamental representation 
and \({\mathsf M}_{i}\) is its co-product \(\Delta^{(L-1)}(e_{ii})\). 
In addition, \(g\) can be written as \(g=\exp (\sum_{i=1}^{N} \log p_{i} e_{ii})=\prod_{i=1}^{N}p_{i}^{e_{ii}}\) with respect to the 
above eigenbasis.}
{\newcommand{\cnt}{\label{Mi}
  {\sf M}_i=\sum_{k=1}^L \left(\I^{\otimes (k-1)}\otimes {\sf m}_i \otimes
  \I^{\otimes (L-k)}\right), \CMPorArticle{\\}{\quad} \textrm{where }
 {\sf m}_i=\mathrm{diag}(\underbrace{0,0,\cdots,0}_{i-1},1,
 \underbrace{0,\cdots,0}_{N-i} )= {\mathsf e}_{i} \, ^{\sf t}{\mathsf
   e}_{i}}\CMPorArticle{\begin{multline}
\cnt
\end{multline}
}{\beq
\cnt
\eeq}}
on the states \(\left | 
\alpha_1 \alpha_2 \ldots \alpha_L\right >\), so 
one can write \( g^{\otimes L}=
p_1^{{\sf M_1}}p_2^{{\sf M_2}}\ldots p_N^{{\sf M_N}}\). It directly
follows from the definition that
\(\displaystyle{\sum_{i=1}^{N}{\sf M}_i = L \, \I ^{\otimes L}}\) and thus
\(\displaystyle{\sum_{i=1}^{L}M_i =L}\).

In \cite{Kazakov:2007na,KLT10}, it is shown that the \(T\)-operators
 can be explicitly written through diagrammatic expressions, and 
 \(T(u,\mathbf{t})\) is then a combination of permutations and diagonal
operators. As a consequence, 
the operators \({\sf M}_i\) commute with all the \(T\)-operators. 
In fact this follows from the \(GL(N)\)-invariance of the 
\(R\)-matrix and means that the eigenstates \(\left |\omega \right >\)
of the \(T\)-operators can be classified according to the set of quantum
numbers \(M_i = M_i (\left |\omega \right >)=
\left < \omega \right | {\sf M}_i \left |\omega \right >\) and
\[
\left < \omega \right | g^{\otimes L}\left |\omega \right >=
\prod_{i=1}^{N}p_{i}^{M_i (\left |\omega \right >)}.
\]
For what follows it is important to note 
that \(M_i (\left |\omega \right >)\) is the multiplicity of zero 
at \(z=p_i\) of 
the eigenvalue of the operator \((z\, \I-g)^{\otimes L}\) on the state
\(\left |\omega \right >\). More precisely, the 
following operator
identity holds:
\beq\label{opind}
(z\, \I-g)^{\otimes L}=(z-p_1)^{{\sf M}_1}(z-p_2)^{{\sf M}_2}
\ldots (z-p_N)^{{\sf M}_N}.
\eeq

\subsubsection{The undressing procedure}

For our purpose in this section the most convenient version 
of the undressing procedure is picking the coefficient in front
of the most singular term of the Laurent
expansion of \(T(u, {\bf t}+[z^{-1}])\) around its poles. 
As is shown in section 4.2, 
this version differs from the one based on taking residues 
by a normalization factor (which is an operator in the present
section).

Applying the undressing procedure based on the highest
poles, we note that
there are \(N\) different ways to define
the \({T}\)-operator at the first level of nesting by
choosing \(i\in\{1,\ldots , N\}\) and picking the coefficient
in front of the highest pole at \(p_i\): 
\begin{equation}
\label{eq:tauBT1a}
  {T}^{(i)}(u,{\bf t})=(-1)^{i-1}\det g \, p_{i}^{-1}
 e^{-\xi ({\bf t},p_i)}\, \mbox{res}_{z=p_i}\left [
    (z-p_i)^{{\sf M}_i}
    {T}(u+1,{\bf t}+[z^{-1}])\right ],
\end{equation}
where \({\sf M}_i\) is the operator defined in (\ref{Mi}).
This ``operator multiplicity'' reflects the fact that the 
order of the highest pole depends on the sector of the Hilbert
space where the diagonalization is performed and is equal to
\(M_i +1\).
This expression \eqref{eq:tauBT1a} is nothing but the equation
\eqref{Back8} with an extra factor 
\begin{align}
\left. 
\left\langle \omega \middle|T^{(i)}(u,{\bf t})
\middle|\omega \right\rangle =
\left( \prod_{k=1, \ne i}^{N}p_k^{-u} \right)
\tau_u^{(i)}({\bf t})\right |_{\left |\omega \right >}\,,
  \label{ruleop0000}
\end{align}
which ensures the polynomiality of \(T^{(i)}(u,{\bf t})\). 
Here we introduced a notation \(|_{\left |\omega \right >}\) 
to denote the state dependence of the function.

As the simplest example, consider the case \(L=1\).
From \eqref{eq:Texplicit1} we find:
\begin{align}
    T_{(L=1)}(u,{\bf t}+[z^{-1}])=&\Bigl (
    (u-\xi_1){\I} +\sum_{k\geq
      1}k t_k g^k+\frac {g} {z-g}\Bigr )
     \frac{e^{\sum_{l\geq 1}t_l {\scriptstyle {\rm tr}}\, 
     g^l}}{\det(\I -z^{-1}g)}\,.
      \label{onespin1}
\end{align}
As \(z\to p_i\), the r.h.s. has a double pole in the subspace spanned
by the basis vector \({\sf e}_i\) and a simple pole when restricted 
to the subspace spanned by \({\sf e}_{\alpha}\) with \(\alpha \neq i\).
In the former subspace the eigenvalue of \((z-p_i)^{{\sf M}_i}\)
is \(z-p_i\) while in the latter it is \(1\).
The explicit calculation of \eqref{eq:tauBT1a} gives: 
\CMPorArticle{\begin{multline}
\label{onespin1-2}
T_{(L=1)}^{(i)}(u,{\bf t}) = (-1)^{i - 1}p_i 
\left [\sum_{j =1, \neq i}^{N}  \Bigl ( u + 1 - 
\xi_1  + \sum_{k\geq
      1}k t_k {p_{j}}^k + \frac {p_{j}} {p_{i} - 
      p_{j}}\Bigr ) e_{jj} \right.\\\left.\vphantom{\sum_{j =1, \neq i}^{N}}
      + p_{i} e_{ii}
\right ]
\prod_{\substack{1\leq \alpha \leq N\\ \alpha\neq i }}
\frac{p_{\alpha}e^{\xi ({\bf t}, p_{\alpha})}}{1 - p_{\alpha}/p_{i}}
\end{multline}}{\beq\label{onespin1-2}
T_{(L=1)}^{(i)}(u,{\bf t})\! =\! (-1)^{i\! -\! 1}p_i 
\!\!\left [\sum_{j =1, \neq i}^{N}\! \! \Bigl ( u\! +\! 1\! -\! 
\xi_1 \! +\!\! \sum_{k\geq
      1}k t_k {p_{j}}^k\! +\! \frac {p_{j}} {p_{i}\! -\! 
      p_{j}}\Bigr ) e_{jj} 
     \! +\! p_{i} e_{ii}
\right ]\!\!
\prod_{\substack{1\leq \alpha \leq N\\ \alpha\neq i }}\!
\frac{p_{\alpha}e^{\xi ({\bf t}, p_{\alpha})}}{1\! -\! p_{\alpha}/p_{i}}
\eeq}
where \(e_{ij}\) are the standard generators of the algebra \(gl(N)\).

At the next level of the B\"acklund nesting chain,
one should choose  \CMPorArticle{\linebreak}{} \(j\in\{1,\ldots , N\}\setminus \{i\}\)
and define:
\beq
\label{eq:tauBT21}
\begin{array}{l}
{T}^{(ij)}(u,{\bf t})=(-1)^{(ij)}
  (\det g)^2 (p_i p_j)^{-2}
e^{-\xi ({\bf t}, p_i)-\xi ({\bf t}, p_j)}
  (p_j-p_i)
  \\ \\
  \hspace{2cm}\times \, \mbox{res}_{\substack{\\z_i=p_i\\z_j=p_j}}
  \left [(z_i-p_i)^{{\sf M}_i}(z_j-p_j)^{{\sf M}_j}
  {T}\left (u+2,{\bf t}+[z_i^{-1}]+[z_j^{-1}]\right )
\right ].
\end{array}
\eeq
The sign factor is \((-1)^{(ij)}=(-1)^{i+j+1}
\varepsilon_{ij}\), where
\(\varepsilon_{ij}=1\) if \(i<j\) and \(-1\) if \(i>j\).
With this sign convention \(%
  {T}^{(ij)}(u,{\bf t})\) is symmetric
in the indices \(i,j\), or, equivalently,
the transformations \(%
  {T}\rightarrow %
  {T}^{(i)}\rightarrow %
  {T}^{(ij)}\)
and \(%
  {T}\rightarrow %
  {T}^{(j)}\rightarrow %
  {T}^{(ji)}\) commute for any
\(i,j\), namely, \( {T}^{(ij)}= {T}^{(ji)} \).
The expression \eqref{eq:tauBT21} is nothing but the \(n=2\) case of 
\eqref{Back9-def}  with an extra factor 
\begin{align}
\left. 
\left\langle \omega \middle|T^{(ij)}(u,{\bf t})
\middle|\omega \right\rangle =
 \left( \prod_{k=1, \ne i,j}^{N}p_k^{-u} \right)
\tau_u^{(ij)}({\bf t})\right |_{\left |\omega \right >}\,.
  \label{ruleop0001}
\end{align}
This factor is necessary to make \(T^{(ij)}(u,{\bf t})\) a polynomial of
the variable \(u\).

In order to proceed to higher levels, one should 
fix a subset   \CMPorArticle{\linebreak}{} \( I=\{i_{1}, i_{2},\dots, i_{n} \} 
\subset \{1,2,\dots, N \} \)
and denote by \(J\) its complement:  \CMPorArticle{\linebreak}{} \( J= \{1,\ldots , N\}
\setminus \{i_1,i_2,\ldots , i_n\}\). The ``nesting
level'' is defined to be the number \(n\) of indices in the set 
\(\{i_{1}, i_{2},\dots, i_{n} \}\).
In complete analogy with the cases 
\(n=1,2\) considered above,
we define \({T}\)-operators 
\({T}^{(i_{1}i_{2}\dots i_{n})}(u,{\bf t})\) 
on all intermediate levels by the formula 
(cf.\ \eqref{Back9-def}):  
\begin{multline}
T^{(i_1 \ldots i_n)} (u, {\bf t})= 
(-1)^{i_{n}-1-
\mathrm{Card} \{ k | i_{k} < i_{n}, 1 \le k \le n-1 \} 
} 
\det g 
\left( \prod_{\alpha =1}^{n} p_{i_{\alpha}}^{-1} \right)
e^{-\xi ({\bf t},p_{i_{n}})}
\\
\times 
\mbox{res}_{ z_{i_{n}}=p_{i_{n}} }
\left[ (z_{i_{n}}-p_{i_{n}} )^{\mathsf{M}_{i_{n}}}
T^{(i_1 \ldots i_{n-1})}
\bigl (u+1, {\bf t}+[z_{n}^{-1}])
\right], 
 \label{genhighest-def}
\end{multline}
which leads to 
\beq\label{genhighest}
\begin{array}{lll}
T^{(i_1 \ldots i_n)} (u, {\bf t})&=&
\displaystyle{
 (-1)^{d_n} \Delta_n(p_{i_1}, \ldots , p_{i_n}) (\det g)^n
\left( 
\prod_{\alpha =1}^{n}p_{i_{\alpha}}^{-n} 
e^{-\xi ({\bf t}, p_{i_{\alpha}})}
\right )
}
\\ &&\\
&& \times \,\,\, 
\displaystyle{
\mbox{res}_{ 
\substack{\\z_{i_k}=p_{i_k}\\1\leq k\leq n}
}
\left [ \Bigl (\prod_{i\in I}(z_i  -p_i )^{{\sf M}_{i}} \Bigr )
T\bigl (u+n, {\bf t}+\sum_{i\in I}[z_{i}^{-1}]\bigr )
\right ]
},
\end{array}
\eeq
where \(\Delta_n \) is the Vandermonde determinant (\ref{Vander})
and \(d_n\) is defined in section 4.2. 
The \(T\)-operator at level \(n=0\) is the master
\(T\)-operator defined in \eqref{master3}.
According to this definition, the correspondence of eigenvalues of the \(T\)-operators at level \(n\) 
and the  \(\tau\)-functions at 
level \(n\) given by \eqref{Back9} is as follows:
\begin{align}
\left. 
\left\langle \omega \middle|T^{(i_1 i_2\ldots i_n)}(u,{\bf t})
\middle|\omega \right\rangle =
\Bigl (\prod_{j\in J} p_j^{-u} \Bigr )
\tau_u^{(i_1 i_2 \ldots i_n)}({\bf t})\right |_{\left |\omega \right >}.
  \label{ruleop}
\end{align}
The \(\tau \)-function in the r.h.s. is obtained by the
undressing procedure from 
\(\tau_u ({\bf t}; \left |\omega \right >)\)
according to the rules presented in section 4.
Note that the eigenstate
\(\left| \omega \right >\) is common for all the 
\(T\)-operators. One can also
introduce ``Pl\"ucker coordinates'' for the \(T\)-operators on
intermediate levels by the expansion
\beq\label{Tlambdaint}
T^{(i_1 i_2 \ldots i_n)}(u, {\bf t})=\sum_{\lambda}
s_{\lambda}({\bf t})\,
T^{(i_1 i_2 \ldots i_n),\lambda}(u)
\eeq
whose form is identical to (\ref{master}). The diagrams
\(\lambda \) correspond here to representations 
of the subalgebra \(gl(N-n) \subset gl(N)\).

It is clear that the bilinear relations of
the form (\ref{Back6}),
(\ref{Back7}) (cf.\ \eqref{bi2a}, \eqref{bi3a}) 
hold for the \({T}\)-operators
\({T}^{(i_1 \ldots i_n)}(u, {\bf t})\) at any level
and relate \({T}^{(i_1 \ldots i_n ij)}(u, {\bf t})\)
to \({T}^{(i_1 \ldots i_n k)}(u, {\bf t})\) and
\({T}^{(i_1 \ldots i_n)}(u, {\bf t})\):
{\newcommand{\cnt}{\label{Back6T}
\varepsilon_{ij}p_{k}^{-1}T^{(i_1\ldots i_n ij)}(u, {\bf t})
T^{(i_1\ldots i_n k)}(u+1, {\bf t})\CMPorArticle{\\+\,\,}{+}
\varepsilon_{jk}p_{i}^{-1}T^{(i_1\ldots i_n jk)}(u, {\bf t})
T^{(i_1\ldots i_n i)}(u+1, {\bf t})
\\ \CMPorArticle{}{\\}
+\,\, 
\varepsilon_{ki}p_{j}^{-1}T^{(i_1\ldots i_n ki)}(u, {\bf t})
T^{(i_1\ldots i_n j)}(u+1, {\bf t})
=0,}\CMPorArticle{\begin{multline}
\cnt
\end{multline}}{\beq \begin{array}{c} 
\cnt
\end{array}
\eeq}}
\beq\label{Back7T}
\begin{array}{c}
p_{j}^{-1}T^{(i_1\ldots i_n i)}(u, {\bf t})
T^{(i_1\ldots i_n j)}(u+1, {\bf t})-
p_{i}^{-1}T^{(i_1\ldots i_n j)}(u, {\bf t})
T^{(i_1\ldots i_n i)}(u+1, {\bf t})
\\ \\
=\,\, \varepsilon_{ij}T^{(i_1\ldots i_n ij)}(u, {\bf t})
T^{(i_1\ldots i_n )}(u+1, {\bf t}),
\end{array}
\eeq
where \(i,j,k \in \{1,2,\dots, N\} \setminus \{i_{1},i_{2},\dots,i_{n}\}\), \(i\ne j,i\ne k, j\ne k\). 

We thus have the undressing chain that terminates at the level \(N\):
\beq\label{TTT}
T(u,{\bf t})\rightarrow T^{(i_1)}(u,{\bf t})\rightarrow 
T^{(i_1 i_2)}(u,{\bf t})\rightarrow  
\ldots \rightarrow T^{(12\ldots N)}(u,{\bf t})
\rightarrow 0.
\eeq
It can be shown that the last (the completely undressed) member 
of this chain is an invertible operator which does 
not depend on \(u, {\bf t}\): \(T^{(12\ldots N)}(u,{\bf t})=
T^{(12\ldots N)}\). An advantage of the undressing at the highest 
poles is that this operator has a simple explicit form:
\beq\label{explicit}
T^{(12\ldots N)}=
\frac{g^{\otimes L}}{\displaystyle{\det_{1\leq i,j\leq N} p_i^{-j}}}
\prod_{i=1}^{N}{\sf M}_i!=
\frac{(\det g)^Ng^{\otimes L}}{\displaystyle{\prod_{1\leq l<k 
\leq N}(p_l-p_k)}}\prod_{i=1}^{N}{\sf M}_i!,
\eeq
which follows from the comparison of the large \(u\) asymptotics of (\ref{rat6}) and (\ref{master3}). Namely, from (\ref{rat6}) and (\ref{A}) it follows that
\beq\label{leadu1}
\tau_u({\bf t}) =(\det g)^u 
u^{\sum_{i=1}^N M_i} \left( \prod_{j=1}^N 
\left( a_{j M_{j}} e^{\xi({\bf t},p_j )}\right)
 \det_{1 \le k,m \le N} 
 \left( p_k^{-m-M_{k}} \right) + O\left( u^{-1 }\right)\right)
\eeq
while for \(T(u,{\bf t})\) we have
\beq\label{leadu2}
T(u, {\bf t}) = u^L  {\I}^{\otimes L} e^{\sum_{k=1}^N \xi({\bf t},p_k)}+ O\left(u^{L-1}\right).
\eeq
Then (\ref{explicit}) follows from (\ref{leadu1}), (\ref{leadu2}) , (\ref{ruleop}) and the expression for the eigenvalues 
of the \(T\)-operator on the last level of nesting 
(follows from \eqref{Back9} and \eqref{ruleop} for \(n=N\)):
\[
\left. \phantom{\int}
\left < \omega \right |  T^{(12\ldots N)}\left | \omega \right >=
\prod_{j=1}^N (M_j ! \, a_{jM_j}) \right |_{\left |\omega \right >}.
\]

Set 
\beq\label{Q1}
Q_j(u, {\bf t})=(T^{(12\ldots N)})^{-1}T^{(12\ldots \not j \ldots N)}
(u, {\bf t}).
\eeq
In the notation of section 4 (see eq. (\ref{Back9})), we can write
\begin{equation}
\left. \phantom{\int}
\left < \omega \right |  Q_j(u, {\bf t})\left | \omega \right >=
p_{j}^{-u}\frac{A_j(u\! -\! 1, {\bf t})}{M_j ! \,
a_{jM_j}}\right |_{\left |\omega \right >},\label{eq:QvsA}
\end{equation}
where the r.h.s. should be evaluated for the \(\tau \)-function 
corresponding to the state \( \left |\omega \right >\). Similarly,
one can define
\beq\label{Q2}
Q_{J}(u, {\bf t})=(T^{(12\ldots N)})^{-1}\, T^{(i_1 \ldots i_n)}
(u, {\bf t}), \quad 
J=\{1,\ldots ,
  N\}\setminus I\,.
\eeq
In this general notation, the previously defined operator is
\(Q_j(u, {\bf t})=Q_{\{j\}}(u, {\bf t})\). From equation 
(\ref{Back9}) we have, for \(J=\{j_1, j_2, \ldots , j_{N-n}\}\):
\beq\label{Q3}
Q_{J} (u, {\bf t})=\det_{1\leq k,l \leq N-n}
\left ( p^{1-l}_{j_k}Q_{j_k}(u-l+1, {\bf t})\right ).
\eeq
In particular, at \(J=\{1,2, \ldots , N\}\) this yields
\beq\label{Q4}
T(u, {\bf t})=g^{\otimes L}\, \Bigl (\prod_{i=1}^{N}{\sf M}_i!\Bigr )
\frac{\displaystyle{\det_{1\leq k,l\leq N}
\left ( p^{N+1-l}_{k}Q_{k}(u-l+1, {\bf t})
\right )}}{\displaystyle{\det_{1\leq k,l\leq N} (p_k^{N-l})}}\,,
\eeq
where we have used (\ref{explicit}).
These Wronskian-like determinants can also be obtained by taking
residues of \eqref{masterdet2} at \(z_{i_k}\to p_{i_k}\).
The Jacobi identity for the determinant \eqref{Q3} gives
the so-called \(QQ\)-relations, from which the Bethe equations follow.

In fact the operators \(Q_{J}(u, {\bf t})\) introduced in the way
explained above are the (specially normalized) 
\(T\)-operators on the intermediate levels of the undressing
procedure. 
The Baxter \(Q\)-operators \(Q_{J}(u)\)
are then defined as their restrictions  
to zero values of \({\bf t}\):
\beq\label{Q5}
Q_{J}(u)=Q_{J}(u,{\bf t}=0).
\eeq

We see from the above formulae that \(Q_{J}(u)\) is expressed as a
residue of the master T-operator. Moreover, if we restrict
\eqref{eq:QvsA} to \({\bf t}=0\), and expand it 
similarly to \eqref{ratQ} then we see that the coefficients \(a_{im}\) in
\eqref{A} can themselves be expressed from the residues of the master
T-operator.

Comparing (\ref{master44}) to the equation \eqref{rat11} 
(restricted to \({\bf t}=0\)) and using the definition
(\ref{Tlambdaint}), we conclude that 
for the symmetric 
representation  \( \lambda=(s)= (s,0,0,\ldots)\) the following
formula holds:
\begin{align}
\label{eq:WronskianForT}
  T^{(i_1 \ldots i_n),\lambda}(u)=&
      T^{(12\ldots N)}\det_{\substack{k\neq
        i_1,i_2,\ldots , i_n\\
1\leq l\leq N-n}}\left (
p_k^{\lambda_l+1-l}Q_{k}(u-l+\lambda_{l}+1)\right )\,.
\end{align}
The Jacobi identity for this determinant
is known as the \(TQ\)-relation. For all other representations this
result can be plugged into an analogue of the 
CBR-formula \eqref{tau4b-intro}: 
\begin{equation}\label{tau4b-nested}
T^{(i_1 \ldots i_n),\lambda}(u)
=\displaystyle{
\Bigl ( \prod_{k=1}^{\lambda_{1}'-1}
T^{(i_1 \ldots i_n)}(u\! -\! k)\Bigr )^{-1}\!\!
\det_{i,j =1,\ldots , \lambda_1^{\prime}}
T^{(i_1 \ldots i_n),(\lambda_i  - i + j)}(u\! -\! j\! +\! 1)},
\end{equation}
 to get that 
the determinant formula \eqref{eq:WronskianForT} actually
holds for arbitrary Young diagrams with at most \(N-n\) rows
(if it has more than \(N-n\) rows, then the \(T\)-operator 
vanishes identically).  In the particular case 
\(I=\emptyset \) equation (\ref{eq:WronskianForT}) yields the Wronskian determinant representation of the \(T\)-operator at fixed representation:
\begin{align}
\label{eq:WronskianForTa}
  T^{\lambda}(u)=&g^{\otimes L}\, \Bigl (\prod_{i=1}^{N}{\sf M}_i!\Bigr )
      \frac{\displaystyle{\det_{1\leq k,l\leq N}
      \left (p_k^{N+1-l+\lambda_l}Q_{k}(u-l+
      \lambda_{l}+1)\right )}}{\displaystyle{\det_{1\leq k,l\leq N}
      (p_{k}^{N-l})}}\,.
\end{align}
At \(L=0\) the pre-factors and the operators \(Q_k (u)\) become
simple (equal to \(1/p_j\)) and
this formula coincides with the determinant 
representation (\ref{characters}) for characters.
By construction, all the \(Q\)-operators are polynomials in 
\(u\) and the determinant formula (\ref{eq:WronskianForTa})
at \(\lambda =\emptyset \) must give the fixed polynomial
\(\phi (u)=\prod_{j=1}^{L}(u-\xi_j)\). This requirement 
implies the system of Bethe equations for roots of the 
\(Q\)-operators (or rather their eigenvalues).

\begin{figure}
\centering
\CMPorArticle{\includegraphics[scale=.9]{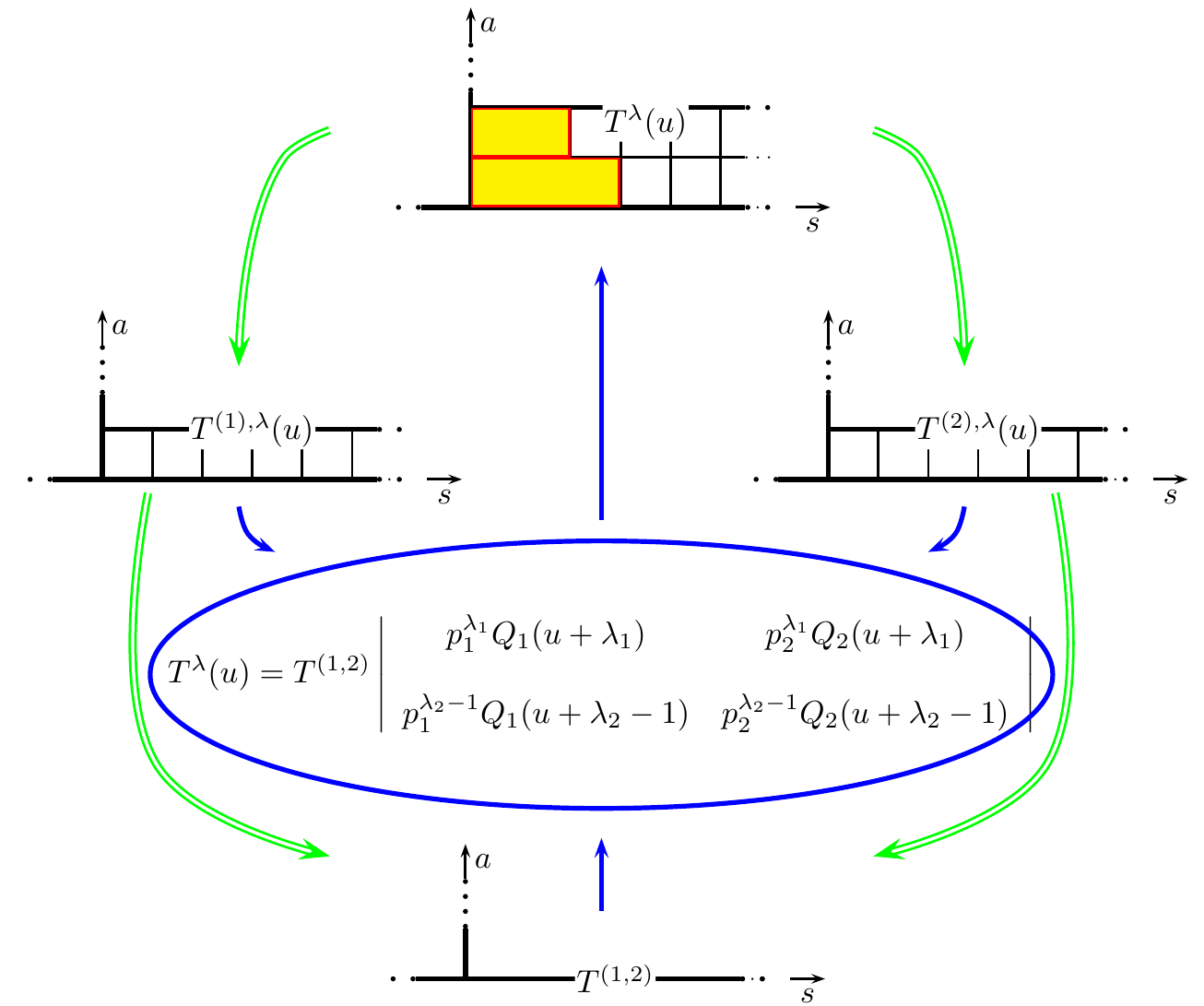}}{\includegraphics{BT2}}
\caption{B\"acklund flow for the \(GL(2)\) spin chain.}
 \label{fig:BT2}
\end{figure}

The undressing procedure is schematically illustrated in
Fig. \ref{fig:BT2} for the case 
\(N=2\). The green double arrow corresponds to the B\"acklund
transformation \eqref{eq:tauBTN2}, which ``undresses'' the 
\(T\)-operators by
gradually decreasing the size of the matrices at each site
of the lattice (which is the maximal possible height for the
Young diagrams giving rise to non-zero \(T\)-functions). The blue
arrow shows how the original \(T\)-operators are 
reconstructed by ``dressing'' with the \(Q\)-operators.

\subsubsection{On other versions of the undressing procedure}

The undressing procedure and the \(Q\)-operators 
can be defined in a number of equivalent 
ways. Here we mention some of them.

\paragraph{Lowest poles.}
\label{sec:lowest-poles}

Instead of picking the highest order pole, one might consider
another version of the undressing procedure, based on taking
residues at other poles (not of the highest order), 
as in section 4.2.
For example, at the first level (cf.\ \eqref{Back1}), we may define:
\begin{equation}
\label{eq:tauBT1}
  {T}^{(i)}(u,{\bf t})\propto
  \,\, \mbox{res}_{z=p_i}\left [
    \left((z/p_i
    \right)^{-u-1}\!\! e^{-\xi ({\bf t},z)}
    {T}(u+1,{\bf t}+[z^{-1}])\right ].
\end{equation}
This definition does not use the operators
\({\sf M}_i\), at the price of  less transparent normalization factors.
In particular, the definition (\ref{eq:tauBT1}) with the unit
pre-factor leads to a more complicated operator 
\(T^{(12\ldots N)}\).

\paragraph{Construction of \cite{KLT10}.}

At last, let us compare our present construction 
with the one introduced in \cite{KLT10}.
On the first level,  noticing  that \((z-p_i)^{{\sf M}_i}=\frac{(z\,
  \I-g)^{\otimes L} }{\prod_{j=1, \neq i} (z -p_j)^{{\sf M}_j}}\),
we can rewrite \eqref{eq:tauBT1a} as
\begin{gather}
    {T}^{(i)}(u,{\bf t}) =
    \det g \, p_{i}^{-1} e^{-\xi
      ({\bf t},p_i)}{\sf N}_i \, \mbox{res}_{z=p_i}\left [
 (z\, \I-g)^{\otimes L}\,
  {T}(u+1,{\bf t}+[z^{-1}])\right ]
\label{eq:tauBT2b}
\,,\\
\textrm{where }
\label{Ni}
{ \sf N}_i = (-1)^{i-1}\prod_{j=1, \neq i} (p_i -p_j)^{-{\sf M}_j}.
\end{gather}
The normalization factor \({ \sf N}_i\) is independent of \(u\) and
\({\bf t}\),  and commutes with all
  the \(T\)-operators. For this reason it was disregarded in
  \cite{KLT10}.

On  higher levels
\eqref{genhighest} can as well be written as:
\beq\label{eq:tauBTN2}
\begin{array}{lll}
{T}^{(i_1\ldots i_n)}(u,{\bf t})& =&\displaystyle{(\det g)^n 
\Bigl (\prod_{\alpha =1}^{n} p_{i_\alpha}^{-n}
e^{-\xi ({\bf t},p_{i_{\alpha}})}\Bigr )\, {\sf N}_{i_1 \ldots i_n}}
\\ &&\\
&&\,\,\,\, \times \,\, \displaystyle{
\mbox{res}_{\substack{\\z_{i_k}=p_{i_k}\\1\leq k\leq n}}\left[
\prod_{\gamma =1}^{n}(z_{i_\gamma}\I -g)^{\otimes L}\,
 {T} \Bigl (u+n,{\bf t}+\sum_{\alpha =1}^{n}
[z_{i_{\alpha}}^{-1}]\Bigr )\right ]}\,,
\end{array}
\eeq
where \({\sf N}_{i_1 \ldots i_n}\) is an operatorial normalization
factor which we don't specify here.
Let us now show that \eqref{eq:tauBTN2} coincides with equation (32) from
\cite{KLT10}.
\begin{enumerate}
\item In \cite{KLT10}, the set of times \(\mathbf{t}\) was not
  introduced. But as it can be seen from \eqref{master3a}, the
  transformation \(\mathbf{t}\mapsto \mathbf{t}+[z^{-1}]\)
  is equivalent to the
  insertion of \(w(z^{-1})=(\det(\I -z^{-1}g))^{-1}\) to the right of the
  co-derivatives.
\item The limit \(z_j\to 1/x_j\) used in \cite{KLT10} has the
  same effect as taking residue at the point \(z_{j}=
  p_{j}\) if we  change   notations \(z\to 1/z\), \(p_j \to x_j\),
  and  notice that \(\mbox{res}_{z=p}f(z)=
  \lim_{z\to p} (z-p)f(z)\) for a function
  \(f\) with a simple pole.
\item In \cite{KLT10}, depending on the
  representation, an overall factor 
  is necessary to
convert characters of \(g=\mathrm{diag}
\left({\left.p_i\right|}_{1\leq i
      \leq N}\right) \in GL(N) \) into characters of
  \(g_{J}=\mathrm{diag}\left({\left.p_j\right|}_{j\in J}\right) 
  \in GL(|J|) \). In our current
  notations, this amounts to changing \(F_g(t)\) to \(F_{g_{J}}(t)\). One
  can easily see that this is exactly what the factor
\(e^{-\sum_{\alpha =1}^{n}\xi ({\bf t}, p_{i_{\alpha}})}\) does. Indeed,
\(e^{-\sum_{\alpha =1}^{n}\xi ({\bf t}, p_{i_{\alpha}})} = F_{g_{J}}(t) / F_g(t)\).
\end{enumerate}
This comparison shows that the undressing procedures of \cite{KLT10} 
and
the one of the present paper are essentially equivalent.
But in the setting of our present paper this undressing procedure has a  transparent connection 
to the classical integrable hierarchy and its polynomial 
\(\tau \)-functions.

Our construction clarifies the
combinatorial structure of the set of the
\(2^{N}\) \(Q\)-operators on the Hasse diagram found in \cite{Tsuboi09}.
Moreover, it naturally takes into account the fact that 
all these \(Q\)-operators, as well as all the
\(T\)-operators, are polynomials in the 
spectral parameter \(u\). As a consequence, the Bethe equations are
automatically 
contained in this analysis, as it was explicitly 
demonstrated in \cite{KLT10}.

%
%
%
%
%
%

%
%
%
%
%
%
%
%
%
%
%
%
%
%
%
%
%
%
%
%
%
%
%

%
%
%
%
%
%
%
%
%
%
%
%
%
%
%
%
%
%
%

\begin{comment}

In this section we started from the fact that  \(T\)-operators rational, 
hence of the
form of section \ref{sec:rati-solut-kp}, and obtained  expressions for the
B\"acklund flow in terms of \(Q\)-operators, obtained as the residues of the
\(\tau\) functions, and these expressions coincide with the results of
\cite{KLT10}. Moreover, the relation
\(Q_{\{i\}}(u)=A_i(u-1,0)/p_i^{u}\) 
allows to identify the coefficients \(a_{i,m}\) in \eqref{A}, to the
coefficients of \(Q_{\{i\}}(u)\) (as polynomial in \(u\)). More
explicitly we get
\begin{align}
  Q_{\{i\}}(u)
=&a_{i,0}+\frac{a_{i,1}}{p_i} u +
\frac{a_{i,2}}{p_i^2}u(u-1)+\frac{a_{i,3}}{p_i^3}u(u-1)(u-2) +\cdots\,,
%
\end{align}
where the left-hand-side is well defined as a residue, which  allows
to deduce 
all coefficients \(a_{i,m}\) as operators. Due to the definition of
\(Q_{\{i\}}\) as a residue, \(Q_{\{i\}}(u)\) commutes with all
T-operators, %
hence
the operatorial coefficients \(a_{i,m}\) also commute
with all T-operators. The Wronskian expression
\eqref{eq:WronskianForT} (which coincides with the expressions
obtained in appendix C of \cite{KLT10}) proves that the
\(\tau\)-functions constructed 
from these values of the coefficients \(a_{i,m}\) coincide with the
T-operators defined in \eqref{QT2}.

\end{comment}

\subsection{Dynamics of zeros of the master \texorpdfstring{$T$}{T}-operator}

As we have seen,
for a \(GL(N)\)-invariant generalized spin chain on \(L\) lattice sites
the master \(T\)-operator \(T(u,{\bf t})\) is essentially
a polynomial in \(u\)
of degree \(L\). It can be represented in the form
\begin{equation}\label{dyn1}
T(u,{\bf t})\propto \prod_{j=1}^{L}(u-u_j({\bf t}))
\end{equation}
where
the roots \(u_j\) depend on all other times \(t_i\)
(we have omitted an irrelevant
exponential function in front of the product).
In some special cases\footnote{For example, when the boundary
twist parameters tend to 1}
one or more roots \(u_j\) can tend to infinity
and so the degree of the polynomial can be less than \(L\).
Note that we treat
equation (\ref{dyn1}) as an operator relation and the roots
\(u_j({\bf t})\) are (commuting) operators. Another possibility is to
treat (\ref{dyn1}) as a relation for eigenvalues of the
master \(T\)-operator, then the roots of each eigenvalue have
their own dynamics in the times \(t_i\). This dynamics is
known \cite{KZ95} to be given by
the rational Ruijsenaars-Schneider model. A solution
of the spectral problem for the master \(T\)-operator in terms
of the integrable dynamics of its zeros and a connection with Bethe
equations will be given elsewhere.
As is well known,
the Ruijsenaars-Schneider model 
contains
the Calogero-Moser system of particles
as a limiting case. In this connection
let us note that a similar relation between the quantum Gaudin
model (which can be regarded as a degeneration
of quantum spin chains or vertex models
with rational $R$-matrices) and classical Calogero-Moser
system was found in \cite{MTV} from a different
reasoning.

An important remark is in order.
As is seen from (\ref{phia}), the inhomogeneity parameters
\(\xi_i\) are coordinates of the Ruijsenaars-Schneider particles
at \(t_i=0\): \(\xi_i =u_i(0)\).
Clearly, they provide only a half of the initial data
necessary to fix the dynamics. The other half is initial values
of momenta which are implicitly determined by the above mentioned
conditions on the \(\tau\)-function (\ref{master}).
Since the quantum transfer matrix for the spin chain is of
the size \(N^L\!\!\times \!\! N^L\), it
has not more than \(N^L\) different eigenvalues.
This means that the system of equations that fix the initial momenta
has only a finite number of solutions, corresponding to common
eigenstates of the transfer matrices.

\section{Concluding remarks}

Our main result in this paper is the identification of
the generating function of commuting integrals of motion in
generalized quantum spin chains (called here the
master \(T\)-operator) with the \(\tau\)-function for
integrable hierarchies of classical soliton equations.
Among other things, this result means that quantum integrable
models have a hidden free fermionic (Grassmann) structure in the auxiliary
(``horizontal'') space.
It would be very interesting to relate it
to the hidden Grassmann structure uncovered in
\cite{BJMST06,JMS08} in the quantum (``vertical'') space of quantum
integrable models. Another context where \(\tau\)-functions
of classical integrable hierarchies enter the theory of
quantum integrable
models and associated 2D lattice models of statistical mechanics
is calculation of scalar products and partition functions
with domain wall boundary conditions
\cite{Izergin,Foda09,Takasaki10}.

Many things remain to be done in order
to further clarify and generalize this classical-quantum
correspondence and, possibly, to use it for
inventing new methods to solve
both kinds of problems.
Here we list some obvious directions of further development:
\begin{itemize}
\item[--]  To construct the Bethe vectors within this formalism (which does not use the standard algebraic Bethe ansatz) and to clarify their role in the classical MKP hierarchy and its polynomial reduction according to Krichever.
\item[--] To extend the results to quantum integrable models
with trigonometric \(R\)-matrices. This is more or less straightforward
and clear how to do.
Whether the approach based on the co-derivative is applicable
is not quite clear at the moment, however.
Instead of rational solutions of the MKP
hierarchy one should consider solutions with trigonometric
dependence on the spectral parameter. The Krichever's construction
is directly extendable to this case. Zeros of the master \(T\)-operator
move as particles of the trigonometric Ruijsenaars-Schneider model.
\item[--] To extend the results to quantum integrable models
with elliptic (Belavin's) \(R\)-matrices. This is the most general
and most interesting case. Zeros of the master \(T\)-operator
move as particles of the elliptic Ruijsenaars-Schneider model.
The relevant class of MKP solutions
is a special subclass of general
elliptic (two-periodic) solutions. One of the interesting problems here
is to characterize this subclass in algebro-geometric terms through
the corresponding algebraic curves.
Elliptic solutions in general position correspond to smooth curves
which are coverings of elliptic curves. Presumably, the
solutions relevant for quantum spin chains are not of general
position and the corresponding algebraic curves are singular.

\item[--] To extend the results to supersymmetric
\(GL(N|M)\)-invariant integrable spin chains.
The general definition of the master \(T\)-operator
and the Hirota equation remain the same. The realization
through the co-derivative is known from \cite{KLT10}, and can certainly
be related to Krichever's rational \(\tau\)-function.

\item[--] The fact that the Baxter \(Q\)-operators and \(T\)-operators on
intermediate levels of the nested Bethe ansatz are contained in
the master \(T\)-operator seems to be general and not restricted to
models with rational \(R\)-matrices.
One can try to find
a general procedure to obtain these objects from the master \(T\)-operator
applicable in all the cases listed above. The example of rational
solutions suggests that such a procedure should be formulated
in terms of Baker-Akhiezer functions on singular algebraic curves.

\item[--]
A development of this work might also contribute to representation theory
of quantum affine algebras and, in particular, to the theory
of \(q\)-characters introduced by Frenkel and
Reshetikhin \cite{FR99} as a spectral parameter dependent
version of usual characters.

\item[--]
We also expect that our result can be extended to
quantum models associated with more general
Lie algebras.  This is in accordance with the fact that the
KP hierarchy has analogs for Lie algebras other than
\(A_{N-1}\) \cite{JM83}.

\item[--]
Our classical interpretation of the quantum integrability should be also applicable to the 2-dimensional integrable quantum sigma models, where the operatorial formalism is almost unexplored. The most interesting case of such applications could be the construction of an operatorial form of the AdS/CFT Y-system (and the associated T-system) \cite{Gromov:2010vb,Gromov:2009tv,Gromov:2010km,Gromov:2011cx} and clarification of its relation to the classical integrable hierarchies.  

\end{itemize}

A more ambitious goal is to suggest an alternative approach to
diagonalization of quantum transfer matrices which
would not use the Bethe ansatz at any step. This does not seem
hopeless because the quantum problem appears to be
reducible to a purely classical integrable dynamics
of Calogero-like systems of particles.

\section*{Acknowledgments}

The authors thank I.Cherednik,  I.Krichever, A.Orlov, T.Ta\-ke\-be
and P.Wiegmann for discussions.
The work of A.A. was supported in part by RFBR grant
10-02-00509, joint RFBR grant 09-02-93105-CNRS, ANR grant GranMA (BLANC-08-1-313695)
and by the ERC Starting Independent Researcher Grant StG No. 204757-TQFT.
The work of V.K. and A.Z. was supported in part by RFBR grant
11-02-01220. The work of V.K. and S.L was  also partly supported by  the ANR grants StrongInt (BLANC-SIMI-4-2011) GranMA (BLANC-08-1-313695) and by the ESF grants HOLOGRAV-09-RNP-092 and ITGP.
The work of Z.T. is supported by SFB647 ``Space-Time-Matter''.
A part of this work was done when he was at Max-Planck-Institut f\"{u}r Gravitationsphysik, Albert-Einstein-Institut.
The work of A.Z. was also partially supported
by joint RFBR grants 09-01-92437-CEa,
09-01-93106-CNRS and 10-01-92104-JSPS.
A.A. and A.Z. were partially supported
by Federal Agency for Science and Innovations of Russian Federation
under contract 02.740.11.0608.
Z.T. and A.Z. thank Ecole Normale Superieure, LPT,
where a part of this work was done, for hospitality.
V.K. and S.L. thank Humboldt-Universit\"{a}t zu Berlin,
where a part of this work was done, for hospitality.
A.A. and A.Z. thank Hausdorff Research Institute for Mathematics,
where a part of this work was done, for hospitality.

\section*{Appendix A: The generalized Wick's theorem}
\label{proof-wick}
\addcontentsline{toc}{section}{Appendix A}
\def\theequation{A\arabic{equation}}
\def\theHequation{\theequation}
\setcounter{equation}{0}

Let \(v_i =\sum_j v_{ij}\psi_j\) be
linear combinations of \(\psi_j\)'s only and
\(w_i^*=\sum_j w^{*}_{ij}\psistar_j\)
be linear combinations of \(\psistar_j\)'s only.
The following version of the Wick's theorem holds:
\begin{equation}\label{Wick1-app}
\frac{\lvacn v_1 \ldots v_m w^{*}_m \ldots w^{*}_1 G\rvacn }{\lvacn
G\rvacn }=\det_{i,j =1,\ldots , m}
\frac{\lvacn v_j w^{*}_iG\rvacn }{\lvacn
G\rvacn },
\end{equation}
where \(G\) is any group-like element. 
This statement can be proved by induction using the bilinear identity
(\ref{bilinear-fermi}).
Indeed, suppose (\ref{Wick1}) is valid
for some \(m \geq 1\) (it is trivially valid at \( m=1\)).
Set
\[
\lbr U\right |=\lvacn w_{1}^{*}, \quad
\lbr U'\right |=\lvacn v_1 v_2
\ldots v_{m+1} w_{m+1}^{*}w_{m}^{*}\ldots
w_{2}^{*},
\quad \left |V\rbr = \left |V'\rbr =\rvacn .
\]
Plugging this in the bilinear identity, we see that its
r.h.s. vanishes while the rest yields, after some
transformations:
\[
\begin{array}{c}
\lvacn G \rvacn \lvacn v_1 \ldots v_{m+1}
w^{*}_{m+1} \ldots w^{*}_{1}G\rvacn
\\ \\
=\,\, \displaystyle{\sum_{j=1}^{m+1} (-1)^{j-1}
\lvacn v_j w^{*}_{1}G\rvacn
\lvacn  v_1 \ldots \not{\!v_{j}} \ldots v_{m+1}
w^{*}_{m+1} \ldots w^{*}_{2}G\rvacn}
\end{array}
\]
(here \(\not{\!v_{j}}\) means that \(v_{j}\) is omitted).
But the r.h.s. is the expansion of the determinant for \(m+1\)
in the first column.

To show that the two determinant formulas (\ref{Wick2})
and (\ref{Wick3}) are equivalent, let us take
\begin{align}
\lbr U \right| =\lbr n+1 \right| \psi_{l}w_{i}^{*} ,
\quad
\lbr U^{\prime} \right| = \lbr n \right| ,
\quad
\left|V \rbr = \left| n \rbr ,
\quad
\left|V^{\prime} \rbr = \left| n+1 \rbr
\end{align}
in the bilinear identity \eqref{bilinear-fermi}.
Using the relations \eqref{actfock1}-\eqref{actfock2}, we see
that only one term in the r.h.s. remains:
\[
\sum_k \lbr n+1\right | \psi_l w_{i}^{*}\psi_k G \rvacn
\lvacn \psistar_k G \left |n+1 \rbr =
\lbr n+1\right | \psi_l w_{i}^{*}G\left |n+1 \rbr
\lvacn G\rvacn
\]
In the l.h.s. we plug \(w_i^*=\sum_j w^{*}_{ij}\psistar_j\)
and move \(\psi_k\) in the first factor to the left position.
Using the anti-commutation relation for the fermion operators,
and transforming the sum \(\sum_k w^{*}_{ik}\psistar_k\)
back to \(w_i^*\) in the second factor, we arrive at the
3-term identity
\begin{multline}
\lbr n \right| G \left| n \rbr
\lbr n+1 \right| \psi_{l} w_{i}^{*} G \left| n+1 \rbr
=
\lbr n+1 \right| G \left| n+1 \rbr
\lbr n\right| \psi_{l} w_{i}^{*} G \left| n \rbr
\\
+
\lbr n+1 \right| \psi_{l} G \left| n \rbr
\lbr n \right| w_{i}^{*} G \left| n+1 \rbr .
\end{multline}
Let us divide it by
\(\lbr n \right | G \left | n \rbr
\lbr n+1 \right | G \left | n+1 \rbr \), take a sum over the values
of \(n\) equal to
\(n-j,n-j+1,\dots, n-1\), and then put \(l = n-j\).
Then we  obtain the relation
\begin{multline}
\frac{\lbr n \right | \psi_{n-j} w^{*}_i G\left | n \rbr }{\lbr n \right |
G \left | n \rbr }
=
\frac{\lbr n\! -\! j\right | w^{*}_i G\left | n \! -\! j\! +\! 1
\rbr }{\lbr n\! -\! j+\!1 \right |
G \left | n\! -\! j +\! 1\rbr }
\\
+
\sum_{k=1}^{j-1}
\frac{\lbr n\! -\! k + \!1 \right | \psi_{n-j} G\left | n \! -\! k
\rbr }{\lbr n\! -\! k \right |
G \left | n\! -\! k \rbr }
\frac{\lbr n\! -\! k \right | w^{*}_i G\left | n \! -\! k \! +\! 1
\rbr }{\lbr n\! -\! k \! +\! 1\right |
G \left | n\! -\! k \! +\! 1\rbr }.
 \label{recwick3}
\end{multline}
which shows that the columns of the matrix in
\eqref{Wick2} are the linear combinations of the
columns of the matrix in \eqref{Wick3} and their determinants
are thus equal.

\section*{Appendix B: Proof of the quantum Jacobi-Trudi
formulas for the Pl\"ucker coordinates \eqref{tau4a}}
\label{proof-jacobi-trudi}
\addcontentsline{toc}{section}{Appendix B}
\def\theequation{B\arabic{equation}}
\setcounter{equation}{0}

In this appendix we prove the ``quantum Jacobi-Trudi'' formulas
(\ref{tau4a}) for the coefficients \( c_{\lambda}(n)\) (\ref{tau2}).

Let us begin with the case when \(\lambda = (\alpha |\beta )\) is a hook.
We have:
\[
\begin{array}{lll}
c_{(\alpha |\beta )}(n)&=& (-1)^{\beta +1}\lvacn \psistar_{n+\alpha}
\psi_{n-\beta -1}G\rvacn
\\ &&\\
&=& (-1)^{\beta +1}
\lbr n\! -\! \beta \! -\! 1\right | \psistar_{n-\beta -1}\ldots
\psistar_{n-1} \psistar_{n+\alpha}\psi_{n-\beta -1}G\rvacn
\\ &&\\
&=&\lbr n\! -\! \beta \! -\! 1\right | \psistar_{n-\beta}\ldots
\psistar_{n-1} \psistar_{n+\alpha}G\rvacn .
\end{array}
\]
Applying the Wick's theorem in the form (\ref{Wick3}), we get
\[
\frac{\lbr n\! -\! \beta \! -\! 1\right | \psistar_{n-\beta}\ldots
\psistar_{n-1} \psistar_{n+\alpha}G\rvacn}{\lvacn G \rvacn}=
\det_{i,j =1,\ldots , \beta +1}
\frac{\lbr n\! -\! j\right | \psistar_{\lambda_i-i+n}
G\left | n\! -\! j\! +\! 1
\rbr }{\lbr n\! -\! j\! +\! 1\right |
G \left | n\! -\! j\! +\! 1\rbr },
\]
where \(\lambda_1 =\alpha +1\), \(\lambda_2 =\lambda_3 =\ldots =
\lambda_{\beta +1}=1\). The numerator in the r.h.s. can be easily
identified with \(c_{(\lambda_i -i+j -1|0)}(n-j+1)\), so we finally
obtain
\begin{equation}\label{tau4}
\begin{array}{lll}
c_{(\alpha |\beta )}(n)&=&
\displaystyle{
\tau_n (0) \det_{i,j =1,\ldots , \lambda_1^{\prime}}
\frac{c_{(\lambda_i -i+j -1|0)}(n-j+1)}{\tau_{n-j+1}(0)}}
\\ &&\\
&=&\displaystyle{
\left ( \prod_{k=1}^{\lambda_{1}'-1}c_{\emptyset}(n-k)\right )^{-1}
\det_{i,j =1,\ldots , \lambda_1^{\prime}}
c_{\lambda_i -i+j}(n-j+1)},
\end{array}
\end{equation}
where \(c_s (n):= c_{(s-1|0)}(n)\) are
the expansion coefficients for one-row diagrams.

In a similar way, one can prove that
the determinant formulas of the Jacobi-Trudi type
hold for arbitrary Young diagrams.
Let us rearrange \eqref{tau2} as follows:
\begin{align}
c_{\lambda}(n)&=(-1)^{b(\lambda)}
\lbr n \right |
 \overrightarrow{\prod_{i=1}^{d(\lambda)}}
\psi_{n+\alpha_{i}}^{*}
\overleftarrow{\prod_{j=1}^{d(\lambda)} }\psi_{n-\beta_{j}-1}
G \rvacn
\nonumber \\
&=
(-1)^{b(\lambda)+d(\lambda)^2+\frac{(d(\lambda)-1)d(\lambda)}{2}}
\lbr n \right |
\overleftarrow{\prod_{j=1}^{d(\lambda)} }\psi_{n-\beta_{j}-1}
\overleftarrow{\prod_{i=1}^{d(\lambda)}} \psi_{n+\alpha_{i}}^{*}
G \rvacn
\nonumber  \\
&=
(-1)^{b(\lambda)+d(\lambda)+\frac{(d(\lambda)-1)d(\lambda)}{2}}
\CMPorArticle{\!\lbr n\!-\!\beta_{1}\!-\!1}{\lbr n-\beta_{1}-1} \right |
\overleftarrow{\prod_{k=1}^{\beta_{1}+1} }\psi_{n-k}^{*}
\overleftarrow{\prod_{j=1}^{d(\lambda)} }\psi_{n-\beta_{j}-1}
\overleftarrow{\prod_{i=1}^{d(\lambda)}} \psi_{n+\alpha_{i}}^{*}
G \rvacn \CMPorArticle{\!}{}.
\end{align}
By using  the
relations \(\psi_{m}^{*}  \psi_{m}=1- \psi_{m} \psi_{m}^{*} \)  and
\CMPorArticle{\(\lbr n  - \beta_{1} - 1 \right |  \psi_{m}= 0 \)
for \linebreak\({m \in \{n - \beta_{j} - 1\}_{j=1}^{d(\lambda)}} \)}{\(\lbr
 n \! -\! \beta_{1}\! -\! 1 \right |  \psi_{m}= 0 \)
for \(m \in \{n\! -\! \beta_{j}\! -\! 1\}_{j=1}^{d(\lambda)} \)}, we obtain
\begin{align}
c_{\lambda}(n)
=
\lbr n-\beta_{1}-1 \right |\!\!\!\!\!\!
\prod_{\genfrac{}{}{0pt}{}{k=1,}{ k
\notin \{\beta_{j}+1 \}_{j=1}^{d(\lambda)}}}^{\overleftarrow{\beta_{1}+1} }
\!\!\!\!\!\!\!\! \psi_{n-k}^{*}\,\,
\overleftarrow{\prod_{i=1}^{d(\lambda)}} \, \psi_{n+\alpha_{i}}^{*}
G \rvacn .
\end{align}
Taking into account the following relations for the
Frobenius notation:
\(\{i-\lambda_{i}\}_{i=1}^{d(\lambda)}=\{-\alpha_{i}\}_{i=1}^{d(\lambda)}\),
\(\{i-\lambda_{i}\}_{i=d(\lambda)+1}^{\beta_{1}+1}=\{1,2,\dots, \beta_{1}+1\} \setminus \{\beta_{i}+1 \}_{i=1}^{d(\lambda)}\),
we can rewrite this as follows:
\begin{align}
c_{\lambda}(n)
=
\lbr n-\beta_{1}-1 \right |
\overleftarrow{ \prod_{ i=1}^{\beta_{1}+1} } \psi_{n-i+\lambda_{i}}^{*}
G \rvacn .
\end{align}
Applying the Wick's theorem in the same way as
for the \(\lambda=(\alpha |\beta)\) case, we finally obtain
the desired result (\ref{tau4a}).

\section*{Appendix C: Proof of the quantum Giambelli formula}
\label{proof-giam}
\addcontentsline{toc}{section}{Appendix C}
\def\theequation{C\arabic{equation}}
\setcounter{equation}{0}

The quantum Giambelli formula was proposed
in \cite{KOS96} in relation to the analytic Bethe ansatz for
finite dimensional \(U_{q}(B_{N}^{(1)})\) modules.
It was also mentioned there
that this formula follows from Sylvester's theorem
without detailed calculations.
To make the text self-contained, we will give a
 proof of \eqref{det2} based on the CBR formula \eqref{det1}.
Here we follow
\cite[page 88, example 22]{Macdonald}, where the classical case
(without spectral parameter) is discussed.
In this section, we will use Frobenius notation for a Young
diagram mentioned in section 2.2.
We introduce the following notation: \(t_{i,j}=T_{j-i}(u-j+1)\), \(n=\lambda_{1}^{\prime}\), \(d=d(\lambda)\),
\(\phi(u-k+1)=T_{0}(u-k+1)=t_{k,k}\).
We also introduce tuples \(I=(-\alpha_{1}, \dots, -\alpha_{d},1,2,\dots, n)\),
\(J=(1,2,\dots, n)\).
We will use the following relations for the Frobenius notations:
\(\{i-\lambda_{i}\}_{i=1}^{d}=\{-\alpha_{i}\}_{i=1}^{d}\),
\(\{i-\lambda_{i}\}_{i=d+1}^{n}=\{1,2,\dots, n\} \setminus \{\beta_{i}+1 \}_{i=1}^{d}\).
Then \eqref{det1} can be written as
\begin{align}
T^{ \lambda }(u)=
 \left(\prod_{k=2}^{n}t_{k,k}\right)^{-1}
 \det_{1 \le i,j \le n}(t_{i-\lambda_{i},j})
=
 \left(\prod_{k=2}^{n}t_{k,k}\right)^{-1}
 \det_{i \in I\setminus \{\beta_{k}+1 \}_{k=1}^{d}, j \in J}(t_{i,j}).
 \label{BR-modi}
\end{align}
Next we introduce the following matrices:
\begin{align}
M_{1}&=(t_{i,j})_{(i,j) \in  I_{1} \times J},
\quad
M_{2}=(t_{i,j})_{(i,j) \in  J \times J},
\quad
U=(\delta_{i,\beta_{j}+1})_{(i,j) \in J\times J_{1} },
\\
A&=
\begin{pmatrix}
{\mathbb I}_{d\times d} & (0)_{d\times n} \\
U & M_{2}
\end{pmatrix}
,
\quad
B=
\begin{pmatrix}
(0)_{d\times d}  & M_{1} \\
U & M_{2}
\end{pmatrix},
\end{align}
where
\(I_{1}=(-\alpha_{1}, \dots, -\alpha_{d})\),
\(J_{1}=(1,2,\dots, d)\), and \({\mathbb I}_{d\times d}\) is a \(d \times d\) unit matrix and
\((0)_{d\times n}\) is a \(d \times n\) zero matrix.  Then we find
\begin{align}
 \det_{i \in I\setminus \{\beta_{k}+1 \}_{k=1}^{d}, j \in J}(t_{i,j})
 =(-1)^{\sum_{j=1}^{d} \beta_{j}+d^2} \det B.
 \label{det-tB}
\end{align}
Let us  replace \(j\)-th row of \(A\) with \(i\)-th row of \(B\) and write the determinant of
this matrix as \(c_{i,j}\).  We will consider a matrix
\(C=(c_{ij})_{1\le i,j \le d+n}\).
Taking note on the relations
\(c_{ij}=\delta_{ij} \det A\) for \(d+1 \le i,j \le d+n\) and
\(c_{ij}=0\) for \(d+1 \le i \le d+n, 1 \le j \le d\), we obtain
\begin{align}
\det_{1\le i,j \le d+n} (c_{i,j}) = \det_{1\le i,j \le d} (c_{i,j}) 
(\det A)^{n}.
 \label{det-cc}
\end{align}
We will use
\( \det A=\prod_{j=1}^{n} t_{j,j}\).
Expanding the \(j\)-th row of the determinant \(c_{i,j}\), we obtain
\begin{align}
c_{i,j}&=\sum_{k=1}^{n+d}(-1)^{j+k}b_{i,k}
 \det_{1 \le p,q \le n+d, \atop p \ne j, q \ne k} (a_{p,q})
 \\
&=\sum_{k=1}^{n+d}b_{i,k} (A^{-1})_{k,j} \det A
=(BA^{-1})_{i,j} \det A
\end{align}
Then we obtain an identity
\begin{align}
\det_{1 \le i,j \le d}(c_{i,j}) = \det B (\det A)^{d-1}.
\label{cBA}
\end{align}
We can calculate \eqref{BR-modi} for the hook diagram \(\lambda =(a+1,b)\):
\begin{align}
T^{(a+1,b)}(u) =T_{a,b}(u)=
 \left(\prod_{k=2}^{b+1}t_{k,k}\right)^{-1}
\sum_{k=1}^{b+1}(-1)^{k+1} t_{-a,k}
 \det_{1 \le p \le b, 1 \le q \le b+1, \atop q \ne k} (t_{p,q})
 \label{hook-ex}
\end{align}
For \(1 \le i,j \le d\),
we can calculate explicitly
\begin{align}
c_{i,j}&=\sum_{k=1}^{n}(-1)^{\beta_{j}+k}t_{-\alpha_{i},k}
 \det_{1 \le p,q \le n, \atop  p \ne \beta_{j}+1, q \ne k} (t_{p,q})
 \\
&=\sum_{k=1}^{\beta_{j}+1}(-1)^{\beta_{j}+k}t_{-\alpha_{i},k}
 \det_{1 \le p \le \beta_{j},
1 \le q \le \beta_{j}+1, \atop  q \ne k} (t_{p,q})
\prod_{p=\beta_{j}+2}^{n} t_{p,p}
\\
&=(-1)^{\beta_{j}+1}
 T_{\alpha_{i}, \beta_{j}} (u)
\prod_{p=2}^{n} t_{p,p} ,
\end{align}
where we used \eqref{hook-ex}.
Based on this identity, we obtain
\begin{align}
\det_{1 \le i,j \le d}(c_{i,j})= (-1)^{\sum_{j=1}^{d} \beta_{j} +d}
\det_{1 \le i,j \le d}(T_{\alpha_{i}, \beta_{j}} (u))
(\prod_{p=2}^{n} t_{p,p})^{d}
\label{cTt}
\end{align}
Hence, from \eqref{det-tB}, \eqref{BR-modi}, \eqref{cBA}, \eqref{cTt} we get
\begin{align}
T^{\lambda }(u)= (t_{1,1})^{-d+1} \det_{1 \le i,j \le d}(T_{\alpha_{i}, \beta_{j}} (u)).
 \label{giambelli}
\end{align}
This is \eqref{det2}.

\section*{Appendix D: The co-derivative}
\label{coderivative}
\addcontentsline{toc}{section}{Appendix D}
\def\theequation{D\arabic{equation}}
\setcounter{equation}{0}

In this appendix
we summarize formulas on the co-derivative introduced in
\cite{Kazakov:2007na}.

The (left) co-derivative
is the operator that acts on any function \(f\) of a group element
\(g\in GL(N)\).  Symbolically, it is defined as follows:
\begin{equation}\label{co1}
\left. \phantom{\int}
\hat D \otimes f(g)=\frac{\p}{\p \phi}\otimes f(e^{\phi \cdot e}g)
\right |_{\phi =0}
\end{equation}
Here
\footnote{In this paper, we assign the left matrix suffix to the
superscript and the right suffix to the subscript.
For example, \((i,j)\) element \(A_{ij}\) of a matrix \(A=(A_{ij})\) is written as \(A^{i}_{j}\).}
\(\phi \cdot e =\sum\limits_{\alpha , \beta =1}^{N}\phi_{\beta}^{\alpha}
e_{\alpha \beta}\) is a general element of the Lie algebra \(gl(N)\)
parametrized by \(\phi_{\beta}^{\alpha}\), with \(e_{\alpha \beta}\) being
the matrix with only one nonzero (equal to \(1\)) matrix element at the place
\((\alpha \beta )\): \((e_{\alpha \beta})^{\mu}_{\nu}=
\delta_{\alpha \mu}\delta_{\beta \nu}\).
To put this more precisely, let us start from scalar functions.
In this case, we do not need  the symbol \(\otimes \) and get:
\begin{equation}\label{co4}
\hat D \, \mbox{tr}g^n =ng^n
\end{equation}
\begin{equation}\label{co5}
\hat D  \det g =\det g \cdot {\I} ,
\end{equation}
where \( {\I}\) is \(N \times N\) unit matrix.
If \(f\) is a \( N \times N \) matrix valued function, the result of \eqref{co1} belongs  to
\(N^{2}\times N^{2}\) matrix.
In particular, if \(f(g)=g\), we have in components:
\[
\left. \phantom{int}
(\hat D \otimes g)^{i_1 i_2}_{j_1 j_2}=D^{i_1}_{j_1}g^{i_2}_{j_2}=
\frac{\p}{\p \phi_{i_1}^{j_1}}\left (e^{\phi \cdot e}
g\right )^{i_2}_{j_2}\right |_{\phi =0}=\delta_{j_1 i_2}g^{i_1}_{j_2}
\]
In invariant form, this formula can be written as
\begin{equation}\label{co2}
\hat D \otimes g ={\cal P}({\I} \otimes g)
\end{equation}
where \({\cal P}\) is the permutation operator defined
as \({\cal P}(A\otimes B)=(B\otimes A){\cal P}\)
for any \(N \times N\) matrices \(A,B\).
Explicitly, we have \( {\cal P} =\sum_{a,b=1}^{N} e_{ab} \otimes e_{ba}\).

In the main text, we considered actions of the co-derivative on
functions in \(({\mathbb C}^{N})^{\otimes L} \).
In this case, it is convenient to define the co-derivative in the following way.
Let us consider an embedding of any matrix \(A\) on \({\mathbb C}^{N}\) into
\(({\mathbb C}^{N})^{\otimes L} \),
and introduce the following notation:
\begin{align}
& A_{i}=\underbrace{\I \otimes \cdots \I}_{i-1} \otimes A \otimes
\underbrace{\I \otimes \cdots \otimes \I}_{L-i}.
\end{align}
The action of \(\hD_{i}\) on any function of \(g \in GL(N)\) in
\(({\mathbb C}^{N})^{\otimes L} \) is defined as
\begin{align}
\hD_{i} f(g)=\sum_{a,b=1}^{N}e_{ab}^{(i)}
\frac{\partial}{\partial \phi^{b}_{a}} f(e^{\phi \cdot e}g)|_{\phi=0},
 \label{co-tensor}
\end{align}
where \(e_{ab}^{(i)}=\I^{\otimes (i-1)} \otimes e_{ab} \otimes \I^{\otimes (L-i)} \).
 In components, (\ref{co-tensor}) reads
\[
\left. \phantom{int}
(\hD_{i} f(g))^{\alpha_1\ldots \alpha_L}_{\beta_1\ldots \beta_L}=
 \sum_{\gamma}\frac{\p}{\p \phi_{\alpha_i}^{\gamma}}
\Bigl (f\left (e^{\phi \cdot e}g\right )
\Bigr )^{\alpha_1\ldots \alpha_{i-1} \gamma \alpha_{i+1} \ldots \alpha_L}_{\beta_1\ldots \ldots  \beta_L }\right |_{\phi =0}
\]
For example for \(f(g)=g_{j}\), we obtain a generalization of \eqref{co2}:
\begin{align}\label{Dg}
\hD_{i} g_{j}&={\mathcal P}_{ij} g_{j} ,
\end{align}
where \({\mathcal P}_{ij} =\sum_{ab}e_{ab}^{(i)}e_{ba}^{(j)} \).
For \(i=j\) case, the permutation operator reduces to the second order Casimir, and
in this case, it is \({\mathcal P}_{ii} =\sum_{ab}e_{ab}^{(i)}e_{ba}^{(i)}=\sum_{ab}e_{aa}^{(i)}=N \I^{\otimes L} \)
since we are considering the fundamental representation.
Then  we obtain \(\hD_{i} g_{i}=N g_{i}\).
In particular for \(L=2\) case, \(\hD_{1} g_{2}={\mathcal P}_{12} g_{2} \) stands for \eqref{co2}, while
\(\hD_{2} g_{1}={\mathcal P}_{21} g_{1} \) stands for
\((\I \otimes \hD)(g \otimes \I)={\mathcal P} (g \otimes \I) \).
Now the Leibniz rule for any functions \(A(g), B(g)\)
of \(g \in GL(N)\) can be written  as
\begin{align}
\overrightarrow{\prod_{i \in I}} \hat D_{i} A(g)B(g)
=\sum_{I=J \sqcup K, J \cap K =\emptyset}
\left[
\overrightarrow{\prod_{i \in J}} \hat D_{i} A(g)
\right]
\left[
\overrightarrow{\prod_{i \in K}} \hat D_{i} B(g)
\right],
\end{align}
where \(I=( i_{1},i_{2},\dots, i_{n} )\), 
\(J=( i_{\alpha_{1}},i_{\alpha_{2}},\dots, i_{\alpha_{k}} )\), 
\(K=( i_{\beta_{1}},i_{\beta_{2}},\dots, i_{\beta_{n-k}} )\)
\( (\alpha_{1} < \dots < \alpha_{k};  
\beta_{1} < \dots < \beta_{n-k}) \)
 are, as sets, subsets of \(\{1,2,\dots,L \}\), 
which satisfy \(I=J \sqcup K, J \cap K =\emptyset\). 
The ordered product on any indexed operator \(\{A_{i}\}_{i \in I}\)
is defined as \(\overrightarrow{\prod}_{i \in I} A_{i}=A_{i_{1}}\cdots A_{i_{n}}\).
The Leibniz rule can be used, for example,
\begin{align}
\hat D_{i} g_{j}^{n}=\sum_{k=0}^{n-1}g_{j}^{k} (\hat D_{i} g_{j}) g_{j}^{n-k-1}
=\sum_{k=0}^{n-1}g_{j}^{k} ({\mathcal P}_{ij} g_{j}) g_{j}^{n-k-1}
=\sum_{k=0}^{n-1}{\mathcal P}_{ij}g_{i}^{k} g_{j}^{n-k},
 \label{D gn}
\end{align}
where \(n \in {\mathbb Z}_{\ge 0}\).
Let us introduce commutative element acting in
\(({\mathbb C}^{N})^{\otimes L}\)
\begin{align}
{\mathbf C}_{I}^{(n)}=\sum_{\sum_{i \in I } m_{i}=n,
m_{i} \in {\mathbb Z}_{\ge 0}}
\prod_{i \in I}g_{i}^{m_{i}},
\end{align}
where \(I  \subset \{1,2,\dots, L\}\), 
\(n \in {\mathbb Z}_{\ge 0}\),
\( {\mathbf C}_{I}^{(0)}=\I^{\otimes L}\) and
\( {\mathbf C}_{I}^{(n)}=0\) for \(n<0\).
Note that this commutes with any permutation indexed by the \(I\).
\eqref{D gn} can be written as
\begin{align}
\hat D_{i} g_{j}^{n}
={\mathbf C}_{\{i,j\}}^{(n-1)} \hat D_{i} g_{j}
={\mathbf C}_{\{i,j\}}^{(n-1)}  {\mathcal P}_{ij} g_{j}
=
 {\mathcal P}_{ij} {\mathbf C}_{\{i,j\}}^{(n-1)}  g_{j}
\end{align}
This can be generalized for example as
 \begin{align}
\hat D_{i} \hat D_{j} g_{k}^{n}
&={\mathbf C}_{\{i,j,k\}}^{(n-2)} (\hat D_{i} g_{j}) (\hat D_{i} g_{k}) +
{\mathbf C}_{\{i,j,k\}}^{(n-1)} \hat D_{i} \hat D_{j} g_{k}
\nonumber
\\
&={\mathbf C}_{\{i,j,k\}}^{(n-2)}   {\mathcal P}_{ij}g_{j}  {\mathcal P}_{ik} g_{k} +
{\mathbf C}_{\{i,j,k\}}^{(n-1)}  {\mathcal P}_{jk}  {\mathcal P}_{ik}g_{k}
\end{align}
Let \(S^{(n)}(J)\) be a subgroup of the permutation group over the set \(J\)
whose elements have exactly \(n\) cycles: any element
\(\sigma\) of \( S^{(n)}(J)\) has a cycle decomposition
\(\sigma= \tau_{1}\cdots \tau_{n}\) (\(\{\tau_{i}\}\) are mutually disjoint cyclic permutations on \(J\)).
Then we have the following formula
{\newcommand{\cntnt}{ \overrightarrow{\prod_{i=1}^{L }}(u_{i}+\hat D_{i}) w(x)^{\alpha}
\CMPorArticle{}{&}=\sum_{I \sqcup J=\{1,2\dots , L\}}
\prod_{i \in I}u_{i} \overrightarrow{\prod_{j \in J}} \hat D_{j} w(x)^{\alpha}
\CMPorArticle{}{\nonumber}
\\
\CMPorArticle{\qquad\qquad=\!\!}{&=}\sum_{I \sqcup J=\{1,2\dots , L\}}
\prod_{i \in I}u_{i}\CMPorArticle{\!}{}
\sum_{n=1}^{\mathrm{Card}(J)}
\alpha^{n}\CMPorArticle{\!}{}
\sum_{\sigma \in S^{(n)}(J)}\CMPorArticle{\!}{}
{\mathcal P}_{\sigma}
\prod_{k \in J}
\frac{(g_{k}x)^{\theta(k-\sigma^{-1}(k))}}{1-g_{k}x}
w(x)^{\alpha}.
\label{coderi-mult}} \CMPorArticle{\begin{multline}
\cntnt
\end{multline}
}{\begin{align}
\cntnt
\end{align}}}
Here
\({\mathcal P}_{\sigma}\) is a matrix in the space
\(({\mathbb C}^{N})^{\otimes L} \), with the matrix elements
 \[[{\mathcal P}_{\sigma}]^{i_{1}\dots i_{L}}_{j_{1}\dots j_{L}}=
\delta_{i_{1}j_{\sigma(1)}}\cdots \delta_{i_{L}j_{\sigma(L)}}.\]
The action of
 \(\sigma \in S^{(n)}(J) \) on
\(i \in \{1,2,\dots, L \} \setminus J \)
is formally defined as \(\sigma(i)=i \).
The summations are taken over all the decompositions of
\(\{1,2,\dots , L\}\) into two disjoint sets \(I\) and \(J\).
The above identity can be proven by induction in \(L\) and
using the Leibniz rule.

\paragraph{Yang-Baxter equation and commutation of co-derivatives}
\label{sec:yang-baxter-equation}

One can check that these co-derivatives obey the following commutation relation
\begin{align}
\label{eq:commuDYB}
\hat D_{i}\hat D_{j}-\hat D_{j}\hat D_{i}=
(\hat D_{i}-\hat D_{j}){\mathcal P}_{ij}
=-{\mathcal P}_{ij}(\hat D_{i}-\hat D_{j}),
\end{align}
where the last equality comes from \(\hat D_{j}{\mathcal P}_{ij} = {\mathcal P}_{ij}\hat D_{i}\).

This relation can be understood from the following Yang-Baxter
equation %
:
\begin{gather}
\label{eq:YangBaxter} 
(u_j-u_i+{\mathcal P}_{ij})~  \overrightarrow{\prod_{k=1}^{L}}
(u_k+\hD_k) = 
 \overrightarrow{\prod_{k=1}^L} (u_{\tau(k)}+\hD_{\tau(k)})~ (u_j-u_i+{\mathcal
   P}_{ij})\\
\textrm{where } \tau(k)=\left\{
  \begin{aligned}
    &i&&\textrm{if }k=j\\
    &j&&\textrm{if }k=i\\
    &k&&\textrm{otherwise}
  \end{aligned}
\right.\qquad\qquad \textrm{and }i<j\,.
\end{gather}
For instance, if we write \eqref{eq:YangBaxter} when \(L=2\), and
\(u_2=0\), then  the
coefficient of degree one in \(u_1\) gives exactly \eqref{eq:commuDYB}.
Multiplying \eqref{eq:YangBaxter} by \({\mathcal P}_{ij}\) from the left, we can rewrite \eqref{eq:YangBaxter} in the notation of the
 tensor product used in the main text: 
\begin{align}
\label{eq:YangBaxter-2} 
\left(1+(u_j-u_i){\mathcal P}_{ij}\right)~  \bigotimes_{k=1}^{L}
(u_k+\hD) = 
 \bigotimes_{k=1}^L
(u_{\tau(k)}+\hD)~ \left(1+(u_j-u_i){\mathcal P}_{ij}\right) .
\end{align}

The equation \eqref{eq:YangBaxter} can either be proven ``from
scratch'' by proving explicitly the property \eqref{eq:commuDYB} of
co-derivatives, or it can be deduced (when the co-derivatives act on
linear combinations of characters, which is sufficient for this paper)
from the usual Yang-Baxter Identity.

\section*{Appendix E: Proof of \eqref{eq:simplestMID}}
\label{simplestproof}
\addcontentsline{toc}{section}{Appendix E}
\def\theequation{E\arabic{equation}}
\setcounter{equation}{0}

The left hand side of \eqref{eq:simplestMID} is a polynomial 
of \(u_{1},u_{2},\dots, u_{L}\). The degree of this polynomial 
with respect to each \(u_{i}\) (\(1 \le i \le L\)) is \(0,1\) or \(2\). 
Taking note on this fact, we will prove \eqref{eq:simplestMID} 
using the induction with respect to the length of the chain \(L\). 
The proof that 
\eqref{eq:simplestMID}
holds for \(L=L_0\),
consists in the following three steps : 

{\bf (i)}
First we will see that the terms of degree two in one of the
  variables \(u_i\) are zero under the hypothesis that
  \eqref{eq:simplestMID} holds for a chain with length \(L=L_0-1\).

  This step is easy since the term of degree two in \(u_j\) is obtained 
  by the \CMPorArticle{}{\linebreak} substitutions \(\hat {\cal D}_{L_0}(u)\rightsquigarrow 
\bigotimes_{i=1}^{j-1}(u_i +\hD)\otimes u_j \I \otimes \bigotimes_{i=j+1}^{{L_0}}(u_i +\hD)
\) and  \CMPorArticle{}{\linebreak}  \(\hat {\cal D}_{L_0}(u+1)\rightsquigarrow 
\bigotimes_{i=1}^{j-1}(u_i+1 +\hD)\otimes u_j \I \otimes \bigotimes_{i=j+1}^{{L_0}}(u_i+1 +\hD)
\). We see that the operator obtained from the action of these
co-derivatives is a trivial operator on the \(j^{\textrm{th}}\) space
times an operator acting on a spin chain with \(L_0-1\) spins
(corresponding to all sites except the site \(j\)). We can then
recognize that the coefficient of \(u_j^2\) in the left-hand side
of \eqref{eq:simplestMID} at \(L=L_0\) is equal to the left-hand side of
\eqref{eq:simplestMID} at \(L=L_0-1\) (up to a relabelling of the spaces), which is zero by the recurrence hypothesis.

{\bf (ii)}
The other terms have degree one in some variables \(u_{i_1},
  u_{i_2}, \ldots, u_{i_n}\) and degree zero in the other variables \(u_{j_1},
  u_{j_2}, \ldots, u_{j_{L_0-n}}\) (where \(\{j_1,\ldots,{j_{L_0-n}}\}\)
  denotes the complement of \(\{i_1,\ldots,{i_{n}}\}\)). We can show
  that all the terms where \(1 \in \{i_1,\ldots,{i_{n}}\}\) vanish,
  assuming that \eqref{eq:simplestMID} holds for a chain with length
  \(L=L_0-1\).

To perform this step, we have to investigate the coefficient of degree
one in \(u_1\), in each term of the left-hand-side of
\eqref{eq:simplestMID}. For instance, in the operator 
\begin{equation*}
\left (\hat {\cal D}_{L_0}(u+1)w(z_1^{-1}) \!\right )
\left (\hat {\cal D}_{L_0}(u)w(z_2^{-1})\! \right )\,,
\end{equation*}
the term of degree one in \(u_1\) is given by
\begin{multline}
\left ((\I+\hD)\otimes \bigotimes_{i=2}^{{L_0}}(u_i+1+\hD)w(z_1^{-1})
\right ) 
\left (u_1 \I\otimes \bigotimes_{i=2}^{{L_0}}(u_i+\hD) w(z_2^{-1})\! \right
) \\+
\left (u_1 \I\otimes \bigotimes_{i=2}^{{L_0}}(u_i+1+\hD)w(z_1^{-1})
\right ) 
\left (\hD\otimes \bigotimes_{i=2}^{{L_0}}(u_i+\hD) w(z_2^{-1})\! \right
) \\=
u_1(\I+\hD)\otimes\left[ \left (\bigotimes_{i=2}^{{L_0}}(u_i+1+\hD)w(z_1^{-1})
\right ) 
\left ( \bigotimes_{i=2}^{{L_0}}(u_i+\hD) w(z_2^{-1}) \right
)\right]\,,
\end{multline}
where the right-hand-side is obtained by the Leibniz rule.
The same analysis \CMPorArticle{}{\linebreak} is easily performed for the other terms \(\left (\hat {\cal D}_{L_0}(u+1)w(z_2^{-1})\! \right )\!
\left (\hat {\cal D}_{L_0}(u)w(z_1^{-1}) \! \right )\) and \CMPorArticle{}{\linebreak} \(\left (\hat {\cal D}_{L_0}(u+1)w(z_1^{-1})w(z_2^{-1})\right )
\left (\hat {\cal D}_{L_0}(u)\right )\), (appearing in \eqref{eq:simplestMID}) and we obtain that the coefficient of
degree one in \(u_1\) is obtained by the action of \((\I+\hD)\)
on the left-hand-side of the
identity with \(L=L_0-1\) spins. %
Hence
the recurrence hypothesis shows that they vanish.

{\bf (iii)}
Finally, we will deduce the vanishing of the  term of
  degree one in the  \CMPorArticle{}{\linebreak} variables \(u_{i_1},
  u_{i_2}, \ldots, u_{i_n}\) and degree zero in the  variables \(u_{j_1},
  u_{j_2}, \ldots, u_{j_{{L_0}-n}}\), where 
  \CMPorArticle{}{\linebreak}  \(\{j_1,\ldots,{j_{{L_0}-n}}\}\)
  denotes the complement of \(\{i_1,\ldots,{i_{n}}\}\). We will prove 
  this vanishing by recurrence over \(n\) (i.e. we will assume that it
  holds for \(n-1\), and we will deduce it for \(n\)). 

  First, let us assume that the  terms with degree one in \(u_k\) 
  and degree zero in
  \(u_{k+1}\) (i.e. the terms where
\(k\in \{i_1,\ldots,{i_{n}}\}\) and 
\(k+1 \in \{j_1,\ldots,{j_{{L_0}-n}}\}\)) do vanish,
  and let us deduce that the term with 
degree zero in \(u_k\) and degree one in
  \(u_{k+1}\) (i.e. the terms where
\(k\in \{j_1,\ldots,{j_{{L_0}-n}}\}\) and \(k+1 \in
\{i_1,\ldots,{i_{n}}\}\)) also vanish. To this end, we can use a
Yang-Baxter equality \eqref{eq:YangBaxter-2} for each term in the
identity \eqref{eq:simplestMID}. For instance for the first term
(i.e. for \(z_2 \!\left (\hat {\cal D}_{L_0}(u+1)w(z_1^{-1}) \!\right )\!
\left (\hat {\cal D}_{L_0}(u)w(z_2^{-1})\! \right )\)) we
write
\begin{multline}
z_2
\left(1\!+\! (u_{k+1}-u_k)\: {\mathcal P}_{k,
      k+1}\right)\left(\bigotimes_{i=1}^{{L_0}}(u_i\!+\!1 \!+\!\hD) %
    w(z_1^{-1})\right)
\left(\bigotimes_{i=1}^{{L_0}}(u_i\!+\!\hD) w(z_2^{-1}) \right)
\\%
- 
z_2
\left(%
  \bigotimes_{i=1}^{{L_0}}(u_{\sigma(i)}\!+\!1 \!+\!\hD)
\CMPorArticle{    w(z_1^{-1})\!\right) \!
\left(\bigotimes_{i=1}^{{L_0}}(u_{\sigma(i)}\!+\!\hD) w(z_2^{-1})\! \right)\!}
{    w(z_1^{-1})\right) 
\left(\bigotimes_{i=1}^{{L_0}}(u_{\sigma(i)}\!+\!\hD) w(z_2^{-1}) \right)}
 \left(1\!+\! (u_{k+1}-u_k)\: {\mathcal P}_{k,
      k+1}\right)\\=0
\label{eq:YB2}
\end{multline}
\begin{equation}
  \textrm{where}\qquad\qquad
 \sigma(i)=\left\{
  \begin{aligned}
    &k+1&&\textrm{if }i=k\\
    &k&&\textrm{if }i=k+1\\
    &i&&\textrm{otherwise}
  \end{aligned}
\right.
\end{equation}
If we sum this Yang-Baxter identity for each term in
\eqref{eq:simplestMID} 
and we keep only the terms of degree one in the variables \(u_{i_1},
  u_{i_2}, \ldots, u_{i_n}\) and degree zero in the variables \(u_{j_1},
  u_{j_2}, \ldots, u_{j_{{L_0}-n}}\), then we get an identity for the sum
  of the following terms
  \begin{itemize}
  \item First, the terms which contain \({\mathcal P}_{k,
      k+1}\). As the total degree in \(u_k\) is one and the total degree
    in \(u_{k+1}\) is zero, these terms do not have any occurrence of
    \(u_k\) or \(u_{k+1}\) in the factors
 containing co-derivatives. One can easily see that (up to multiplication by \({\mathcal P}_{k,
      k+1}\)) these terms are exactly the terms of
    \eqref{eq:simplestMID} with degree zero in the 
    variables \(u_{j_1},
  u_{j_2}, \ldots, u_{j_{L_0-n}}\) and \(u_{k}\), and degree one in the other
  variables. By the hypothesis of the recurrence over \(n\), these terms
  vanish. 
\item Next, the terms of the first line in \eqref{eq:YB2} which do not
  contain \({\mathcal P}_{k, k+1}\). If we sum them for the three terms
  of \eqref{eq:simplestMID}, we exactly obtain the terms of degree one
  in the variables \(u_{i_1}, 
  u_{i_2}, \ldots, u_{i_n}\) and degree zero in the variables \(u_{j_1},
  u_{j_2}, \ldots, u_{j_{{L_0}-n}}\) of \eqref{eq:simplestMID} (with \(k\in
  \{i_1,\ldots,{i_{n}}\}\) and \(k+1 \in
  \{j_1,\ldots,{j_{{L_0}-n}}\}\)). By hypothesis these terms do vanish.
\item Finally, the terms of the second line in \eqref{eq:YB2} which do not
  contain \({\mathcal P}_{k, k+1}\). If we sum them for the three terms
  of \eqref{eq:simplestMID}, we exactly obtain the coefficient of
  degree one in  \(u_{k+1}\) and degree zero in \(u_k\) (and with degree
  one, resp zero in the other variables \(i_m\) resp \(j_m\)). These last
  terms vanish (because the sum of expressions of the form
  \eqref{eq:YB2} is zero, and because the other terms vanish). 
Here we used the fact that the coefficient of 
\(u_{k}^{1}u_{k+1}^{0}\) in a polynomial 
\(f(u_{\sigma(1)},u_{\sigma(2)},\dots, u_{\sigma(L)})\) 
coincides with the coefficient of \(u_{k}^{0}u_{k+1}^{1}\) in a polynomial 
\(f(u_{1},u_{2},\dots, u_{L})\), where 
\(f(u_{1},u_{2},\dots, u_{L})\) is any polynomial in 
\(u_{1},u_{2},\dots, u_{L}\). 
\end{itemize}
  This shows that 
if the  terms with degree one in \(u_k\) and degree zero in
  \(u_{k+1}\) (i.e. the terms where
\(k\in \{i_1,\ldots,{i_{n}}\}\) and \(k+1 \in \{j_1,\ldots,{j_{{L_0}-n}}\}\)) do vanish,
then the term with degree 
 zero in \(u_k\) and degree one in
  \(u_{k+1}\) (i.e. the terms where
\(k\in \{j_1,\ldots,{j_{{L_0}-n}}\}\) and 
\(k+1 \in \{i_1,\ldots,{i_{n}}\}\)) 
also vanish.

The same result is easily obtained if replace \(k+1\) with \(k-1\), and by
iterating this argument, we obtain that the  term of
  degree one in any set of \(n\) variables \(u_{i_1},
  u_{i_2}, \ldots, u_{i_n}\) (and degree zero in the other variables)
  do vanish. Indeed, the above argument allows to reduce these term
  where \(1 \in \{i_1,\ldots,{i_{n}}\}\), treated above. This proves the
  equation \eqref{eq:simplestMID} by 
  recurrence.

\end{document}